\newcommand\SEKIusersusepackages{
\usepackage{named,makeidx}
\usepackage{url}
\usepackage{amssymb}
\usepackage{pstricks}
\usepackage{graphicx}
\usepackage{makeidx}
}
\newcommand\SEKIREVIEWSTATUS{1}
\newcommand\SEKIREVIEWER{%
\anellisname
\\Indiana University -- Purdue University Indianapolis
\\E-mail: {\tt irvanellis@lycos.com}
\\WWW: \url{http://irvinganellis.blogspot.com}
\\
\\\bussottiname
\\Centro Studi Enriques, Livorno
\\E-mail: {\tt paolo.bussotti@alice.it}
\\
\\\paulsonname
\\Computer Laboratory, University of Cambridge, UK
\\E-mail: {\tt lp15@cam.ac.uk}
\\WWW: \url{http://www.cl.cam.ac.uk/~lp15}
}
\newcommand\SEKIDOCUMENTCLASS{0}
\newcommand\SEKIYEAR{2009}
\newcommand\SEKINUMBEROFYEAR{01}
\newcommand\SEKITYPE{0}
\newcommand\SEKIPUBLISHEDTYPE{0}
\title{Lectures on\\\herbrandname\\as a Logician}
\author{\wirthname, \siekmannname,\\\benzmuellername, \autexiername}
\date{\small\mbox{}\\\mbox{}\\
\SEKIedition\smallfootroom
Submitted: \ December\,13, 2007
\\\mbox{}\\[-.9ex]
First Printing: \ February 25, 2009
\\\mbox{}\\[-.9ex]
\noteref{new note} and 
\figurefs{figure herbrand two}{figure herbrand three}
added,
\\\herbrand's Bibliography corrected and extended,
\\References corrected and extended: \ June 30, 2011%
\\\mbox{}\\[-.9ex]
Thoroughly updated, corrected, and improved: \ \May\,24, 2014%
}%

\makeindex
\input headerhot
\mathcommand\ident[1]{\mathsf{#1}}
\newcommand\plussymbol  {\ident{+}}
\newcommand\minussymbol {\ident{-}}
\newcommand\dividesymbol{\ident{/}}
\newcommand\timessymbol {\ident{*}}

\newcommand\nat     {\ident{nat}}

\newcommand\set     {\ident{set}}

\newcommand\naturalssymbol{\ident{naturals}}
\newcommand\gensymsymbol{\ident{gensym}}
\mathcommand\mbpsymbol{\ident{m\hspace{-0.055em}b\hspace{-0.045em}p}}

\newcommand\csymbol     {\ident c}
\newcommand\esymbol     {\ident e}
\newcommand\fsymbol     {\ident f}
\newcommand\gsymbol     {\ident g}
\newcommand\hsymbol     {\ident h}
\newcommand\ksymbol     {\ident k}
\newcommand\psymbol     {\ident p}
\newcommand\ssymbol     {\ident s}
\newcommand\Everysymbol {\ident{Every}}
\newcommand\Permsymbol {\ident{Perm}}
\newcommand\RExistssymbol{\ident{Rexists}}
\newcommand\invertsymbol{\ident{invert}}
\newcommand\invsymbol{\ident{inv}}
\newcommand\abssymbol   {\ident{abs}}
\newcommand\cnssymbol   {\ident{cons}}
\mathcommand\cnsindexsymbol[1]{\ident{cons}_{#1}}
\newcommand\carsymbol   {\ident{car}}

\newcommand\cdrsymbol   {\ident{cdr}}
\newcommand\lengthsymbol{\ident{length}}
\newcommand\sizesymbol{\ident{size}}
\newcommand\dlsymbol    {\ident{dl}}
\newcommand\dloncesymbol{\ident{delfirst}}
\newcommand\rcsymbol    {\ident{rc}}
\newcommand\brsymbol    {\ident{br}}
\newcommand\revtailsymbol{\ident{revtail}}
\newcommand\revsymbol{\ident{rev}}
\newcommand\appendsymbol {\ident{append}}
\newcommand\zeropredicatesymbol{\ident{zerop}}
\newcommand\eqsymbol        {\ident{eq}}
\newcommand\ifthensymbol    {\mbox{\ident{If{}Then}}}
\newcommand\ifthenelsesymbol{\mbox{\ident{If{}ThenElse}}}
\mathcommand\eqindexsymbol        [1]{\eqsymbol        _{#1}}
\mathcommand\ifthenindexsymbol    [1]{\ifthensymbol    _{#1}}
\mathcommand\ifthenelseindexsymbol[1]{\ifthenelsesymbol_{#1}}
\newcommand\orsymbol    {\ident{or}}
\newcommand\andsymbol   {\ident{and}}
\newcommand\leqsymbol   {\ident{leq}}
\newcommand\lessymbol   {\ident{less}}
\newcommand\lexlessymbol{\ident{lexless}}
\newcommand\lexlimlessymbol{\ident{lexlimless}}
\newcommand\lexsymbol   {\ident{lex}}
\newcommand\acksymbol   {\ident{ack}}
\newcommand\switchsymbol{\ident{switch}}
\newcommand\swatchsymbol{\ident{swatch}}
\newcommand\diveinssymbol{\ident{div1}}
\newcommand\divzweisymbol{\ident{div2}}
\newcommand\divrestsymbol{\ident{divrest}}
\newcommand\diveinstailsymbol{\ident{div1tail}}
\newcommand\divzweitailsymbol{\ident{div2tail}}
\newcommand\remsymbol{\ident{rem}}
\newcommand\divsymbol{\ident{div}}

\newcommand\turingmachinesymbol{\ident T}
\newcommand\terminatespsymbol  {\ident{terminatesp}}
\newcommand\statesymbol        {\ident{state}}
\newcommand\cmdsymbol          {\ident{cmd}}
\newcommand\nthsymbol          {\ident{nth}}
\newcommand\doublesymbol       {\ident{double}}

\newcommand\ppsymbol           {\ident{p}}
\newcommand\qpsymbol           {\ident{q}}
\newcommand\Epsymbol           {\ident{E}}
\newcommand\Ppsymbol           {\ident{P}}
\newcommand\Qpsymbol           {\ident{Q}}
\newcommand\Marriessymbol      {\ident{Marries}}
\newcommand\Lovessymbol        {\ident{Loves}}
\newcommand\StolenBysymbol     {\ident{StolenBy}}
\newcommand\Humansymbol        {\ident{Human}}
\newcommand\Evensymbol         {\ident{Even}}
\newcommand\Oddsymbol          {\ident{Odd}}
\newcommand\Primesymbol        {\ident{Prime}}
\newcommand\EveryPairsymbol   {\ident{EveryPair}}
\newcommand\Givesymbol         {\ident{Give}}
\newcommand\Fathersymbol       {\ident{Father}}
\newcommand\Elephantpsymbol    {\ident{Elephant}}
\newcommand\Flowerpsymbol    {\ident{Flower}}
\newcommand\Germanpsymbol      {\ident{German}}
\newcommand\Bicyclepsymbol     {\ident{Bicycle}}
\newcommand\Hugepsymbol        {\ident{Huge}}
\newcommand\Animalpsymbol      {\ident{Animal}}
\newcommand\Malepsymbol        {\ident{Male}}
\newcommand\Boypsymbol         {\ident{Boy}}
\newcommand\Girlpsymbol        {\ident{Girl}}
\newcommand\Femalepsymbol      {\ident{Female}}
\newcommand\Roundpsymbol       {\ident{Round}}
\newcommand\Quadrangularpsymbol{\ident{Quadrangular}}
\newcommand\Metpsymbol         {\ident{Met}}
\newcommand\Kissedpsymbol      {\ident{Kissed}}
\newcommand\Bishopsymbol       {\ident{Bishop}}
\newcommand\mindexsymbol[1]{\existsvari w{#1}}

\newcommand\nonnegpsymbol      {\ident{nonnegp}}
\newcommand\wellsymbol         {\ident{well}}
\newcommand\welltailsymbol     {\ident{welltail}}
\newcommand\varsymbol          {\ident{var}}
\newcommand\aritysymbol        {\ident{arity}}

\newcommand\whilesymbol        {\ident{while}}

\newcommand\nullsymbol         {\ident{null}}
\newcommand\hdsymbol           {\ident{hd}}
\newcommand\tlsymbol           {\ident{tl}}
\newcommand\insymbol           {\ident{in}}
\newcommand\applysymbol        {\ident{app}}
\newcommand\termsymbol         {\ident{term}}
\newcommand\russellsymbol      {\ident{russell}}
\newcommand\sqrtindordsymbol[1]{\ident{sqrtio#1}}
\mathcommand\tightim{\longrightarrow}
\mathcommand\im{\ \tightim\ }
\mathcommand\rs{\:\rulesugar\:\:}
\mathcommand\rulesugar{\longleftarrow}

\mathcommand\doublepp[1]      {\doublesymbol   \beginargs{#1}\allargs}
\mathcommand\aritypp[1]      {\aritysymbol   \beginargs{#1}\allargs}
\mathcommand\lengthpp[1]      {\lengthsymbol   \beginargs{#1}\allargs}
\mathcommand\sizepp[1]      {\sizesymbol   \beginargs{#1}\allargs}
\mathcommand\wellpp[1]      {\wellsymbol   \beginargs{#1}\allargs}
\mathcommand\welltailpp[1]      {\welltailsymbol   \beginargs{#1}\allargs}
\mathcommand\varpp[1]      {\varsymbol   \beginargs{#1}\allargs}
\mathcommand\rempp[2]    {\remsymbol\beginargs{#1}\separgs{#2}\allargs}
\mathcommand\divpp[2]    {\divsymbol\beginargs{#1}\separgs{#2}\allargs}
\mathcommand\divrestpp[2]    {\divrestsymbol\beginargs{#1}\separgs{#2}\allargs}
\mathcommand\diveinspp[2]    {\diveinssymbol\beginargs{#1}\separgs{#2}\allargs}
\mathcommand\divzweipp[3]    {\divzweisymbol\beginargs{#1}\separgs{#2}
\separgs{#3}\allargs}
\mathcommand\diveinstailpp[4]    {\diveinstailsymbol\beginargs{#1}\separgs{#2}
\separgs{#3}\separgs{#4}\allargs}
\mathcommand\divzweitailpp[6]    {\divzweitailsymbol\beginargs{#1}\separgs{#2}
\separgs{#3}\separgs{#4}\separgs{#5}\separgs{#6}\allargs}
\mathcommand\mbppp[2]         {\mbpsymbol   \beginargs{#1}\separgs{#2}\allargs}
\mathcommand\revpp[1]     
{\revsymbol\beginargs{#1}\allargs}
\mathcommand\revppi[2]     
{\revsymbol^{#1}\beginargs{#2}\allargs}
\mathcommand\revtailpp[2]     
{\revtailsymbol\beginargs{#1}\separgs{#2}\allargs}
\mathcommand\revtailppi[3]
{\revtailsymbol^{#1}\beginargs{#2}\separgs{#3}\allargs}
\mathcommand\Permpp[2]     
{\Permsymbol\beginargs{#1}\separgs{#2}\allargs}
\mathcommand\Permppi[3]
{\Permsymbol^{#1}\beginargs{#2}\separgs{#3}\allargs}
\mathcommand\appendpp[2]      
{\appendsymbol \beginargs{#1}\separgs{#2}\allargs}
\mathcommand\appendppi[3]      
{\appendsymbol^{#1}\beginargs{#2}\separgs{#3}\allargs}
\mathcommand\Everypp[2]      
{\Everysymbol \beginargs{#1}\separgs{#2}\allargs}
\mathcommand\RExistspp[1]      
{\RExistssymbol \beginargs{#1}\allargs}
\mathcommand\appendlongpp[2]      
{\appendsymbol\left(\begin{array}{@{}l@{}}{#1}\separgs\\{#2}\end{array}\right)}
\mathcommand\cnspp[2]         {\cnssymbol   \beginargs{#1}\separgs{#2}\allargs}
\mathcommand\cnsppi[3]       {\cnssymbol^{#1}\beginargs{#2}\separgs{#3}\allargs}
\mathcommand\cnsindexpp[3]
{\cnsindexsymbol{#1}\beginargs{#2}\separgs{#3}\allargs}
\mathcommand\dlpp[2]          {\dlsymbol    \beginargs{#1}\separgs{#2}\allargs}
\mathcommand\dloncepp[2]      {\dloncesymbol\beginargs{#1}\separgs{#2}\allargs}
\mathcommand\dlonceppi[3]{\dloncesymbol^{#1}\beginargs{#2}\separgs{#3}\allargs}
\mathcommand\rcpp[2]          {\rcsymbol    \beginargs{#1}\separgs{#2}\allargs}
\mathcommand\brpp[2]          {\brsymbol    \beginargs{#1}\separgs{#2}\allargs}
\mathcommand\orpp[2]          {\orsymbol    \beginargs{#1}\separgs{#2}\allargs}
\mathcommand\andpp[2]         {\andsymbol   \beginargs{#1}\separgs{#2}\allargs}
\mathcommand\shortcnspp[2]    {\csymbol     \beginargs{#1}\separgs{#2}\allargs}
\mathcommand\tightshortcnspp[2]
{\csymbol\beginargs{#1}\tightsepargs{#2}\allargs}
\mathcommand\spp[1]           {\ssymbol     \beginargs{#1}\allargs}
\mathcommand\sppiterated[2]   {\ssymbol^{#1}\beginargs{#2}\allargs}
\mathcommand\sqrtindordpp[3]
                       {\sqrtindordsymbol{#1}\beginargs{#2}\separgs{#3}\allargs}
\mathcommand\ppp[1]           {\psymbol     \beginargs{#1}\allargs}
\mathcommand\pppiterated[2]   {\psymbol^{#1}\beginargs{#2}\allargs}
\mathcommand\zeropp           {\ident 0}
\mathcommand\Julietpp         {\ident{Juliet}}
\mathcommand\Romeopp          {\ident{Romeo}}
\mathcommand\Ipp              {\ident I}
\mathcommand\onepp            {\ident1}
\mathcommand\twopp            {\ident2}
\mathcommand\threepp          {\ident3}
\mathcommand\invertpp[1]      {\invertsymbol\beginargs{#1}\allargs}
\mathcommand\invpp[1]         {\invsymbol\beginargs{#1}\allargs}
\mathcommand\abspp[1]         {\abssymbol\beginargs{#1}\allargs}
\mathcommand\naturalspp[1]    {\naturalssymbol\beginargs{#1}\allargs}
\mathcommand\gensympp[1]      {\gensymsymbol\beginargs{#1}\allargs}
\mathcommand\nilpp            {\ident{nil}}
\mathcommand\falsepp          {\ident{false}}
\mathcommand\truepp           {\ident{true}}
\mathcommand\FALSEpp          {\ident{FALSE}}
\mathcommand\TRUEpp           {\ident{TRUE}}
\mathcommand\UNDEFpp          {\ident{UNDEF}}
\mathcommand\weirdppp         {\ident{weirdp}}
\mathcommand\ambigppp         {\ident{ambigp}}
\mathcommand\zeropredicatepp[1]{\zeropredicatesymbol\beginargs{#1}\allargs}
\mathcommand\cppeins       [1]{\csymbol     \beginargs{#1}\allargs}
\mathcommand\cppzwei       [2]{\csymbol\beginargs{#1}\separgs{#2}\allargs}
\mathcommand\eppeins       [1]{\esymbol     \beginargs{#1}\allargs}
\mathcommand\fppeins       [1]{\fsymbol     \beginargs{#1}\allargs}
\mathcommand\fppeinsindex  [2]{\fsymbol_{#1}\beginargs{#2}\allargs}
\mathcommand\fppeinsiterated[2]{\fsymbol^{#1}\beginargs{#2}\allargs}
\mathcommand\gppeins       [1]{\gsymbol     \beginargs{#1}\allargs}
\mathcommand\gppzwei       [2]{\gsymbol     \beginargs{#1}\separgs{#2}\allargs}
\mathcommand\hppeins       [1]{\hsymbol     \beginargs{#1}\allargs}
\mathcommand\kppeins       [1]{\ksymbol     \beginargs{#1}\allargs}
\mathcommand\appzero          {\ident a}
\mathcommand\bppzero          {\ident b}
\mathcommand\cppzero          {\ident c}
\mathcommand\dppzero          {\ident d}
\mathcommand\eppzero          {\ident e}
\mathcommand\eqindexpp[3]{\eqindexsymbol{#1}\beginargs{#2}\separgs{#3}\allargs}
\mathcommand\ifthenindexpp
[3]{\ifthenindexsymbol{#1}\beginargs{#2}\separgs{#3}\allargs}
\mathcommand\ifthenelseindexpp
[4]{\ifthenelseindexsymbol{#1}\beginargs{#2}\separgs{#3}\separgs{#4}\allargs}
\mathcommand\eqpp[2]{\eqsymbol\beginargs{#1}\separgs{#2}\allargs}
\mathcommand\leqpp[2]{\leqsymbol\beginargs{#1}\separgs{#2}\allargs}
\mathcommand\lespp[2]{\lessymbol\beginargs{#1}\separgs{#2}\allargs}
\mathcommand\lexlespp[2]{\lexlessymbol\beginargs{#1}\separgs{#2}\allargs}
\mathcommand\lexlimlespp[3]
               {\lexlimlessymbol\beginargs{#1}\separgs{#2}\separgs{#3}\allargs}
\mathcommand\lexpp[3]{\lexsymbol\beginargs{#1}\separgs{#2}\separgs{#3}\allargs}
\mathcommand\ackpp[2]{\acksymbol\beginargs{#1}\separgs{#2}\allargs}
\mathcommand\switchpp[1]{\switchsymbol\beginargs{#1}\allargs}
\mathcommand\swatchpp[1]{\swatchsymbol\beginargs{#1}\allargs}
\mathcommand\whilepp[2]{\whilesymbol\beginargs{#1}\separgs{#2}\allargs}
\mathcommand\nullpp[1]{\nullsymbol\beginargs{#1}\allargs}
\mathcommand\nullppiterated[2]{\nullsymbol^{#1}\beginargs{#2}\allargs}
\mathcommand\hdpp[1]{\hdsymbol\beginargs{#1}\allargs}
\mathcommand\hdppiterated[2]{\hdsymbol^{#1}\beginargs{#2}\allargs}
\mathcommand\carpp[1]{\carsymbol\beginargs{#1}\allargs}
\mathcommand\cdrpp[1]{\cdrsymbol\beginargs{#1}\allargs}
\mathcommand\tlpp[1]{\tlsymbol\beginargs{#1}\allargs}
\mathcommand\tlppiterated[2]{\tlsymbol^{#1}\beginargs{#2}\allargs}
\mathcommand\inpp[2]{\insymbol\beginargs{#1}\separgs{#2}\allargs}
\mathcommand\inppiterated[3]{\insymbol^{#1}\beginargs{#2}\separgs{#3}\allargs}
\mathcommand\applypp[2]{\applysymbol\beginargs{#1}\separgs{#2}\allargs}
\mathcommand\applyppiterated
[3]{\applysymbol^{#1}\beginargs{#2}\separgs{#3}\allargs}
\mathcommand\termpp[2]{\termsymbol\beginargs{#1}\separgs{#2}\allargs}
\mathcommand\setpp[1]{\set\beginargs{#1}\allargs}
\mathcommand\russellpp[1]{\russellsymbol\beginargs{#1}\allargs}

\mathcommand\Tpp[6]{\turingmachinesymbol\beginargs{#1}\separgs{#2}\separgs
{#3}\separgs{#4}\separgs{#5}\separgs{#6}\allargs}
\mathcommand\Tppseven[7]{\turingmachinesymbol\beginargs{#1}\separgs{#2}\separgs
{#3}\separgs{#4}\separgs{#5}\separgs{#6}\separgs{#7}\allargs}
\mathcommand\foreverppp[6]{\ident{foreverp}\beginargs{#1}\separgs{#2}\separgs
{#3}\separgs{#4}\separgs{#5}\separgs{#6}\allargs}
\mathcommand\terminatesppp[6]{\terminatespsymbol\beginargs{#1}\separgs
{#2}\separgs{#3}\separgs{#4}\separgs{#5}\separgs{#6}\allargs}
\mathcommand\terminatespppone[1]{\terminatespsymbol \beginargs{#1}\allargs}
\mathcommand\statepp
[3]{\statesymbol\beginargs{#1}\separgs{#2}\separgs{#3}\allargs}
\mathcommand\tightstatepp
[3]{\statesymbol\beginargs{#1}\tightsepargs{#2}\tightsepargs{#3}\allargs}
\mathcommand\cmdpp
[3]{\cmdsymbol  \beginargs{#1}\separgs{#2}\separgs{#3}\allargs}
\mathcommand\tightcmdpp
[3]{\cmdsymbol  \beginargs{#1}\tightsepargs{#2}\tightsepargs{#3}\allargs}
\mathcommand\stoppp           {\ident{stop}}
\mathcommand\leftpp           {\ident{left}}
\mathcommand\rightpp          {\ident{right}}
\mathcommand\nthpp         [2]{\nthsymbol  \beginargs{#1}\separgs{#2}\allargs}
\mathcommand\pppp          [1]{\ppsymbol\beginargs{#1}            \allargs}
\mathcommand\qppp          [2]{\qpsymbol\beginargs{#1}\separgs{#2}\allargs}
\mathcommand\Eppp          [1]{\Epsymbol\beginargs{#1}            \allargs}
\mathcommand\Epppzwei      [2]{\Epsymbol\beginargs{#1}\separgs{#2}\allargs}
\mathcommand\Pppp          [1]{\Ppsymbol\beginargs{#1}            \allargs}
\mathcommand\Ppppeinsindex [2]{\Ppsymbol_{#1}\beginargs{#2}\allargs}
\mathcommand\Ppppdrei      
[3]{\Ppsymbol\beginargs{#1}\separgs{#2}\separgs{#3}\allargs}
\mathcommand\Ppppvier
[4]{\Ppsymbol\beginargs{#1}\separgs{#2}\separgs{#3}\separgs{#4}\allargs}
\mathcommand\Qppp          [2]{\Qpsymbol\beginargs{#1}\separgs{#2}\allargs}
\mathcommand\Qpppeins      [1]{\Qpsymbol\beginargs{#1}\allargs}
\mathcommand\Qpppeinsindex [2]{\Qpsymbol_{#1}\beginargs{#2}\allargs}
\mathcommand\Qpppdrei      
[3]{\Qpsymbol\beginargs{#1}\separgs{#2}\separgs{#3}\allargs}
\mathcommand\Fatherpp      [2]{\Fathersymbol\beginargs{#1}\separgs{#2}\allargs}
\mathcommand\Marriespp     [2]{\Marriessymbol\beginargs{#1}\separgs{#2}\allargs}
\mathcommand\Lovespp       [2]{\Lovessymbol\beginargs{#1}\separgs{#2}\allargs}
\mathcommand\StolenBypp    [2]
{\StolenBysymbol\beginargs{#1}\separgs{#2}\allargs}
\mathcommand\Humanpp       [1]{\Humansymbol\beginargs{#1}\allargs}
\mathcommand\Evenpp        [1]{\Evensymbol\beginargs{#1}\allargs}
\mathcommand\Evenppi       [2]{\Evensymbol^{#1}\beginargs{#2}\allargs}
\mathcommand\Oddpp         [1]{\Oddsymbol\beginargs{#1}\allargs}
\mathcommand\Primepp       [1]{\Primesymbol\beginargs{#1}\allargs}
\mathcommand\EveryPairpp  [2]{\EveryPairsymbol\beginargs{#1}\separgs
{#2}\allargs}
\mathcommand\mindexppeins  [2]{\mindexsymbol{#1}\beginargs{#2}\allargs}
\mathcommand\Givepp        [3]{\Givesymbol
\beginargs{#1}\separgs{#2}\separgs{#3}\allargs}
\mathcommand\mindexppzwei  [3]{\mindexsymbol
{#1}\beginargs{#2}\separgs{#3}\allargs}
\mathcommand\mindexppdrei  [4]{\mindexsymbol
{#1}\beginargs{#2}\separgs{#3}\separgs{#4}\allargs}

\mathcommand\nonnegppp     [1]{\nonnegpsymbol\beginargs{#1}\allargs}

\mathcommand\anonymouscsymbol{c}
\mathcommand\anonymouscindexsymbol[1]{\anonymouscsymbol_{#1}}
\mathcommand\anonymousfsymbol{f}
\mathcommand\anonymouscpp
[2]{\anonymouscsymbol\beginargs{#1}\separgs\ldots\separgs{#2}\allargs}
\mathcommand\anonymouscindexpp
[3]{\anonymouscindexsymbol{#1}\beginargs{#2}\separgs\ldots\separgs{#3}\allargs}
\mathcommand\anonymousfpp
[2]{\anonymousfsymbol\beginargs{#1}\separgs\ldots\separgs{#2}\allargs}
\mathcommand\coerceindexpp[3]{[#3]_{#1}^{#2}}

\mathcommand\Elephantppp    [1]{\Elephantpsymbol\beginargs{#1}\allargs}
\mathcommand\Flowerppp      [1]{\Flowerpsymbol  \beginargs{#1}\allargs}
\mathcommand\Bicycleppp     [1]{\Bicyclepsymbol \beginargs{#1}\allargs}
\mathcommand\Germanppp      [1]{\Germanpsymbol  \beginargs{#1}\allargs}
\mathcommand\Hugeppp        [1]{\Hugepsymbol    \beginargs{#1}\allargs}
\mathcommand\Animalppp      [1]{\Animalpsymbol  \beginargs{#1}\allargs}
\mathcommand\Maleppp        [1]{\Malepsymbol    \beginargs{#1}\allargs}
\mathcommand\Boyppp         [1]{\Boypsymbol     \beginargs{#1}\allargs}
\mathcommand\Girlppp        [1]{\Girlpsymbol    \beginargs{#1}\allargs}
\mathcommand\Femaleppp      [1]{\Femalepsymbol  \beginargs{#1}\allargs}
\mathcommand\Roundppp       [1]{\Roundpsymbol   \beginargs{#1}\allargs}
\mathcommand\Bishoppp       [1]{\Bishopsymbol   \beginargs{#1}\allargs}
\mathcommand\Quadrangularppp[1]{\Quadrangularpsymbol  \beginargs{#1}\allargs}
\mathcommand\Kissedppp[2]{\Kissedpsymbol\beginargs{#1}\separgs{#2}\allargs}
\mathcommand\Metppp[2]   {\Metpsymbol   \beginargs{#1}\separgs{#2}\allargs}

\newcommand\bound     {{\rm bound}}
\newcommand\free      {{\rm free}}

\mathcommand\Vtripleindex[3]{\V\!_{{#1},\,{#2},\,{#3}}}
\mathcommand\Vdoubleindex[2]{\V\!_{{#1},\,{#2}}}
\mathcommand\Vsingleindex[1]{\V\!_{{#1}}}

\mathcommand\Erel[1]{\Gammaoffont\!_{#1}}
\mathcommand\Urel[1]{\Deltaoffont_{#1}}

\mathcommand\theRprimefromstrongtoweak{
  \inparenthesesinlinetight{
     \domres\id{\Vwall\cup\Vsome\setminus\RAN\varsigma}
     \nottight{\nottight\uplus}
     \reverserelation\varsigma
  }
  \nottight{\circ}
  \ranres
    {\transclosureinline R}
    {\Vwall\cup\Vsome\setminus\RAN\varsigma}
  \nottight{\nottight{\nottight{\uplus}}}
  \Vsome\tighttimes\Vsall
}

\mathcommand\deltaminus{\delta^-}
\mathcommand\deltaplus{\delta^+}
\mathcommand\deltaplusplus{\delta^{+^+}}
\mathcommand\deltastar{\delta^*}
\mathcommand\deltastarstar{\delta^{*^*}}
\newcommand\varihelper[1]{free \discretionary
{\mbox{\math{#1}-vari-}}{\mbox{able}}{\mbox{\math{#1}-variable}}}
\newcommand\Varihelper[1]{Free \discretionary
{\mbox{\math{#1}-vari-}}{\mbox{able}}{\mbox{\math{#1}-variable}}}
\newcommand\fev {\varihelper\gamma}
\newcommand\fuv {\varihelper\delta}

\newcommand\Fev {\Varihelper\gamma}

\mathcommand\Vall     {\Vsingleindex\indexdelta         }
\mathcommand\Vwall    {\Vsingleindex\indexdeltaminu     }
\mathcommand\Vsall    {\Vsingleindex\indexdeltaplus     }
\mathcommand\Vgsome   {\Vsingleindex\indexgammaplus     }
\mathcommand\Vsome    {\Vsingleindex\indexgamma         }
\mathcommand\Vfree    {\Vsingleindex\indexfree          }
\mathcommand\Vbound   {\Vsingleindex\indexbound         }
\mathcommand\Vsomesall{\Vsingleindex\indexgammadeltaplus}

\mathapplycommand\VARall      {\VARsingleindex\indexdelta         }
\mathapplycommand\VARwall     {\VARsingleindex\indexdeltaminu     }
\mathapplycommand\VARsall     {\VARsingleindex\indexdeltaplus     }
\mathapplycommand\VARgsome    {\VARsingleindex\indexgammaplus     }
\mathapplycommand\VARsome     {\VARsingleindex\indexgamma         }
\mathapplycommand\VARfree     {\VARsingleindex\indexfree          }
\mathapplycommand\VARbound    {\VARsingleindex\indexbound         }
\mathapplycommand\VARsomesall {\VARsingleindex\indexgammadeltaplus}
\mathcommand\displayVARsall[1]{\VARsingleindex\indexdeltaplus
\!\!\!\:\left(\begin{array}{@{}c@{}}#1\end{array}\right)}

\mathcommand\rigidvari     [2]{#1_{#2}^\indexgammadeltaplus}
\mathcommand\existsvari    [2]{#1_{#2}^\indexgamma    }
\mathcommand\forallvari    [2]{#1_{#2}^\indexdelta    }
\mathcommand\freevari      [2]{#1_{#2}^\indexfree     }
\mathcommand\wforallvari   [2]{#1_{#2}^\indexdeltaminu}
\mathcommand\sforallvari   [2]{#1_{#2}^\indexdeltaplus}
\mathcommand\gexistsvari   [2]{#1_{#2}^\indexgammaplus}
\mathcommand\boundvari     [2]{#1_{#2}}
\mathcommand\vari          [2]{#1_{#2}}
\mathcommand\wforallvarilow[2]{#1_{#2}^
{\raisebox{-.82ex}{\math\indexdeltaminu}}}

\newcommand\indexhelper[1]{{\scriptscriptstyle#1\:\!\!}}
\newcommand\indexdeltaplus
{\indexhelper{\delta^{\raisebox{-.17ex}{\fvesf\hskip-0.14em +}}}}
\newcommand\indexdeltaminu
{\indexhelper{\delta^{\mbox{\fvesf\hskip-0.14em\rule[.2ex]{.7em}{.15ex}}}}}
\newcommand\indexgammaplus
{\indexhelper{\gamma^{\mbox{\fvesf\hskip-0.14em +}}}}
\newcommand\indexgammadeltaplus
{\indexhelper{\gamma\delta^{\raisebox{-.17ex}{\fvesf\hskip-0.14em +}}}}

\newcommand\indexdelta{\indexhelper\delta}
\newcommand\indexgamma{\indexhelper\gamma}
\newcommand\indexfree
{{\scriptscriptstyle\free}}
\newcommand\indexbound
{{\scriptscriptstyle\bound}}

\newcommand\Wellfsymb{\ident{Wellf}}
\mathapplycommand\Wellfpp{\Wellfsymb}
\mathcommand\beginargs{(}
\mathcommand\allargs  {)}
\mathcommand\separgs  {,\,}
\mathcommand\tightsepargs{,}

\mathcommand\minusppnoparentheses  [2]{{#1}\,\minussymbol\,{#2}}
\mathcommand\tightminusppnoparentheses  [2]{{#1}\minussymbol{#2}}
\mathcommand\divideppnoparentheses [2]{{#1}\,\dividesymbol\,{#2}}
\mathcommand\plusppnoparentheses   [2]{{#1}\,\plussymbol \,{#2}}
\mathcommand\plusppnoparenthesesi  [3]{{#2}\,\plussymbol^{#1}\,{#3}}
\mathcommand\tightplusppnoparentheses   [2]{{#1}\plussymbol{#2}}
\mathcommand\timesppnoparentheses  [2]{{#1}\,\timessymbol\,{#2}}
\mathcommand\undppnoparentheses    [2]{{#1}\und            {#2}}
\mathcommand\oderppnoparentheses   [2]{{#1}\oder           {#2}}
\mathcommand\impliesppnoparentheses[2]{{#1}\implies        {#2}}
\mathcommand\leqinfixppnoparentheses[2]{{#1}\,\tight\leq\,{#2}}
\mathcommand\geqinfixppnoparentheses[2]{{#1}\,\tight\geq\,{#2}}
\mathcommand\dividepp [2]{(\divideppnoparentheses {#1}{#2})}
\mathcommand\minuspp  [2]{(\minusppnoparentheses  {#1}{#2})}
\mathcommand\pluspp   [2]{(\plusppnoparentheses   {#1}{#2})}
\mathcommand\timespp  [2]{(\timesppnoparentheses  {#1}{#2})}
\mathcommand\undpp    [2]{(\undppnoparentheses    {#1}{#2})}
\mathcommand\oderpp   [2]{(\oderppnoparentheses   {#1}{#2})}
\mathcommand\impliespp[2]{(\impliesppnoparentheses{#1}{#2})}

\newcommand\englishtextelevenone
{\englishtexteleventwo{Not only the notorious golden mountain is of gold, 
 but also\\the}.}
\newcommand\englishtexteleventwo[1]
{#1 round quadrangle is just as certainly round as it is quadrangular}

\newcommand\germantextfifty
{Diese\es\ gelingt nach einer Methode, welche unabh\ae ngig voneinander
J. \herbrand\ und M. \presburger\ au\esi gebildet haben.
Diese Methode besteht darin, da\ses\ man Formeln, welche gebundene Variablen
enthalten, \glq Reduzierte\grq\ zuordnet, in denen keine gebundenen Variablen
mehr au\efi tre\-ten und welche im Sinne der inhaltlichen Deutung jenen Formeln
gleichwertig sind.}
\newcommand\germantextfiftyquotation{%
\cite[\p\,234]{grundlagen-first-edition-volume-one}
(or \cite[\p\,233]{grundlagen-second-edition-volume-one})
(note-mark omitted, orthography modern\-ized)%
}

\newcommand\germantextskolemeins{{%\germanfont
 Er mu\ses\ also sozusagen einen Umweg \ue ber da\es\ 
 Nichtabz\ae hlbare machen.}}
\newcommand\germantexthilbertsicherheit{{\germanfont
Und wo sonst soll Sicherheit und Wahrheit zu finden sein,
wenn sogar da\es\ mathematische Denken versagt?}}
\newcommand\germantexthilbertignorabimus{{\germanfont
E\es\ bildet ja gerade einen Hauptreiz bei der 
Besch\ae ftigung mit einem mathematischen Problem, da\ses\
wir in un\es\ den steten Zuruf h\oe ren:
da ist ein Problem, suche die L\oe sung;
du kannst sie durch reine\es\ Denken finden;
denn in der Mathematik gibt e\es\ kein {\rm Ignorabimus}.}}
\newcommand\germantexthilbertcantorsparadise{{\germanfont 
Au\es\ dem Paradie\es, da\es\ un\es\ \cantor\ geschaffen,
soll un\es\ niemand ver\-treiben.}}
\newcommand\germantexthilbertcantorsparadisecitation{\cite[\p 170]{unendliche}}
\newcommand\germantexthilbertonkant{{\germanfont 
Schon \kant\ hat gelehrt -- und zwar bildet die\es\ einen
integrierenden Bestandteil seiner Lehre \nolinebreak--,
da\ses\ die Mathematik \ue ber einen unabh\ae ngig von aller Logik 
gesicherten Inhalt verf\ue gt und daher nie und nimmer allein durch Logik
begr\ue ndet werden kann,
we\esi halb auch die Bestrebungen von \frege\ und \dedekind\ scheitern
mu\sesi ten. \ 
Vielmehr ist al\es\ Vorbedingung f\ue r die Anwendung logischer Schl\ue sse
und f\ue r die Bet\ae tigung logischer Operationen schon etwa\es\ in der 
Vorstellung gegeben:
gewisse, au\sz er-logische konkrete 
Objekte, die anschaulich al\es\ unmittelbare\es\
Erlebni\es\ vor allem Denken da sind.}}
\newcommand\germantextkantsnotionofproblematisch
{Ich nenne einen Begriff problematisch, der
   keinen Widerspruch enth\ae lt, der auch al\es\ eine Begrenzung
   gegebener Begriffe mit anderen Erkenntnissen zusammenh\ae ngt,
   dessen objektive Realit\ae t aber auf keine Weise erkannt werden kann.}
\newcommand\germantextartineins
{Eine rationale Funktion \nlbmath{F(x_1,x_2,\cdots,x_n)} von 
 \math n \nolinebreak Ver\ae nderlichen hei\sz e definit,
 wenn sie f\ue r kein reelle\es\ Wertsy\esi tem der \nlbmath{x_i}
 negative Werte annimmt.}
\newcommand\germantextartinzwei
{Satz\,4: E\es\ sei \math R ein reeller Zahlk\oe rper, der sich nur auf eine
 Weise ordnen l\ae sst, wie zum Beispiel der K\oe rper der rationalen Zahlen,
 oder der aller reellen algebraischen Zahlen oder der aller reellen Zahlen.\\
 Dann ist jede rationale definite Funktion von \nlbmath{x_1,x_2,\cdots,x_n}
 mit Ko\ewithtrema 
 ffizienten au\es\ \nlbmath R\\Summe von Quadraten von rationalen
 Funktionen der \nlbmath{x_i} mit Ko\ewithtrema ffizienten au\es\ \nlbmath R.}

\newcommand\frenchtextninty{{\frenchfont 
Nous entendons par raisonnement intuitionniste, un raisonnement qui
 satisfait aux conditions suivantes:
 on n'y consid\`ere jamais qu'un nombre fini
 d\'etermin\'e d'objets et de fonctions;
 celles-ci sont bien d\'efinies, leur d\'efinition
 permettant de calculer leur valeur d'une mani\`ere univoque;
 on n'affirme
 jamais l'existence d'un objet sans donner le moyen de le construire;
 on n'y consid\`ere jamais l'en\-semble de tous les objets \nlbmath x d'une
 collection infinie;
 et quand on dit qu'un raisonnement (ou un th\'eor\`eme) est vrai pour tous ces
 \nlbmath x,
 cela signifie que pour chaque \nlbmath x pris en particulier, 
 il est possible de r\'ep\'eter le raisonnement g\'en\'eral en question
 qui ne doit \^etre consid\'er\'e que comme le prototype de ces raisonnements 
 particuliers.}}
\newcommand\frenchtextnintylocationoriginal
{\cite[\p\,3, \litnoteref 3]{herbrand-consistency-of-arithmetic}}
\newcommand\frenchtextnintylocationoriginalmodifierforreprint
{Without the comma after
  ``{\frenchfont intuitionniste}\closequotecomma
  but with an additional comma after
  ``{\frenchfont question}''}
\newcommand\frenchtextnintylocationreprint
{\cite[\p\,225, \litnoteref 3]{herbrand-ecrits-logiques}}
\newcommand\frenchtextonehundred{{\frenchfont
Nous pouvons dire que la d\'emonstration de {\em\loewenheim}\/
  \'etait suf\-fi\-sante
  en Math\'e\-matiques; 
  mais il nous a fallu, dans ce travail, 
  la rendre `m\'eta\-math\'e\-ma\-ti\-que'
  (voir l'introduction),
  pour qu'elle nous soit de quelque utilit\'e.}}
\newcommand\frenchtextonehundredsourcelocationoriginal
{\cite[\p 118]{herbrand-PhD}} 
\newcommand\frenchtextonehundredsourcelocationmodifierforreprint
{Without the emphasis on ``{\frenchfont\em\loewenheim}\,'' and with
 ``{\frenchfont math\'e\-matiques}'' instead of 
 ``{\frenchfont Math\'e\-matiques}'' and
 ``{\frenchfont l'Introduction)}''
 instead of ``{\frenchfont l'introduction),}''}
\newcommand\frenchtextonehundredsourcelocationreprint
{\cite[\p 144]{herbrand-ecrits-logiques}}
\newcommand\frenchtextonehundredandone{{\frenchfont
(On remarquera que cette d\'efinition
diff\`ere de la d\'efinition qu'on pourrait croire la plus naturelle seulement
par le fait que, quand le nombre \nlbmath p augmente, 
le nouveau champ \nlbmath{C'} et les nouvelles valeurs ne peuvent
pas forc\'ement \^etre consid\'er\'es comme un `prolongement' des anciens;
mais cependant, la connaissance de \nlbmath{C'} et des valeurs pour un nombre
\nlbmath p d\'etermin\'e, % dieses Komma steht nur im Pariser Original,
entra\^\i ne celle d'un champ et de valeurs 
convenant pour tout nombre inf\'erieur;
seul donc un `principe de choix' pourrait conduire \`a prendre un syst\`eme
de valeurs fixe dans un champ infini).}}
\newcommand\frenchtextonehundredandonesourcelocationoriginal
{\cite[\p 109]{herbrand-PhD}} 
\newcommand\frenchtextonehundredandonesourcelocationmodifierforreprint
{Without the comma after ``{\frenchfont d\'etermin\'e}'' also in:}
\newcommand\frenchtextonehundredandonesourcelocationreprint
{\cite[\p 135\f]{herbrand-ecrits-logiques}}
\newcommand\frenchtextonehundredandtwo{{\frenchfont
Il est absolument n\'ecessaire de
prendre de telles d\'efinitions, pour donner un sens pr\'ecis aux mots:
`vrai dans un champ infini\closesinglequotecomma
qui ont souvent \'et\'e employ\'es sans explication suffisante,
et pour justifier une proposition \`a laquelle on fait souvent allusion,
d\'emontr\'ee par {\em\loewenheim}\commanospace\footnotemark[1]
sans bien remarquer que cette proposition
n'a aucun sens pr\'ecis sans d\'efinition pr\'ealable, et que la d\'emonstration
de {\em\loewenheim} est au surplus totalement insuffisante pour notre but 
(voir 6.4).}}
\newcommand\frenchtextonehundredandtwosourcelocationoriginal
{\cite[\p 110]{herbrand-PhD}} 
\newcommand\frenchtextonehundredandtwosourcelocationmodifierforreprint
{Without the signs after ``{\frenchfont d\'efinitions}\closequotecomma
``{\frenchfont mots}\closequotecomma and
``{\frenchfont pr\'ealable}'' and the emphasis on 
``{\frenchfont\em\loewenheim}\,\closequotecomma
but with the correct ``{\frenchfont 6.2}''
instead of the misprint ``{\frenchfont 6.4}''}
\newcommand\frenchtextonehundredandtwosourcelocationreprint
{\cite[\p 136\f]{herbrand-ecrits-logiques}}
\newcommand\frenchtextonehundredeleven{{\frenchfont
Cette expression signifie: traduites en language ordinaire, 
consid\'er\'ees comme une propri\'et\'e des entiers, 
et non comme un pur symbole.}}

\newcommand\frenchtextonehundredtwelve{{\frenchfont
On pourra aussi introduire un nombre quelconque de 
\\fonctions \bigmathnlb{\fsymbol_i x_1 x_2\cdots x_{n_i}}{}
avec des hypoth\`eses telles que:
\begin{enumerate}\notop\item[a)]{\em
Elles ne contiennent pas de variables apparentes.}
\item[b)]\sloppy{\em
Consid\'er\'ees intuitionnistiquement\commanospace\footnotemark[5] 
elles permettent de faire effectivement le calcul de\ 
\math{\fsymbol_i x_1 x_2\cdots x_{n_i}},
pour tout syst\`eme particulier de nombres;
et l'on puisse d\'emontrer intuitionnistiquement que
l'on obtient un r\'esultat bien d\'etermin\'e.}
 \ {\bf (Groupe\,C.)}
\noitem\end{enumerate}}}

\newcommand\frenchtextonehundredandthirteenwithquotes{``{\frenchfont
\frenchtextonehundredtwelve
\noindent{\footnotesize\frenchfont
\footnotemark[5]\frenchtextonehundredeleven{\normalsize''}\noitem\par}}
\getittotheright
{\frenchtextonehundredandthirteentwosourcelocationoriginal}}

\newcommand\frenchtextonehundredandthirteentwosourcelocationoriginal
{\citep[\p\,5]{herbrand-consistency-of-arithmetic}}
\newcommand
\frenchtextonehundredandthirteensourcelocationmodifierforreprint
{Without the colon after ``{\frenchfont que}'' and the comma after
 ``\math{\fsymbol_i x_1 x_2\cdots x_{n_i}}\closequotecomma
 with a semi-colon instead of a full-stop after 
``{\frenchfont apparentes}\closequotecomma
 and with ``{\frenchfont langage}'' instead of ``{\frenchfont language}''
 also in: \ \citep[\p\,226\f]{herbrand-ecrits-logiques}}

\newcommand\frenchtextfourhundred{{\frenchfont
\citet{loewenheim-1915} a publi\'e du r\'esultat \'enonc\'e dans ce paragraphe
  une d\'e\-mon\-stra\-tion dont nous avons montr\'e 
  les graves lacunes dans notre
  travail d\'ej\`a cit\'e (\litchapwithsectref 5{6.2}).}}
\newcommand\frenchtextfivehundredwithfrontandend[2]{
\begin{enumerate}\item[#1{\frenchfont\math{1^0}.}]{\frenchfont
Si une des \'egalit\'es \`a satisfaire \'egale une variable
restreinte \nlbmath x \`a un autre individu;
ou bien cet individu contient \nlbmath x,
et on ne peut y satisfaire;
ou bien il ne contient pas \nlbmath x;
cette \'egalit\'e sera alors une des \'egalit\'es normales cherch\'ees;
et on remplacera \nlbmath x par cette fonction dans les autres \'egalit\'es
\`a satisfaire.}
\item[\frenchfont\math{2^0}.]{\frenchfont
Si une des \'egalit\'es \`a satisfaire \'egale une variable g\'en\'erale
\`a un autre individu, qui ne soit pas une variable restreinte,
il est impossible d'y satisfaire.}
\item[\frenchfont\math{3^0}.]{\frenchfont
Si une des \'egalit\'es \`a satisfaire \'egale
\\\LINEnomath{
\math{f_1(\varphi_1,\varphi_2,\ldots,\varphi_n)} \`a
\math{f_2(\psi_1,\psi_2,\ldots,\psi_m)},}\\
ou bien les fonctions \'el\'ementaires \math{f_1} et \math{f_2} 
sont diff\'erentes, auquel cas il es impossible d'y satisfaire;
ou bien les fonctions \math{f_1} et \math{f_2} sont les m\^emes; 
auquel cas on remplace l'\'egalit\'e
par celles obtenues en \'egalant 
\math{\varphi_i} \`a \nlbmath{\psi_i}.}#2\end{enumerate}}
\newcommand\frenchtextfivehundredsource
{\cite[\p\,96\f]{herbrand-PhD}}
\newcommand\frenchtextfivehundredsourcemodifier
{Note that the reprint in \citep[\p124]{herbrand-ecrits-logiques}
has \mbox{``{\frenchfont associ\'ees}''} instead of 
``{\frenchfont normales}\closequotecomma
which is not an improvement,
and the English translation of \englishtextfivehundredsource\
is based on this contorted version.}

\newcommand\englishtextskolemeins
{Thus he must make a detour, so to speak, through the non-denumerable.}
\newcommand\englishtextfivehundredwithoutfrontandend
{If one of the equations to be satisfied equates a 
\math\gamma-variable \nlbmath{\existsvari x{}} to another term;
either this term contains \nlbmath{\existsvari x{}},
and then the equation cannot be satisfied;
or else the term does not contain \nlbmath{\existsvari x{}},
and then the equation will be one of the normal-form
equations we are looking for,
and we \nolinebreak replace \nlbmath{\existsvari x{}} 
with the term in the other equations to be satisfied.
\item[2.] 
If one of the equations to be satisfied 
equates a \math\delta-variable to another term that is not a 
\math\gamma-variable, the equation cannot be satisfied.
\item[3.]
If one of the equations to be satisfied equates
\math{f_1(\varphi_1,\varphi_2,\ldots,\varphi_n)} to 
\math{f_2(\psi_1,\psi_2,\ldots,\psi_m)};
either the function symbols \math{f_1} and \math{f_2}
are different, and then the equation cannot be satisfied;
or \nolinebreak they are the same, 
and then we replace this equation
with those that 
equate \math{\varphi_i} to \nlbmath{\psi_i}.}

\newcommand\englishtextfivehundredsource
{\cite[\p148]{herbrand-logical-writings}}
\newcommand\englishtextfifty
{This is achieved by a method developed independently by 
 J. \herbrand\ and M. \presburger. \ 
 The method consists in associating `reduced forms' to formulas with
 bound variables, in which no bound variables occur anymore and 
 which are semantically equivalent to the original formulas.}
\newcommand\englishtextonehundred
{We could say that \loewenheim's proof was sufficient in mathematics;
    but, in the present work, 
    we had to make it `metamathematical' (see Introduction) 
    so that it would be of some use to us.}
\newcommand\englishtextonehundredsourcelocationwithoutpagenumber
{herbrand-logical-writings}
\newcommand\englishtextonehundredsourcelocationpagenumber
{\p 176}
\newcommand\englishtextonehundredandone{We observe that this definition
differs from the definition that would seem the most natural only in that, 
as the number \nlbmath p increases, 
the new domain \nlbmath{C'} and the new values need not be regarded as forming
an `extension' of the previous ones.
Clearly, if we know \nlbmath {C'} and the values for a given number \nlbmath p,
then for each smaller number we know a domain and values that answer to the
number;
but only a `principle of choice' could 
lead us to take a fixed system of values in an infinite domain.}

\newcommand\englishtextonehundredandtwo{It is absolutely necessary to adopt
such definitions if we want to give a precise sense to the words
`true in an infinite domain\closesinglequotecomma
words that have frequently been used without sufficient explanation,
and also if we want to justify a proposition proved by \loewenheim,
a proposition to which many refer without clearly seeing that \loewenheim's
proof is totally inadequate for our purposes (see \nolinebreak 6.4) and that, 
indeed,
the proposition has no precise sense until such a definition has been given.}
\newcommand\englishtextonehundredandthirteen
{We can also introduce any number of functions 
 \bigmathnlb{\fppeinsindex i{x_1,x_2,\ldots,x_{n_i}}}{}
 together with some hypotheses such that 
 \begin{enumerate}\item[(a)]{\em
 The hypotheses contain no bound variables.}\item[(b)]\sloppy{\em
 Considered intuitionistically\commanospace\footnotemark[5] they make the 
 effective computation of the 
 \math{\fppeinsindex i{x_1,x_2,\ldots,x_{n_i}}}
 possible for every given set of numbers, and it is 
 possible to prove intuitionistically that we obtain a 
 well-defined 
 result.}
 \ {\bf (Group\,C.)}\end{enumerate}
 {\footnotesize\footnotemark[5]This expression means: 
  translated into ordinary language,
  considered as a property of integers and not as a mere symbol.}}
\newcommand\englishtextonehundredandthirteenquotation
{our translation}
\newcommand\englishtexttwohundred
{If \math\phi\ denotes an unknown function and \math{\psi_1,\ldots,\psi_k}
are known functions, and if the \math\psi's and the \math\phi\ are 
substituted in one another in the most general fashions and certain
pairs of the resulting expressions are equated, then, if the resulting set of
functional equations has one and only one solution for \nlbmath\phi, \ \ 
\math\phi\ \nolinebreak is a recursive function.}
\newcommand\englishtextfourhundred
{\citet{loewenheim-1915} published a proof of the result stated in this section.
In \cite[\litchapwithsectref 5{6.2}]{herbrand-PhD}, we pointed out that there 
are serious gaps in his proof.}
\newcommand\englishtextgoedelonherbrand
{I have never met \herbrand.
His suggestion was made in a letter in 1931, 
and it was formulated {\em exactly}\/ as on \p\,26 of my lecture notes, that is,
without any reference to computability. \ 
However, since \herbrand\ was an intuitionist, 
this definition for him evidently meant that there is a
{\em constructive}\/ proof for the existence and uniqueness of \nlbmath\phi.}
\newcommand\propertyA{Property\,A} %%% Two macros for freeness of change
\newcommand\propertyB{Property\,B} 
\newcommand\propertyC{Property\nolinebreak\hskip.2em\nolinebreak C} 
 
\newcommand\englishherbrandribettheorem
{``An odd prime $p$ is called irregular if the class number of the field ${\bf
    Q}(\mu_p)$ is divisible by $p$ ($\mu_p$ being, as usual, the group of $p$-th
  roots of unity). According to Kummer's criterion, $p$ is irregular if and only
  if there exists an even integer $k$ with $2\leq k \leq p-3$ such that $p$
  divides (the numerator of) $k$-th Bernoulli number $B_k$, given by the
  expansion
  $$
   \frac{t}{e^t-1}+\frac{t}{2}-1=\sum_{n\geq 2} \frac{B_n}{n!} t^n.
  $$
  The purpose of this paper is to strengthen Kummer's criterion.
  
  Let $A$ be the ideal class group of $\Q(\mu_p)$, and let $C$ be the
  $F_p$-vector space $A/A^p$. The Galois group $\mbox{Gal}(\overline\Q/\Q)$
  acts on $C$ through its quotient $\Delta = \mbox{Gal}(\Q(\mu_p)/\Q)$. Since
  all characters of $\Delta$ with values in $\overline{{\bf F }}^*_p$ are powers of the
  standard character
  $$
  \chi: \mbox{Gal}(\overline\Q/\Q) \rightarrow \Delta
  \tilde{\longrightarrow} {\bf F}^*_p
  $$
  giving the action of $\mbox{Gal}(\overline\Q/\Q)$ on $\mu_p$, the
  vector space $C$ has a canonical decomposition
  $$
  C = \bigoplus_{i\bmod(p-1)} C(\chi^i),
  $$
  where 
  $$
  C(\chi^i) = \{c\in C|\sigma c = \chi^i(\sigma)\, c \mbox{ for all }
  \sigma\in\Delta\}. 
  $$\yestop\par
  (1.1) {\bf Main Theorem.}
  \\Let $k$ be even, $2\leq k\leq p-3$.\\Then $p|B_k$ if
  and only if $C(\chi^{1-k}) \not= 0$

  In fact, the statement that $C(\chi^{1-k}) \not= 0$ implies $p|B_k$ is well known \citep[Th.\,3]{herbrand-1932}%
''\getittotheright{\citep[\littheoref{1.1}, \p\,151]{ribet76:_q}}}
\newcommand\englishtexthilbertonkant{\kant\ already taught ---~and indeed it 
is part and parcel of his doctrine~--- that mathematics has 
at its disposal a content secured independently
of all logic and hence can never be provided with a foundation by means of
logic alone;
that is why the efforts of \frege\ and \dedekind\ were bound to fail. \ 
Rather, as a condition for the use of logical inferences
and the performance of logical operations, 
something must already be given to our faculty of representation,
certain extra-logical concrete objects that are intuitively present as 
immediate conceptions prior to all thought.}

\newcommand\englishtextninehundred
{Thus, in mathematics, we have no reasons to assume any meaning of
 `existence' that would be fundamentally different from that of 
 `the validity of axiomatic relations\closesinglequotefullstopnospace}

\newcommand\englishquoteforstertext
{Early twentieth century mathematicians used the expression
 `The Crisis in Foundations\closesinglequotefullstop
 This crisis had many causes and 
 ---~despite the disappearance of the expression from contemporary speech~---
 has never really been resolved.
 One of its many causes was the increasing formalisation of mathematics,
 which brought with it the realisation that the paradox of the liar could
 infect even mathematics itself.
 This appears most simply in the form of \russell's paradox,
 appropriately in the heart of set theory.}
\newcommand\englishquoteforstertextplace{\cite[\p\,838]{forster-years-on}}

\input seki-deckblatt-5
\begin{document}
\makecover
\maketitle
\begin{abstract}%
We give some lectures on the work on formal logic of \herbrandname,
and sketch his life and his influence on automated theorem proving.
The intended audience ranges from students interested in logic 
over historians to logicians.
Besides the well-known correction of \herbrandsfalselemma\ by
\goedel\ and \dreben,
we also present the hardly known unpublished correction of \heijenoort\
and its consequences on \herbrand's {\em Modus Ponens} Elimination.
Besides \herbrandsfundamentaltheorem\ and its relation to the
\loewenheimskolemtheorem,
we carefully investigate \herbrand's notion of intuitionism
in connection with his notion of falsehood in an infinite domain.
We sketch \herbrand's two proofs of the consistency of arithmetic
and his notion of a recursive function, 
and last but not least, present the correct original text of 
his unification algorithm with a new translation.
\end{abstract}

\catcode`\@=11
\renewcommand*\l@section[2]{%
  \ifnum \c@tocdepth >\z@
    \addpenalty\@secpenalty
    \addvspace{0.5em \@plus\p@}%
    \setlength\@tempdima{2.5em}%
    \begingroup
      \parindent \z@ \rightskip \@pnumwidth
      \parfillskip -\@pnumwidth
      \leavevmode \rm
      \advance\leftskip\@tempdima
      \hskip -\leftskip
      #1\nobreak\hfil \nobreak\hb@xt@\@pnumwidth{\hss #2}\par
    \endgroup
  \fi}
\catcode`\@=12

\vfill\pagebreak\setcounter{tocdepth}{1}\tableofcontents\cleardoublepage

\newcommand\Firstsectionname{Introductory Lecture}
\newcommand\firstsectionname{introductory lecture}
\section{\Firstsectionname}%
\label{sec:preface}%

Regarding the work on formal logic of \herbrandname\ \herbrandlifetime,
our following lectures will provide a lot of useful information for the 
student interested in logic 
as well as a few surprising insights for the experts in the fields
of history and logic.

As \herbrandname\ 
is an idol of many scholars
today,
% right from the start  
% we \nolinebreak could not 
we cannot
help asking ourselves the following questions: 
Is there still something to learn from
his work on logic 
which has not found its way into the standard textbooks on logic? 
Has everything already been published
which should be said or written on him? 
Should we treat him just as an icon? 

Well, the lives of 
% young 
mathematical prodigies who 
% tragically passed away early
passed away very early
after ground-breaking work invoke a fascination for later generations: \
% This is now sorted by year of death !!!
The early death of \abelname\ \abellifetime\ from ill health after a
sled trip to visit his \fiance\ for Christmas; \hskip .2em the obscure
circumstances of \galoisname' \galoislifetime\ duel; \hskip .2em the
deaths of consumption of \eisensteinname\ \eisensteinlifetime\ (who
sometimes lectured his few students from his bedside) and of
\rochname\ \rochlifetime\ in Venice; \hskip .2em the drowning of the
topologist {\urysohnname} {\urysohnlifetime} on vacation; \hskip .2em
the burial of \paleyname\ \paleylifetime\ in an avalanche at Deception
Pass in the Rocky Mountains; \hskip .2em as well as the fatal
imprisonment of \gentzenname\ \gentzenlifetime\ in
\Prag\footnote{\Cf\ \citep{last-months-gentzen}, \makeaciteoftwo
{menzler-gentzen-german}{menzler-gentzen-english}.} --- these are 
tales most scholars of logic and mathematics have heard in their 
student days.

\herbrandindexbegin\herbrandname, a young prodigy admitted to the
{\frenchfont\EcoleNormaleSuperieure} as the best student of the
year\,1925, when he was\,17, died only six years later in a
mountaineering accident in {\frenchfont La B\'erarde, Is\`ere, France}. \   
He left a legacy in logic and mathematics that is outstanding.

Despite his very short life,
% and his youth, 
% CP: Youth is a pro in mathematics!
\herbrand's contributions
were of great significance at his time and they had a strong impact on the
work by others later in 
mathematics, proof theory, computer science, and artificial intelligence. \ 
% As we have already mentioned, amongst
% those who were (at least) inspired in their own foundational
% contributions by him we find such a prominent name as \goedelname's. \
Even today the name ``\herbrand'' can be found astonishingly often
 in research papers in fields that did not even exist at his time.\footnote
{To \nolinebreak wit, the search in any online library (\eg\ citeseer) 
reveals that astonishingly many authors
dedicate parts of their work directly to \herbrandname. \
A ``Google Scholar'' search gives a little less than ten thousand hits and the 
phrases we find by such an experiment include:
\herbrand\ agent language, \herbrand\ analyses, \herbrand\ automata,
\herbrand\ base, \herbrand\ complexity, \herbrand\ constraints,
\herbrand\ disjunctions, \herbrand\ entailment, \herbrand\ equalities,
\herbrandexpansion, \herbrandsfundamentaltheorem, \herbrand\
functions, \herbrand--\gentzen\ theorem, \herbrand\ interpretation,
\herbrand--\kleene\ universe, \herbrand\ model, \herbrand\ normal
forms, \herbrand\ procedures, \herbrand\ quotient, \herbrand\
realizations, \herbrand\ semantics, \herbrand\ strategies, \herbrand\
terms, \herbrandribet\ theorem, \herbrand's theorem, \herbrand\
theory, \herbranduniverse. \ Whether and to what extend these
references to \herbrand\ are justified is sometimes open for debate.
This list shows, however, that in addition to the
foundational importance of his work at the time, his insights still
have an impact on research even at the present time. \ \herbrand's
name is therefore not only frequently mentioned among the most important 
mathematicians and logicians of the \nth{20} century but also among
the pioneers of modern computer science and artificial intelligence.}

Let us start this \firstsectionname\ by sketching a preliminary
%(non-conclusive) 
list of topics that were influenced by \herbrand's work.%

\vfill\pagebreak

\subsection{Proof Theory}\noindent
\dalenname\ \dalenlifetime\ \hskip.2em
begins his review on 
\citep{herbrand-logical-writings} \hskip.2em
%\herbrand's 'Logical Writings' 
as follows:%
\index{consistency!proof of|(}%
\index{consistency!of arithmetic|(}%
\notop\halftop\begin{quote}
``Much of the logical activity in the first half of this century was inspired by
\index{program@\programme!Hilbert's|(}%
\hilbertsprogram, which contained, besides fundamental reflections on the
   nature of mathematics, a number of clear-cut problems for technically gifted
   people. \ 
   In particular the quest for so-called ``consistency proofs'' was
   taken up by quite a number of logicians. \
   Among those, two men can be singled
   out for their imaginative approach to logic and mathematics: \ 
   \herbrandname\ and \gentzenname. \hskip.3em
   Their contributions to this specific area of logic, called
   ``proof theory'' ({\germanfont Bewei\esi theorie}) \hskip.3em
   following 
\hilbertindex\hilbert, are so fundamental that
   one easily recognizes their stamp in modern proof theory.''\footnote
   {\Cfnlb\ \citep[\p\,544]{dalen74:_review}.}
\notop\halftop\end{quote} 

\noindent
\dalenindex\dalen\ continues:
\notop\halftop\begin{quote}
  ``When we realize that 
  \herbrand's activity in logic took place in just a few
  years, we cannot but recognize him as a giant of proof theory. \
  He discovered
  an extremely powerful theorem and experimented with it in proof theory. \ 
  It is
  fruitless to speculate on the possible course \herbrand\ would have chosen,
  had he not died prematurely; \hskip.2em
  a \nolinebreak consistency proof for full arithmetic would
  have been within his reach.''\footnote
{\Cfnlb\ \citep[\p\,548]{dalen74:_review}.}
\notop\halftop\end{quote} 

\noindent
The 
% major 
thesis of 
\anellisindex\cite{anellis-loewenheim}
is that, building on the \loewenheimskolemtheorem,
it \nolinebreak was \herbrand's work in elaborating 
\hilbertindex\hilbert's concept
of ``being a proof'' that gave rise to the development
of the variety of \firstorder\ calculi in the 1930s,
such as the ones of the
\index{Hilbert!school}%
\hilbert\ school,
and such as Natural Deduction and Sequent calculi in 
\index{Gentzen!'s calculi}%
\citep{gentzen}.

As will be shown in \sectref{section lemma}, \hskip.2em
\herbrandsfundamentaltheorem\ has directly influenced 
\bernaysname' work on proof theory. \ 

The main inspiration in the 
\index{program@\programme!unwinding|(}%
{\em unwinding \programme},\footnote 
{\Cfnlb\ 
 \makeaciteoffour{kreisel-1951}{kreisel-1952}{kreisel-1958}{kreisel-1982},
 and do not miss the discussion in 
\fefermanindex\citep{unwinding}!\notop\halftop\begin{quote}
 ``{\em To determine the constructive (recursive) content or the 
       constructive equivalent of the non-constructive concepts and
       theorems used in mathematics},
       particularly arithmetic and analysis.''\getittotheright
 {\cite[\p\,155]{kreisel-1958}}\notop\notop\end{quote}}
which 
--- \nolinebreak to save the merits of proof theory \nolinebreak---
\kreiselname\ \kreisellifetime\ \hskip.2em
suggested as a replacement for 
\index{program@\programme!Hilbert's|)}%
\hilbert's failed \programme,
is \herbrandsfundamentaltheorem,
especially for \kreisel's notion of a {\em recursive interpretation}\/
of a logic calculus in another one,\footnote
{\Cfnlb\ \cite[\p\,160]{kreisel-1958} and \citep[\p\,259\f]{unwinding}.}
such as given by \herbrandsfundamentaltheorem\
for his \firstorder\ calculus in the
\sententialtautologies\ over the language \signatureenlargedby\
\skolem\ functions.

\herbrand's approach to consistency proofs, as we will sketch in
\sectrefs{section 1 Proof}{section 2 Proof}, has a
semantical flavor and is inspired by \hilbert's evaluation method of 
\index{Hilbert!'s epsilon}%
\mbox{\math\varepsilon-substitution},
whereas it avoids the dependence on \hilbert's $\varepsilon$-calculus. \
The main idea (\cfnlb\ \sectref{section 2 Proof}) \hskip .2em
is to replace the induction axiom 
by 
\index{function!recursive|(}%
recursive functions of finitistic character. \
% \pagebreak
\herbrand's
approach is in contrast to the purely syntactical style of 
\index{Gentzen!'s consistency proof}%
\gentzen\footnotemark\
and \schuette\footnotemark\
in which semantical interpretation plays no \role.\pagebreak

\addtocounter{footnote}{-1}\footnotetext{\Cfnlb\ 
\makeaciteofthree{gentzenfirstconsistent}{gentzenconsistent}{gentzenepsilon}.}%
\addtocounter{footnote}{1}\footnotetext{\Cfnlb\ \citep{schuette60:_beweis}.}%
So-called 
\index{Herbrand-style consistency proof}%
{\em \herbrand-style consistency proofs}\/
follow \herbrand's idea of constructing finite sub-models to 
imply consistency by \herbrandsfundamentaltheorem.
During the early 1970s, this technique was used by 
\scanlonname\ \scanlonlifetime\ \hskip.2em
in collaboration with 
\drebenindex\dreben\ and \goldfarbindex\goldfarb.\footnote
{\Cfnlb\ \citep{herbrand-style-consistency-proofs},
 \citep{scanlon73:_consis_number_theor_via_theor},
 \citep{goldfarb-herbrand-consistency}.}
These consistency proofs for arithmetic
roughly follow 
\ackermannindex\ackermann's previous proof,\footnote
{\Cfnlb\ \citep{ackermann-consistency-of-arithmetic}.}   
but they apply \herbrandsfundamentaltheorem\ in advance
%\pagebreak
and consider \skolemizedform\ instead of 
\index{Hilbert!'s epsilon}%
\hilbert's \nlbmath\varepsilon-terms.\footnote 
%% Moreover,
%% in \citep{scanlon73:_consis_number_theor_via_theor} we find
%% a very similar proof 
%% %% More precisely, \dreben\ and
%% %% \denton\ tie their proof very closely to the standard model as also
%% % % used by \ackermann, which complicates matters unnecessarily and also
%% %% restricts the application of the technique to systems with induction
%% %% Das schreibt der Scanlon, aber das stimmt nicht so ganz!
{Contrary to the proofs of \citep{herbrand-style-consistency-proofs}
and \citep{ackermann-consistency-of-arithmetic},
the proof of \citep{scanlon73:_consis_number_theor_via_theor},
which is otherwise 
similar to the proof of \citep{herbrand-style-consistency-proofs},
admits 
the inclusion of induction axioms over {\em any}\/
recursive \wellordering\ on the natural numbers: \par 
By an application of 
\herbrandsfundamentaltheorem,
from a given derivation of an inconsistency,
we can compute a positive natural number 
\nlbmath n such that 
\index{Property C}%
\propertyC\ of order \nlbmath n holds. \ 
Therefore, in his analog of \hilbert's and 
\ackermannindex\ackermann's 
\index{Hilbert!'s epsilon}%
\math\varepsilon-substitution method,
\scanlonindex\scanlon\ can effectively pick a minimal counterexample on the 
\index{champ fini}%
{\frenchfont champ fini} 
\nlbmath{\termsofdepthnovars n} 
from a given critical counterexample, even if this 
neither has a direct predecessor nor a finite initial segment. \ 
%% Das folgende muss wohl nicht noch mal extra gesagt werden!
%% Thus, \scanlon's proof is\begin{quote}
%% ''%both simpler and
%% % darueber kann man streiten!
%% more general in that it applies to 
%% systems \nlbmath{Z_R} of number theory with
%% induction on arbitrary recursive well-orderings \nlbmath R.''\footnote
%% {\Cfnlb\ \citep[\p\,29]{scanlon73:_consis_number_theor_via_theor}.}
%% \end{quote}
\par
This result was then further generalized in 
\citep{goldfarb-herbrand-consistency}
to $\omega$-consistency of arithmetic.%
}

\index{Gentzen!'s {\germanfont Hauptsatz}}%
\gentzen's and \herbrand's insight on Cut and 
\index{modus ponens}%
\index{modus ponens!elimination}%
\index{Cut elimination}%
{\em modus ponens} elimination
and the existence of normal form derivations
with mid-sequents had a strong influence
on \craigname's work on interpolation\fullstopnospace\footnote
{\Cfnlb\ \citep{craig57:_linear_reason,craig57:_three_uses_herbr_gentz_theor}.} 
The impact of {\em\craigsinterpolationtheorem}\/ to various
disciplines in turn has recently been discussed at the Interpolations
Conference in Honor of \craigname\ in May 2007.\footnote
{\Cfnlb\ \url{http://sophos.berkeley.edu/interpolations/}.}

% See \cite{girard-herbrand-1982} for more on \herbrand's proof theory.

\subsection{Recursive Functions and \\\protect
\goedelssecondIncompletenessTheorem}%
\index{Goedel@G\"odel!'s Second Incompleteness Theorem}%

\noindent
As will be discussed in detail in \nlbsectref{section recursive functions},
in his 1934 \Princetonnostate\ lectures,
\goedelindex\goedel\ introduced the notion of
\index{function!recursive|)}%
{\em (general) recursive functions} and mentioned that this notion 
had been proposed to him in a letter from \herbrand,
\cfnlb\ \nlbsectref{sec:life}. \ 
This letter, however, seems to have had more influence on \goedel's thinking,
namely on the consequences of \goedelssecondincompletenesstheorem:
%% which are uniquely
%% determined by equations between terms 
%% involving known (recursive) functions and
%% the unknown function. \ 
%% \goedel\ thought 
%% that the proposal for such a general notion of recursive
%% function had been suggested to him by \herbrand\ 
%% \footnote{\goedel\ did respond on July\,25, 1931, 
%% in a letter sent to the address of \herbrand's parents in Paris; 
%% since this was two days before
%% \herbrand's accident we can assume that \herbrand\ never read it.}   
%% \goedel\ mentioned this in his 1934 \Princetonnostate\ lectures
%% and confirmed it also in 1963 when questioned about this significant issue by
%% \heijenoortname.  Unfortunately, however, he was unable to find the letter
%% of \herbrand\ and only later, after the \herbrand\ letter was found in 1986 by John
%% Dawson in \goedel's {\em Nachlass}, it became clear that \goedel\ apparently
%% did not remember its detailed content correctly.
\notop\halftop\begin{quote}
``Nowhere in the correspondence 
%[between \herbrand\ and \goedel] 
does the issue of
{\em general}\/ computability arise. \
\herbrand's discussion, in particular, is solely
trying to explore the limits of 
\index{consistency!proof of|)}%
\index{consistency!of arithmetic|)}%
consistency proofs that are imposed by the second theorem. \ 
%[\goedelssecondincompletenesstheorem]
\goedel's response also
focuses on that very topic. \
It seems that he subsequently developed 
a more critical perspective on the 
very character and generality of this theorem.''\getittotheright
{\citep[\p\,180]{sieg05:_only}}
\notop\halftop\end{quote} 
% \herbrand, in fact, had previously achieved a characterization of classes of
% finitistically calculable functions, the attribution of the notion of general
% recursive functions to him, as brought up by \goedel's misremembering of the
% details of \herbrand's letter, is, as we know today, (at least partly) mistaken and
% this notion has to be attributed solely to \goedel. 
\noindent
\citet{sieg05:_only} argues that the letter of \herbrand\ to
\goedel\ caused a change of \goedel's perception of the impact of his 
own \secondincompletenesstheorem\ on 
\index{program@\programme!Hilbert's}%
\hilbertsprogram: 
Initially \goedel\ did assert that it would not
contradict \hilbert's viewpoint. \ 
Influenced by \herbrand's letter, however, he accepted the
more critical opinion of \herbrand\ on this matter.%

\vfill\pagebreak

\yestop\begin{figure}[h]
  \framebox{%%
    \begin{minipage}{.995\linewidth}
      \small
      \englishherbrandribettheorem
    \end{minipage}}
  \caption{\herbrandribet\ Theorem\label{figure herbrand ribet}}
\end{figure}

\yestop\yestop\yestop\yestop\subsection{Algebra and Ring Theory}

\yestop\noindent
In 1930--1931, within a few months, \herbrand\ wrote several papers
on algebra and ring theory. \ 
During his visit to Germany he met and
briefly worked on this topic with 
\noetherindex\noether, \hasseindex\hasse, and \artinindex\artin;
\cfnlb\ \sectref{sec:life}\@. \ 
He contributed several new theorems of his own and
simplified proofs of results by 
\kroneckername\ \kroneckerlifetime, 
\webername\ \weberlifetime, 
\takaginame\ \takagilifetime, 
\hilbertindex\hilbert, and 
\artinindex\artin, thereby generalizing some of these results.
The 
\index{Herbrand--Ribet Theorem|(}%
\herbrandribet\ Theorem
is a result on the class number of certain number fields and it strengthens 
\kummer's convergence criterion; \cfnlb\ \figuref{figure herbrand ribet}.
%% to the effect that the prime p divides the class number of the
%% cyclotomic field of p-th roots of unity if and only if $p$ divides the numerator
%% of the nth Bernoulli number $B_n$ for some $n, 0 < n < p-1$. The \herbrandribet\ 
%% theorem specifies what, in particular, it means when $p$ divides such an $B_n$.
%% One direction of the theorem has been contributed by \herbrand. 
% For illustration see the \herbrandribet\ theorem (which attributes one
% direction to Herbrand) in \figuref{figure herbrand ribet}.
%% on p.{\pageref{figure herbrand ribet}}.
% This is not the place, however, to 
% discuss this significant contribution of \herbrand\ further.

\vfill\pagebreak

Note that there is no direct connection between \herbrand's work on logic and
his work on algebra. 
Useful applications of proof theory to mathematics are very rare.
\index{program@\programme!unwinding|)}%
\kreisel's ``unwinding'' of 
\index{Artin's Theorem!Artin's proof}%
\artin's proof of 
\index{Artin's Theorem|(}%
\artin's Theorem
into a constructive proof seems to be one of the few exceptions.\footnote
{With \artin's Theorem we mean:
\begin{quote}``{\germanfontfootnote\germantextartineins}''
 \getittotheright{\citep[\p\,100]{artin-1927}}
\end{quote}
\begin{quote}``{\germanfontfootnote\germantextartinzwei}''
 \getittotheright{\citep[\p\,109, modernized orthography]{artin-1927}}
\notop\end{quote}
\Cf\ \cite{unwinding-artin} for the unwinding
 of \artin's proof of \artin's Theorem.
\\\Cf\ \cite{unwinding} for a discussion of the application of proof theory
 to mathematics in general. %
\index{Herbrand--Ribet Theorem|)}%
\index{Artin's Theorem|)}%
}%

\subsection{A first r\'esum\'e}

Our following lectures will be more self-contained than this 
\firstsectionname. \
But already on the basis of this first overview,
we just have to admit that 
\herbrandname's merits are so outstanding 
that he has no chance to escape idolization. \
Actually, he has left a world heritage in logic in a very short time. \
But does this mean that we should treat him just as an icon?

On a more careful look, 
we will find out that this genius had his flaws,
just as everybody of us made of this strange protoplasmic variant of matter,
and that he has left some of them in his scientific writings.
And he can teach us not only to be less afraid
of logic than of mountaineering; \hskip.2em
he can also provide us with a surprising amount of insight
that partly still lies to be rescued from contortion 
in praise and faulty quotations.

\subsection{Still ten minutes to go}\label
{section Still ten minutes to go}

Inevitably, when all introductory words are said, 
% in the annual postgraduate course on logic and automated deduction, 
%the professor
we will feel the urge to point out to the young students that
there are things beyond the 
% here and now of the current courses on the 
latest developments of computer % compiler 
technology or the fabric
of the Internet: eternal truths valid on planet Earth but 
in all those far away galaxies just as \nolinebreak well. 
 
And as there are still ten minutes to go till the end of the lecture,
the students listen in surprise to the strange tale about the unknown
flying objects from the far away, now visiting planet Earth and being
welcomed by a party of human dignitaries from all strata of society.
Not knowing what to make of all this, the little green visitors will
ponder the state of evolution on this strange but beautiful planet:
obviously life is there --- but can it think?

The Earthlings seem to have flying machines, they are all connected
planet-wide by communicators --- but can they really think? 
Their gadgets and little pieces of machinery appear impressive --- but is
there a true civilization on planet Earth? 
How dangerous are they,
these Earthlings made of a strange protoplasmic variant of matter?%
\pagebreak

And then cautiously looking through the electronic windows of their
flying unknown objects, they notice that strange little bearded
Earthling, being pushed into the back by the more powerful
dignitaries, who holds up a sign post with
\par\yestop\noindent\LINEnomath{\fbox{$\ \models\quad\equiv\ \ \yields$}}
\par\yestop\noindent
written on it. 

Blank faces, not knowing what to make of all this, the
oldest and wisest scientist is slowly moved out through the e-door of
the flying object, slowly being put down to the ground, and
now the bearded Earthling is asked to come forward and the two begin
that cosmic debate about syntax and semantics, proof theory and model
theory, while the dignitaries stay stunned and silent. 

And soon there is a sudden flash of recognition and 
a warm smile on that green and wrinkled old face, 
who has seen it all and now waves back to his
fellow travelers who remained safely within the flying object:
``Yes, they have minds --- yes \nolinebreak oh \nolinebreak yes!''

And this is why the name ``{\herbrandname}'' is finally written among others
with a piece of chalk onto the blackboard --- and now that the introductory
lecture is coming to a close, 
we promise
%the professor promises 
to tell in the following lectures,
what this name stands for and what that young scientist found
out when he was only 21\,years old.

\vfill
\begin{center}
  \includegraphics
[bb=20 20 575 420,width=1.0\linewidth]
{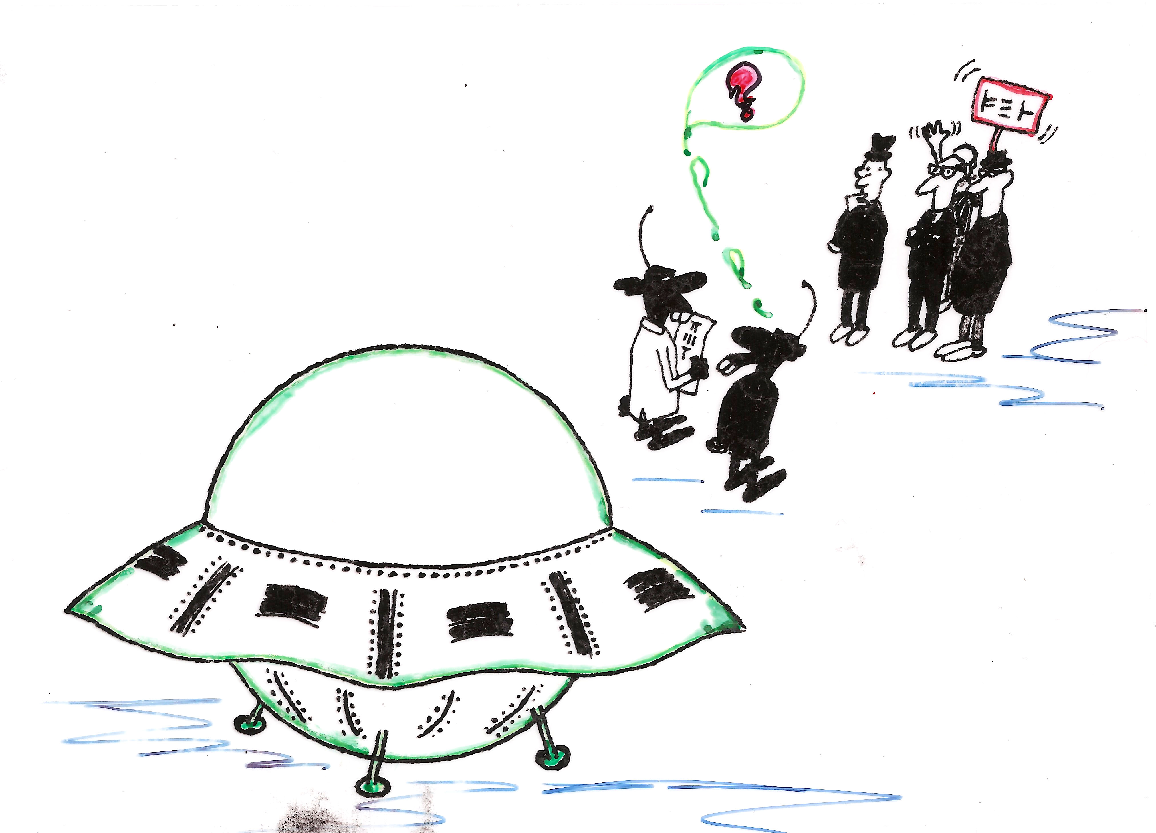}
\end{center}
\vfill

%And when the professor had grown old and frail, having told that tale
%more often than he cared to remember, the students presented him with
%a framed little birthday drawing that looked almost like this:
%\yestop\yestop\yestop

\vfill\pagebreak
%%%%%%%%%%%%%%%%%%%%%%%%%%%%%%%%%%%%%%%%%%%%%%%%%%%%%%%%%%%%%%
\section{\herbrand's Life}
\label{sec:life}

\begin{sloppypar}%\notop\noindent\headroom 

\halftop\indent
He was born on \herbrandbirthday, in \Paris, France,
where his father, 
\herbrandsfatherindexbegin\herbrandsfathername, 
worked as a trader in
antique paintings\fullstopnospace\footnote{%
 \label{note chevalley1982-herbrand-colloquium}%
 \Cfnlb\ \citep{chevalley1982-herbrand-colloquium}.} 

\halftop\halftop\indent
His parents were of Belgian origin. 

\halftop\halftop\indent
He died --- 23
years old --- in a mountaineering accident on \herbranddeathday, in
{\frenchfont La B\'erarde, Is\`ere}, France.

\halftop\halftop\indent
He remained the only
child of his parents. ---

\halftop\halftop\halftop\halftop\halftop\halftop\halftop\indent
This brief r\'esum\'e of \herbrandname's life
focuses on his entourage and the people he met\fullstopnospace\footnote{%
  \majorheadroom
  More complete accounts of \herbrand's life and personality can be found in
  \makeaciteoftwo{herbrand-thoughts}{chevalley1982-herbrand-colloquium},
  \citep{herbrand-praise}, \citep[\Vol\,V, \PP{3}{25}]{goedelcollected}.} \
All in all, very little is known about his personality and life.

\halftop\halftop\indent
In 1925, only 17 years old, he was ranked first at the entrance
examination to the 
prestigious 
{\frenchfont\em\EcoleNormaleSuperieure}
({\frenchfont\em ENS}\/)\footnote{%
  \majorheadroom
  \Cf\ \url{http://www.ens.fr} for the {\frenchfont ENS} in
  general and \url{http://www.archicubes.ens.fr} for the former
  students of the {\frenchfont ENS}.}  
in \Paris\ --- but he showed little interest for
the standard courses at the Sorbonne,
% and those of the school director,
% This is a confusing detail, isn't it? CP
which he considered a waste of time\fullstopnospace
%\footnotemark[\ref{note chevalley1982-herbrand-colloquium}] \
\arXivfootnotemarkref{note chevalley1982-herbrand-colloquium} \
However, he closely
followed the famous ``{\frenchfont S\'eminaire \hadamard}'' at the
{\frenchfont Coll\`ege de France}, organized by 
\index{Hadamard, Jacques S.|(}%
\hadamardname\ \hadamardlifetime\ from 1913 until 1933.\footnote{%
 \majorheadroom 
 \Cfnlb\ \citep[\p\,82]{hadamard-RS-memories}, 
 \citep[\p\,107]{chevalley-praise}.}
That seminar attracted many students.
At \herbrand's time, 
among these students, prominent in their later lives,
were:%
\weilindex\dieudonneindex\lautmanindex\chevalleyindex%
\par\halftop\halftop\noindent\LINEnomath{
  \begin{tabular}[t]{l|l|l}\footroom name
 &lifetime
 &year of entering the {\frenchfont ENS}
\\\hline
  \majorheadroom
  \weilname       
 &\weillifetime     
 &1922
\\\headroom
  \dieudonnename
 &\dieudonnelifetime
 &1924
\\\headroom
  \herbrandname
 &\herbrandlifetime
 &1925
\\\headroom
  \lautmanname 
 &\lautmanlifetime 
 &1926
\\\headroom
  \chevalleyname
 &\chevalleylifetime
 &1926
\\\end{tabular}}
\par\halftop\halftop\halftop\halftop\noindent\weil, 
\dieudonneindex\dieudonne, and \chevalleyindex\chevalley\ 
would later be known among the
eight founding members of the renowned \bourbaki\ group:
the French mathematicians who published the book series
on the formalization of mathematics, starting with \cite{bourbaki}.
\pagebreak

\par\indent
\weilindex\weil, \lautmanindex\lautman, and \chevalleyindex\chevalley\ 
were \herbrand's friends. 
\chevalley\ and
\herbrand\ became particularly close friends\footnote{%
 % \majorheadroom
 \index{Chevalley, Catherine}\catherine\ \chevalley, 
 the daughter of \chevalleyname\ has written to \roquettename\ on \herbrand: 
 ``he was maybe my father's dearest friend'' 
 \citep[\p\,36, \litnoteref{44}]{roquette-artin}.}  
and they worked together on algebra.\footnote{%
 \majorheadroom
 \Cfnlb\ 
 \citep{chevalley-herbrand-lost-one} and
 \citep{herbrand-chevalley}.} \hskip.4em
%% Regarding logic and proof theory, however, \chevalley's cautious
%% memories\footnote {\Cfnlb\ \citep{herbrand-thoughts}.} on \herbrand\ 
%% suggest that the two years younger friend was not on a par with
%% \herbrand\ during the time of \herbrand's life.

\halftop\halftop\indent
\chevalley\ depicts \herbrand\ as an adventurous, passionate, and often
perfectionistic personality who was not only interested in
mathematics, but also in poetry and sports.\arXivfootnotemarkref
{note chevalley1982-herbrand-colloquium} \
In particular, he
seemed to have liked extreme sportive challenges: mountaineering,
hiking, and long distance swimming. 
His interest in philosophical
issues and foundational problems of science was developed well beyond
his \nolinebreak age.

\halftop\halftop\indent
At that time, the {\frenchfont ENS} 
did not award a diploma, but the students had to
prepare the {\frenchfont\em agr\'egation},
an examination necessary to be promoted
to {\frenchfont\em professeur agr\'eg\'e},\footnote{%
 \majorheadroom
 This corresponds to a high-school teacher. 
 The original \role\ of the {\frenchfont ENS} was to educate students 
 to become high-school teachers. \ 
 Also Jean-Paul Sartre
 %, probably the most famous scholar of {\frenchfont ENS}, 
 started his career like this.}
even though most students engaged into research.  

\halftop\halftop\indent
\herbrand\ passed the {\frenchfont agr\'egation} in\,\,1928, 
again ranked first, and he prepared his doctoral thesis under the
direction of \vessiotname\ \vessiotlifetime, who was the director of
the {\frenchfont ENS} since\,1927.%
\footnote{%
 \majorheadroom
 It is interesting to note that \vessiotname\ and 
 \index{Hadamard, Jacques S.|)}\hadamardname\ where the two top-ranked students
 at the examination for the {\frenchfont ENS} in 1884.}%
\end{sloppypar}

\halftop\halftop\indent
\herbrand\ submitted his thesis \citep{herbrand-PhD}, entitled
{\frenchfont\em\herbrandPhDtitle}, on \herbrandPhDdating. \
It was approved for publication on \herbrandPhDapprovedforpublicationdate. \

\halftop\halftop\indent
In \nolinebreak 
October~1929, he had to join the army for his military service
which lasted for one year in those days.

\halftop\halftop\indent
He finally defended his
thesis on \herbrandPhDdefensedate. \ %\footnote{
  One reason for the late defense is that because of the minor \role\
  mathematical logic played at that time in France, \herbrand's
  supervisor \vessiotname\ had problems finding examiners for the
  thesis.  The final committee consisted of 
  \vessiotname,
  \denjoyname\ \denjoylifetime\ and \frechetname\ \frechetlifetime.
%}

\halftop\halftop\indent
After completing his military service in September\,1930, awarded
with a Rocke\-feller Scholarship, he spent the academic year
1930--1931 in Germany and planned to stay in \Princetonnostate\footnote{%
 \majorheadroom
 \Cfnlb\ \citep[\Vol\,V, \p\,3\f]{goedelcollected}. \ 
 \Cfnlb\ \citep{hasse-herbrand-correspondence}
 for the letters
 between \hasseindex\hasse, \herbrand, and \wedderburnname\ \wedderburnlifetime\ 
 (\Princetonnostate) on \herbrand's visit to \Princetonnostate.%
}
for the year after.

\pagebreak

\noindent
\herbrand\ visited the following mathematicians:%
\par\halftop\halftop\halftop\noindent\noetherindex\neumannindex\artinindex
\LINEnomath{
  \begin{tabular}{l|l|l|l}hosting scientist\footroom
    &lifetime
    &place
    &time of \herbrand's stay
    \\\hline
    \begin{tabular}[c]{@{}l@{}}
      \neumannname
      \\\artinname
      \\\noethername\footnotemark
      \\\end{tabular}
    &\begin{tabular}[c]{@{}c@{}}
      \neumannlifetime
      \\\artinlifetime
      \\\noetherlifetime   
      \\\end{tabular}
    &\begin{tabular}[c]{@{}l@{}}
      \Berlin\ 
      \\\Hamburg
      \\\Goettingen
      \\\end{tabular}
    &\begin{tabular}[c]{@{}l@{}}
        \math\{      
      \\\math\{    
      \\\math\{    
      \\\end{tabular}
      \begin{tabular}[c]{@{}l@{}}
      \Oct\,20, 1930\,\footnotemark
      \\middle of \May\,1931\footnotemark
      \\middle of \Jun\,1931
      \\middle of \Jul\,1931\footnotemark
      \\\end{tabular}
    \\\end{tabular}}\addtocounter{footnote}{-3}\footnotetext{%
 % \majorheadroom
 \herbrand\ had met \noether\ already very early in\,1931 in \Halle. \
 \Cfnlb\ \citep[\p 106, \litnoteref{10}]{noether-hasse-correspondence}.}%
\addtocounter{footnote}{1}\footnotetext{\label{new note}%
 \majorheadroom
 See \noteref{note Herbrand leave Germany} for the not completely
 reliable source of this date.
 The following sentence from 
 \herbrand's letter to \chevalleyname\ of December\,3, 1930,
 however,
 may indicate that he arrived in \Berlin\
 not before middle of November (\cf\ \cite[\Vol\,V, \p\,4, \litnoteref i]
 {goedelcollected}):\begin{quote}
 ``{\frenchfont 
 Les math\'ematiciens sont une bien bizarre chose;
 voici une quinzaine de jours que chaque fois que je vois \neumann\
 nous causons d'un travail d'un certain \goedel, 
 qui a fabriqu\'e de bien curieuses fonctions; 
 et tout cela d\'etruit quelques notions solidement ancre\'es.}''%
 \notop\end{quote}}%
\addtocounter{footnote}{1}\footnotetext{%
 \majorheadroom
 The third letter of \herbrand\ to \hasseindex\hasse\ is dated May\,18, 1931,
 in Hamburg; \cfnlb\ \cite{hasse-herbrand-correspondence}.%
}%
\addtocounter{footnote}{1}\footnotetext{\label{note Herbrand leave Germany}%
 \majorheadroom
 In \nolinebreak his report to the Rockefeller
 Foundation, \herbrand\ wrote that his stay in Germany lasted from \Oct\,20,
 1930, until the {\em end}\/ of \Jul\,1931, which is unlikely because
 he died on \herbranddeathday, in France,
 and because \herbrand's sixth letter to \hasseindex\hasse\ is dated 
 \Jul\,23, 1931, in Paris, France,
 \cfnlb\ \cite{hasse-herbrand-correspondence}. \hskip.3em
  According to \citep[\p 73]{Dubreil-1983-a}, 
 \herbrand\ stayed in 
 \index{Goettingen@G\"ottingen}\Goettingen\ 
 until the {\em beginning}\/ of \Jul\,1931.%
}%
\par\halftop\halftop\halftop\noindent\herbrand\ discussed his ideas with 
\bernaysname\ \bernayslifetime\ 
in \Berlin, and he met \bernaysname, 
\hilbertindex\hilbertname\ \hilbertlifetime, and 
\courantname\ \courantlifetime\
later in 
\index{Goettingen@G\"ottingen}%
\Goettingen.\footnote{%
 \majorheadroom
 That \herbrand\ met \bernays, \hilbert, and \courantindex\courant\ in 
 \index{Goettingen@G\"ottingen}%
 \Goettingen\ is most likely, but we cannot document it. \ 
 \hilbert\ was still lecturing regularly in\,1931, \cfnlb\
 \citep[\p 199]{reid-hilbert}. \ 
 \courantindex\courant\ wrote a letter to 
 \herbrandsfatherindexend\herbrand's father,
 \cfnlb\ \citep[\p\,25, \litnoteref 1]{herbrand-logical-writings}.} 

\halftop\indent
On \herbrandtogoedeldate, \herbrand\ wrote a letter to 
\goedelname\ \goedellifetime, who answered with some delay on \Jul\,25, most
probably too late for the letter to reach \herbrand\ before his early
death two days later\fullstopnospace\footnote{%
 \majorheadroom
 \Cfnlb\ \citep[\Vol\,V, \PP{3}{25}]{goedelcollected}.}

\halftop\indent
In other words, although {\herbrandname} was still a relatively
unknown young scientist, he was well connected to the best
mathematicians and logicians of his time, particularly to those
interested in the foundations of mathematics. ---

\begin{sloppypar}
\halftop\halftop\halftop\halftop\halftop\halftop\halftop\indent
\herbrand\ met \hassename\ \hasselifetime\ at the {\germanfont
  Schie\fki\oe rper-Kongre\ses} 
(\Feb\,26~--~\Mar\,1, 1931) \hskip.2em
in \Marburg, \hskip.2em
and he wrote six letters including plenty of mathematical ideas to
\hasse\ afterward.\footnote{%
 \majorheadroom
 \Cfnlb\ \cite{hasse-herbrand-correspondence}.%
} \
After exchanging several
compassionate letters with 
\herbrandsfatherindex\herbrand's father,\footnote{%
 \majorheadroom
 \Cfnlb\ \citep{hasse-herbrand-senior-correspondence}.} \hskip.3em  
\hasseindex\hasse\ wrote \herbrand's obituary which is printed as the 
foreword to
\herbrand's article on the consistency of arithmetic
\citep{herbrand-consistency-of-arithmetic}.\par\end{sloppypar}

\vfill\pagebreak
%%%%%%%%%%%%%%%%%%%%%%%%%%%%%%%%%%%%%%%%%%%%%%%%%%%%%%%%%%%%%%%%%%%%%%%%%%%%%%% 
%\section{Logic}\label{sec:logic}
\section{Finitistic 
%Classical \FirstOrder\ Proof Theory}
Proof Theory of Classical \FirstOrder\ Logic}
\label
{section Subject Area and Methodological Background}%
\index{finitism}%
\index{logic!classical}%
\index{logic!two-valued}%

\noindent \herbrand's work on logic falls into the area of what is
called {\em proof theory}\/ today. \ 
More specifically, he is concerned
with the {\em finitistic} analysis of {\em two-valued} (\ie\ {\em
  classical}\/), {\em\firstorder}\/ logic and its relationship to
sentential, \ie\ propositional logic.

Over the millennia, logic developed as {\em proof theory}. \hskip.1em
% The 
%CP 20140420:
A
key
observation of the ancient Greek schools
% , first formulated by \aristotlename\ \aristotlelifetime, \hskip .2em 
%CP 20140420
is that certain
patterns of reasoning are valid irrespective of their actual
denotation. 
From ``all men are mortal'' and ``Socrates is a man'' we
can conclude that ``Socrates is mortal''
% irrespectively
%CP:
---~irrespective
of Socrates' most interesting personality and the contentious meaning
of ``being mortal'' in this and other possible worlds. 
The discovery
of 
% those
%CP 20140420
patterns of reasoning, 
% called 
%CP 20140420:
such as these
{\em syllogisms}, where
meaningless symbols 
% are
%CP 20140420:
can be
used instead of everyday words, was the
starting point of the known history of 
% mathematical 
%CP 20140420
logic in the
ancient 
world. \hskip.2em
For over two millennia, the development of these rules
for drawing conclusions from given assumptions just on the basis of
their {\em syntactical form}\/ was the main subject of logic.

{\em Model theory} --- on the other hand --- the study of formal languages
and their interpretation, became a respectable way of logical
reasoning through the seminal works of 
\loewenheimname\ \loewenheimlifetime\ and \tarskiname\ \tarskilifetime. \ \ 
Accepting the
actual infinite, model theory considers the %possible 
set-theoretic semantical structures of a given language. \ 
With \tarski's work, the relationship between these two areas of logic
assumed overwhelming importance --- as
captured in our little anecdote 
of the \firstsectionname\ (\sectref{section Still ten minutes to go}),
where `\math\models' \nolinebreak signifies model-theoretic validity and
`\tightyields' \nolinebreak
denotes proof-theoretic derivability.\loewenheimindex\footnote{%
 As we will discuss in \sectref{section herbrand loewenheim skolem},
 \herbrandname\ still had problems in telling `\tightyields' and 
 `\math\models' apart: \ 
 For instance, he blamed 
 \loewenheim\ for not showing 
 \index{consistency!of first-order logic}%
 consistency of \firstorder\ logic,
 which is a property related to 
 \herbrand's \nolinebreak `\tightyields\closesinglequotecommaextraspace 
 but not to \loewenheim's \nolinebreak
 `\math\models\closesinglequotefullstopextraspace}

\herbrand's 
% Joerg wrote :lifetime and 
% But this is grammatically a problem in the following sentence.
% Moreover, however contentious the meaning of being mortal may ever be,
% the dead do not work scientifically.
scientific work 
% does not only fall into the time the development 
coincided with the maturation
of modern logic,
%culminated, 
as marked {\it inter alia}\/ by \goedel's 
% The completeness theorem should not be mentioned here 
% because it is not a culmination.
% If someone thinks it should be mentioned,
% please add also the citation!
\index{Goedel@G\"odel!'s First Incompleteness Theorem}%
\index{Goedel@G\"odel!'s Second Incompleteness Theorem}%
\incompletenesstheorem s of 1930--1931.\footnote
{\label{note incompleteness theorems}\Cfnlb\ \citep{goedel}, 
 \citep{rosser-incompleteness}. \ 
 For an interesting discussion of the reception of 
 the \incompletenesstheorem s \cfnlb\ 
\dawsonindex\cite{dawson-reception-goedel}.}  
It was strongly influenced by the
foundational crisis in mathematics as well. \ 
\russellsparadox\ was not only a personal calamity to 
\fregename\ \fregelifetime,\footnote 
{\Cfnlb\ \citep[\Vol\,II]{frege-grundgesetze}\@.} 
but it jeopardized the whole
enterprise of set theory and thus the foundation of modern mathematics. \ 
% into jeopardy. \ 
From an epistemological point of view, maybe there
was less reason for getting scared as it appeared at the time: \ 
As \wittgensteinname\ \wittgensteinlifetime\ reasoned later,\footnote{%
 \Cfnlb\ \eg\ \citep{wittgenstein-lectures-fundations-of-mathematics}.%
} 
the detection of 
inconsistencies is an inevitable element of human learning, and many
logicians today would be happy to live at such an interesting time of
a raging\footnote
{This crisis has actually never been resolved in the sense that we 
 would have a single set theory that suits all the needs of a
 working mathematician. 
 \notop\halftop\begin{quote}
 ``\englishquoteforstertext''
 \getittotheright{\englishquoteforstertextplace}
 \end{quote}} 
foundational crisis. \ 
\pagebreak

\yestop\yestop\noindent
The development of mathematics, however, more
often than not attracts the intelligent young men looking for clarity
and reliability in a puzzling world, threatened by social complexity. \ 
As 
\hilbertindex\hilbertname\ put it:
\begin{quote}
  ``\germantexthilbertsicherheit''\footnote
{\Cfnlb\ \citep[\p 170]{unendliche}.\begin{quote}
 ``And where else are security and truth to be found,
 if even mathematical thinking fails?''% 
% \opt{our translation}
\end{quote}}%\pagebreak
\end{quote}
\begin{quote}
  ``\germantexthilbertignorabimus''\footnote
{\Cfnlb\ \citep[\p 180, modernized orthography]{unendliche}.\begin{quote}
  ``After all, one of the things that attract us most when we apply
  ourselves to a mathematical problem is precisely that within us we
  always hear the call: here is a problem, search for the solution;
  you can find it by pure thought, for in mathematics there is no {\em
    ignorabimus}.''\getittotheright 
  {\translationnotewithlongcite
   {\p\,384}{heijenoort-source-book}{translation by \bauermengelbergname}}%
\end{quote}}
\end{quote}

\yestop\halftop\noindent
Furthermore, 
\hilbertindex\hilbert\ did not want to surrender to the new
\index{intuitionism}%
``intuitionistic'' movements of 
\brouwername\ \brouwerlifetime\ and
\weylname\ \weyllifetime, who suggested a restructuring of mathematics
with emphasis on the problems of existence and 
\index{consistency|(}%
consistency rather than 
elegance,
giving up many previous achievements, especially in analysis and in
the set theory of \cantorname\ \cantorlifetime:\footnote 
{\Cfnlb\ \makeaciteofthree{brouwer1}{brouwer2}{brouwer3}, \
 \makeaciteoftwo{weyl-1921}{weyl-grundlagenvortrag}, \
 \citep{cantor-collected}.}
\begin{quote}
  ``\germantexthilbertcantorsparadise''\footnote
{\Cfnlb\ \germantexthilbertcantorsparadisecitation.\begin{quote}
 ``No one shall drive us from the paradise \cantorindex\cantor\ has created.''
 \notop\notop\end{quote}%
% \opt{our translation}
}
\end{quote}

\begin{sloppypar}\yestop\halftop\noindent
Building on the works
of \dedekindname\ \dedekindlifetime, \hskip.3em
\peircename\ \peircelifetime, \hskip.3em
\schroedername\ \schroederlifetime, \hskip.3em
\fregename\ \fregelifetime, \hskip.3em
and 
\peanoname\ \peanolifetime, \hskip.3em
the celebrated three volumes of 
\index{Principia Mathematica}%
{\em\PM}\/ \citep{PM} \hskip.3em
of 
\whiteheadname\ \whiteheadlifetime\ \hskip.3em
and
\russellname\ \russelllifetime\ \hskip.3em
had provided evidence
that --- \nolinebreak in principle \nolinebreak--- mathematical
proofs could be reduced to logic, \hskip.3em
using only a few rules of inference and appropriate axioms.
\end{sloppypar}

\yestop\yestop\noindent
The goals of 
\hilbertindexbegin\index{program@\programme!Hilbert's|(}%
\index{finitism|(}%
{\em\hilbertsprogram}\/ on the foundation of
mathematics, however, extended well beyond this: \ 
% in the 1920s: 
His contention was that the reduction of mathematics to formal theories of
logical calculi would be 
% nice and well, but of no help out of 
insufficient to resolve 
the foundational crisis of mathematics, neither would it protect against
\russellsparadox\ and other inconsistencies in the future --- unless the
consistency of these theories could be shown formally by simple
means.

\yestop\yestop\noindent
Let us elaborate on what was meant by these ``simple means\closequotefullstop
\\\mbox{}\majorfootroom
% Joerg wrote: And ``simple means'' were taken to mean ``finitistic'' means.
% This does not help the reader here and is explained later at the proper
% place in the text anyway. CP
\vfill\pagebreak

Until he moved to \Goettingen,
\hilbert\ lived in \Koenigsberg, \hskip.2em
and his view on mathematics in the 1920s 
was partly influenced by 
\kantindexbegin\kant's {\em Critique of pure reason}.\footnote
{\label{note objective reality}The transcendental philosophy of 
{\em pure speculative\/\footnotemark\ reason}\/ 
is developed in the main work on 
epistemology, the
{\em Critique of pure reason}\/ \makeaciteoftwo{KrVA}{KdrV},
of \kantname\ \kantlifetime, who spent most of his life in \Koenigsberg\
and strongly influenced the education 
% on 
at
\hilbert's 
high school and university in \Koenigsberg. \ 
The {\em Critique of pure reason}\/ elaborates how little we can know about
things independent of an observer
({\em things in themselves}, {\germanfontfootnote\em Dinge an sich selbst}\/)
in comparison to our conceptions, 
\ie\ the representations of the things within our thinking 
({\germanfontfootnote Erscheinungen und sinnliche Anschauungen, 
 Vorstellungen}). \ 
In \nolinebreak what he compared\footnotemark\
to the \kopernikan\ revolution 
\cite[\p\,XVI]{KdrV}, \hskip.2em
\kant\ \nolinebreak 
considered the conceptions gained in connection with sensual experience
to be real and partly objectifiable,
and accepted the
%  materialistic 
things in themselves only as limits of 
our thinking, about which nothing can be known for certain. \
%% For instance, besides a material horse,
%% a thing in itself which common sense would claim to exist,
%% for \kant\ there are the concrete real horse of our perception
%% and the horseness as an {\em ideal}\/ thing in itself.
}\addtocounter{footnote}{-1}\footnotetext{%
 \majorheadroom
 The term ``{\em pure (speculative) reason}\/'' is opposed to ``{\em (pure) 
 practical reason}\/\closequotefullstop}\addtocounter{footnote}{1}%
 \footnotetext{%
 \majorheadroom
 Contrary to what is often written, \kant\ never wrote
 of ``his \kopernikan\ revolution of philosophy\closequotefullstop} \
% In the world of material objects and 
% their sensual perception, this is a surprising point of view. \ 
% In mathematics, however, such a view is 
% straightforward and can be very helpful. \ 
Mathematics as directly and intuitionally perceived by a mathematician is called
{\em contentual}\/\footnote{% 
 \majorheadroom
 The word 
 \index{contentual!history of the English word}%
 ``contentual'' did not use to be part of the English language
 until recently. \ 
 For instance, 
 it \nolinebreak is not listed in the most complete Webster's \cite{webster}. \
 According to \cite[\p\,viii]{heijenoort-source-book}, \hskip .3em
 this \nolinebreak neologism was 
 % especially 
 introduced by \bauermengelbergname\
 as a translation for the word
 ``{\germanfontfootnote inhaltlich}'' in German texts on mathematics and logic,
 because there was no other way to reflect the special intentions
 of the \hilbert\ school when using this word. \
 In \nolinebreak January\,2008, \hskip .2em
 ``contentual'' got 6350 Google hits, 5600 of which, however,
 contain neither the word ``\hilbert'' nor the word 
 ``\bernays\closequotefullstopextraspace
 As these hits also include a pop song, \hskip .2em
 ``contentual'' is likely to become an English word outside science in 
 the nearer future. \
 For a comparison, 
 there were 4\,million Google hits for ``contentious\closequotefullstop} 
({\germanfont\em inhaltlich}\/) by \nolinebreak\hilbert. \
According to \citep{unendliche}, \hskip.2em
the notions and methods of contentual mathematics are partly abstracted from 
finite symbolic structures 
(such as explicitly and concretely given natural numbers, proofs, and 
 algorithms) \hskip.2em
where we can effectively {\em decide}\/ (in
finitely many effective steps) \hskip.2em
whether a given object has a certain
property or not. \
Beyond these
{\em aposterioristic}\/ abstractions from phenomena, 
contentual mathematics also has a
{\em synthetic}\/\footnote{\label{note wrong}%
  \majorheadroom
  ``synthetic'' is the opposite of
  ``analytic'' and means that a statement provides new information
  that cannot be deduced from a given knowledge base. \ Contrary to
  \kant's opinion that all mathematical theorems are synthetic
  \citep[\p 14]{KdrV}, \ (contentual) mathematics also has
  analytic sentences. \ 
  In \nolinebreak particular, \kant's example
  \bigmaths{7+5=12}{} becomes analytic when we read it as
  \bigmaths{\plusppnoparentheses{\sppiterated 7\zeropp} {\sppiterated
      5\zeropp}=\sppiterated{12}\zeropp}{} and assume the
  non-necessary, synthetic, aprioristic axioms \bigmathnlb{\plusppnoparentheses
    x\zeropp=x}{} and \bigmaths{\plusppnoparentheses x{\spp
      y}=\spp{\plusppnoparentheses x y}}. \ 
\fregeindex\Cfnlb\ \citep[\litsectref{89}]{frege-grundlagen}.}  
{\em aprioristic}\/\footnote{%
  \majorheadroom
  ``{\em a priori}\/'' is the opposite of
  ``{\em a posteriori}\/'' and means that a statement does not depend
  on any form of experience. \ For instance, all necessary 
  \citep[\p\,3]{KdrV} and all analytic \citep[\p 11]{KdrV}
  statements are {\em a priori}. \ 
  Finally, \kant\ additionally 
  assumes that all aprioristic statements are
  necessary \citep[\p\,219]{KdrV}, which seems to be wrong, 
  \cfnlb\ \noteref{note wrong}.}  
aspect, 
which depends neither
on experience nor on deduction, and which cannot be reduced to logic,
but which is transcendentally related to intuitive conceptions. \
Or, as \nolinebreak\hilbert\ put \nolinebreak it:
\notop\halftop\begin{quote}
\noindent``\germantexthilbertonkant''\footnotemark\end{quote}
\pagebreak\par
\footnotetext{%
  \Cfnlb\ \citep[\p 170\f, modernized orthography]{unendliche}.%
  \begin{quote}
    ``\englishtexthilbertonkant''\getittotheright 
   {\translationnotewithlongcite
     {\p\,376}
     {heijenoort-source-book} 
     {transl.\ by \bauermengelbergname, 
      modified\footnotemark}}\end{quote}}%
\addtocounter{footnote}{1}%
\footnotetext{%
  \majorheadroom
  We have replaced ``\mbox{extralogical}'' with
       ``extra-logical\closequotecomma
       and --- more importantly ---
       ``experience'' with ``conception\closequotecomma for
       the following reason: 
       Contrary to ``{\germanfontfootnote Erfahrung}'' (experience),
       the German word ``{\germanfontfootnote Erlebni\es}'' 
       does not suggest an {\em aposterioristic}
       intention, which would contradict the obviously {\em aprioristic}\/ 
       intention of \hilbert's sentence.%
 \majorfootroom
}%
\noindent
To refer to intellectual concepts
which are not directly related to sensual perception
or intuitive conceptions, \hskip.2em
both \kant\ and \hilbert\ use the word 
``ideal\closequotefullstopextraspace
{\em Ideal}\/ objects and methods in mathematics ---~as opposed to
contentual ones~--- may involve the 
\index{infinite!the actual}%
actual infinite; \hskip.3em 
such \nolinebreak as quantification, \math\varepsilon-binding, set theory, and
non-terminating computations.

According to both \citep{KdrV} and \citep{unendliche}, \hskip.2em
the only possible criteria for the acceptance of {\em
  ideal}\/ notions are {\em consistency}\/ and {\em usefulness}. \ \
Contrary to 
\kantindexend\kant,\footnote{%
 \majorheadroom
 \kant\ considers ideal notions to be problematic,
 because they transcend what he considers to be the area
 of objective reality; \cfnlb\ \noteref{note objective reality}. \ 
 For notions that are consistent, useful, and ideal,
 \kant\ actually introduces the technical term {\em problematic}\/
  ({\germanfontfootnote problematisch}):\begin{quote}
 ``{\germanfontfootnote\germantextkantsnotionofproblematisch}''\nopagebreak
\getittotheright
  {\citep[\p\,310, modernized orthography]{KdrV}}\end{quote}}
however, \hilbert\
is willing to accept useful ideal theories, provided that
their consistency can be shown with contentual and intuitively clear
methods --- \ie\ with ``simple means\closequotefullstop

These ``simple means'' that may be admitted here must be, 
on the one hand, 
sufficiently expressive and powerful to show the 
\index{consistency!of arithmetic}%
consistency of arithmetic,
but, on the other hand, 
simple, \ie\ intuitively clear and contentually reliable. \
The notion of 
\index{finitism}%
{\em\hilbert's finitism}\/ was born out of the conflict
of these two goals. 

Moreover, \hilbert\ expresses the hope that the new proof theory, 
primarily developed to show the 
\index{consistency!proof of}%
consistency of ideal mathematics with
contentual means, would also admit (possibly ideal, \ie\
non-finitistic) proofs of 
\index{completeness!{\em definition}}%
{\em completeness}\/\footnote{%
 \majorheadroom
 A theory is {\em complete} \udiff\ for any formula \nlbmath A
 without free variables 
 (\ie\ any closed formula in the given language) \hskip.2em
 which is not part of this theory, 
 its negation \nlbmath{\neg A} is part of this theory.}
for certain mathematical theories. \ 
If this goal of \hilbertsprogram\ had been achieved, then ideal proofs
would have been justified as convenient short-cuts for constructive,
contentual, and intuitively clear proofs, 
% so that 
and
---~even under the
threat of \russellsparadox\ and others~--- 
there would be no reason to
give up the paradise of axiomatic mathematics and abstract set theory.

And these basic convictions of the \hilbert\ school constituted the
most important influence on the young student of mathematics \herbrandname.

\pagebreak

As \goedel\ showed with his famous 
\index{Goedel@G\"odel!'s First Incompleteness Theorem}%
\index{Goedel@G\"odel!'s Second Incompleteness Theorem}%
\incompletenesstheorem s in 
1930--1931,\arXivfootnotemarkref{note incompleteness theorems}
however, the
consistency of any (reasonably conceivable)
recursively enumerable
mathematical theory that
includes arithmetic excludes both its 
\index{completeness}%
completeness and
the existence of a finitistic 
\index{consistency|)}%
\index{consistency!proof of}%
consistency proof.

Nevertheless, the contributions of \ackermannname\ \ackermannlifetime,
\bernaysindex\bernays, \herbrand, and 
\gentzenindex\gentzen\ within 
\hilbertindexend\index{program@\programme!Hilbert's|)}%
\index{finitism|)}%
\hilbertsprogram\ gave {\em
  proof theory}\/ a new meaning as a field in which proofs are the
objects and their properties and constructive transformations are the
field of mathematical study, just as in arithmetic the numbers are the
objects and their properties and algorithms are the field of study.%
\notop\halftop\pagebreak

\section
{\herbrand's Main Contributions to Logic and\\his Notion of Intuitionism}\label
{section Contributions and his Notion of Intuitionism}%
\index{intuitionism|(}%

\noindent
The essential 
% Joerg wrote: "most essential", but there is neither a 
% comparative nor a superlative to this word.
works of {\herbrand} on logic are his \PhDthesis\
\citep{herbrand-PhD} \hskip.1em
and the subsequent journal article
\citep{herbrand-consistency-of-arithmetic}, \hskip .3em
both to be found in \citep{herbrand-logical-writings}.\footnote 
{\label{note on herbrand-logical-writings}%
 This book is not just an English translation 
 of \herbrand's complete works on logic:
 It contains additional annotation, brief introductions, and extended notes by
 \heijenoortname, \drebenname, and \goldfarbname.
 \textonherbrandlogicalwritings\
 We would appreciate to include also 
 \herbrand's mathematical writings outside of logic; some remarks on
 the two \repair s of \herbrandsfalselemma\ by \goedel\ and
\heijenoortindex\heijenoort, respectively,  \cfnlb\ \sectrefs
 {section lemma}{section herbrand fundamental theorem} below; \ 
 and \herbrand's correspondence. \ 
 The correspondence with \goedelindex\goedel\ is published in
 \citep[\Vol\,V, \PP{14}{25}]{goedelcollected}. \ 
 \herbrand's letters to \hasseindex\hasse\ 
 are published in \cite{hasse-herbrand-correspondence}.
 The whereabouts of the rest of his correspondence is unknown.%
}  

The main contribution is captured in what is called today
{\em\herbrandsfundamentaltheorem}. \ 
Sometimes it is simply
called ``\herbrand's Theorem\closequotecomma but the longer name is
preferable as there are other important
``\herbrand\ theorems\closequotecomma such as the {\herbrand--\ribet}
Theorem.  \
Moreover, \herbrand\ himself calls it
``{\frenchfont Th\'eor\`eme fondamental}\/\closequotefullstop
%extraspace
% Finally, it deserves the name ``fundamental'' because of its similarity to
% \gentzensHauptsatz\ (\ie\ fundamental (or main) theorem) and
% {\germanfont \verschaerfterHauptsatz}.

The subject of \herbrandsfundamentaltheorem\ is the effective reduction
% CP wrote: 
% "\herbrandsfundamentaltheorem\ expresses the effective reduction". 
% Joerg did not like that.
% Joerg wrote: 
% "\herbrandsfundamentaltheorem\ shows the effective reduction". 
% But only proofs show.
% Alternative: 
% "\herbrandsfundamentaltheorem\ states the effective reducibility".
% But the above is better than all these because it also expresses 
% the necessary vagueness of the statement without an additional
% "roughly speaking" or "inter alia".
of (the semi-decision problem of) provability in \firstorder\ logic to 
provability in sentential logic. 

Here we use the distinction well-known to \herbrand\ and 
his contemporaries between {\em\firstorder\ logic}\/ 
(where quantifiers bind variables ranging over individual objects of 
 the universe, \ie\ of the domain of reasoning or discourse) \hskip .2em
and {\em sentential logic}\/ without any quantifiers. \ 
Validity of a formula in sentential logic is effectively decidable,
for instance with the truth-table method. \ 
%% The question is: Can we do the same for \firstorder\ logic? \ 
%% If so, there would be an affirmative answer to \hilbertsprogram,
%% provided all of mathematics could be formulated within the frame of 
%% \firstorder\ logic.  
% I have removed this because it sounds perfectly silly.
% Moreover, it is hard to understand, what is meant,
% because it requires the historic context on decidability,
% while the previous is a complete modern statement of the theorem
% followed by naive explanation.
% This is an uneven development par excellence!
% Moreover, it is absolutely superfluous at this early stage.
% Finally, the removal improves the flow to the next sentence. CP

Although \herbrand\ spends \litchapref 1 
of his thesis on the subject, \hskip.1em
he actually ``shows no interest for the sentential work''
\heijenoortindex\citep[\p 120]{heijenoort-work-herbrand},
and takes it for granted.
\begin{enumerate}
\item[A.]  Contrary to 
\index{Gentzen!'s {\germanfont Hauptsatz}}%
\gentzensHauptsatz\ \citep{gentzen}, \hskip.3em
  \herbrandsfundamentaltheorem\ starts right with a 
  {\em single \sententialtautology}\/ 
  (\cfnlb\ \sectref{section herbrands calculi} below). \ 
  He treats this property as given and does not fix a
  concrete method for establishing it.
\item[B.]  The way \herbrand\ presents his sentential logic in terms
  of `\math\neg' and `\math\vee' indicates that he is not concerned
  with 
\index{logic!intuitionistic}%
{\em intuitionistic logic}\/ as we understand the term today.\footnote
{\Cfnlb\ \eg\ 
\heytingindexbegin\makeaciteoftwo{heyting-1930-logic}{heyting}, \citep{gentzen}.}
\end{enumerate}
%%
% Indeed, \herbrand\ is not occupied with intuitionistic logic as 
% an object of his studies, nor does he restrict his meta logic to
% intuitionistic logic. \ 
Contrary to 
\index{Gentzen!'s calculi}%
\gentzen's sequent calculus \nolinebreak\LK\
\citep{gentzen}, in \herbrand's calculi
we do not find something like a sub-calculus \nolinebreak\LJ\
for intuitionistic logic. \ 
Moreover, there is no way to generalize \herbrandsfundamentaltheorem\
to include intuitionistic logic: \ 
Contrary to the Cut elimination in 
\index{Cut elimination}%
\index{Gentzen!'s {\germanfont Hauptsatz}}%
\gentzensHauptsatz,
% (\cfnlb\ \citep{gentzen}), \ 
the \nolinebreak elimination of 
\index{modus ponens}%
\index{modus ponens!elimination}%
{\em modus ponens}\/ according to
\herbrandsfundamentaltheorem\ does not hold for intuitionistic logic.
\begin{quote}
  ``All the attempts to generalize \herbrand's theorem in that
  direction have only led to partial and unhandy results (see
\heijenoortindex\citep{heijenoort-mints}).'' \getittotheright
  {%
\heijenoortindex\citep[\p 120\f]{heijenoort-work-herbrand}}
\end{quote} 

\vfill\pagebreak

\noindent
When \herbrand\ uses the term ``intuitionism\closequotecomma this
typically should be understood as referring to something closer to the
\index{finitism}%
finitism of \hilbert\ than to the intuitionism of
\brouwerindex\brouwer.\footnote
{%
\goedelindex\goedel, however, expressed a
  different opinion in a letter to 
\heijenoortindex\heijenoort\ of \Sep\,18, 1964:\begin{quote}
  ``In \litnoteref 3 of \citep{herbrand-consistency-of-arithmetic} 
    he does {\em not}\/ require the enumerability of mathematical objects, and
    gives a definition which fits 
    \brouwerindex\brouwer's intuitionism 
    very well'' % There is no full-stop here!
    \getittotheright
    {\citep[\Vol\,V, \p\,319\f]{goedelcollected}}%\vspace*{-2ex}
\end{quote}} \
This ambiguous usage of the term ``intuitionism'' ---~centered around the
partial rejection of the Law of the Excluded Middle, the
\index{infinite!the actual}%
actual infinite,
as well as quantifiers and other binders~--- was common 
in the 
\index{Hilbert!school}%
\hilbert\ school at \herbrand's time.\footnote
{\Cfnlb\ \eg\ \citep[\p\,283\f]{herbrand-logical-writings}, \ 
\citep[\p\,82\ff]{tait-2006}.\majorfootroom} \
\herbrand's view on what he calls ``intuitionism'' 
is best captured in the following quote:
% Please, keep in mind, 
% however, that we have to replace ``intuitionistic'' with ``finitistic''!
\begin{quote}%
%\selectlanguage{french}%
``\frenchtextninty''\footnote{%
 \majorheadroom
 \Cfnlb\ \frenchtextnintylocationoriginal. \ 
 \frenchtextnintylocationoriginalmodifierforreprint\ also in:
 \frenchtextnintylocationreprint.
 % \selectlanguage{USenglish}%
 \begin{quote}``By an intuitionistic argument 
 we understand an argument satisfying the following conditions: in it
 we never consider anything but a given finite number of objects and
 of functions; these functions are well-defined, their definition
 allowing the computation of their value in a univocal way; we never
 state that an object exists without giving the means of constructing
 it; we never consider the totality of all the objects \nlbmath x of
 an infinite collection; and when we say that an argument (or a
 theorem) is true for all these \nlbmath x, we mean that, for each
 \nlbmath x taken by itself, it is possible to repeat the general
 argument in question, which should be considered to be merely the
 prototype of the particular arguments.'' \getittotheright
 {\translationnotewithlongcite{\litnoteref 5, \p\,288\f}
  {herbrand-logical-writings}{translation by \heijenoort}}%\notop
\end{quote}}
\end{quote}
% It seems hardly possible to bring more clarity to the historical
% subject of \herbrand's exact notion of
% ``intuitionism''
Contrary to today's precise meaning of the term ``intuitionistic
logic\closequotecomma the terms 
\index{intuitionism|)}%
``intuitionism'' and ``finitism''
denote slightly different concepts, which are related to the
philosophical background, differ from person to person, and vary
over times.\footnote 
{Regarding intuitionism,
  besides 
\brouwerindex\brouwer, 
\weylindex\weyl, and \hilbert, we may count
\kroneckername\ \kroneckerlifetime\ and
\poincarename\ \poincarelifetime\ among the ancestors, and have to mention
\heytingindexend\heytingname\ \heytinglifetime\ for his major differing view, 
  \cfnlb\ \eg\
  \makeaciteofthree{heyting-1930-logic}{heyting-1930-mathematics}{heyting}. \ 
  Deeper discussions of \herbrand's notion of ``intuitionism'' can be found in
\heijenoortindex\citep[\PP{113}{118}]{heijenoort-work-herbrand}  
  and in \citep[\p\,82\ff]{tait-2006}\@. \
  Moreover, we briefly discuss it in \noteref{note intuitionism}. \ 
  For more on finitism \cfnlb\ \eg\
  \citep{parsons-finitism}, \citep{tait-finitism}, \citep{zach-finitism}. \ 
  For more on \herbrand's background in
  philosophy of mathematics, \cfnlb\ \citep{herbrand-thoughts},
  \citep{DubucsEgre-Herbrand-PUF}.}  

\pagebreak

While \herbrand\ is not concerned with 
\index{logic!intuitionistic}%
intuitionistic logic,
he is a finitist with respect to the following two aspects:
\begin{enumerate}
\item[1.]\herbrand's work is strictly contained within 
\hilbertindexbegin\index{program@\programme!Hilbert's|(}%
\index{finitism|(}%
\hilbert's
  finitistic \programme\ and he puts ample emphasis on his finitistic
  standpoint and the finitistic character of his theorems.
\majorfootroom
\noitem\item[2.]\herbrand\ does not accept any
  model-theoretic semantics unless the models are finite. \ In this
  respect, \herbrand\ is more finitistic than \hilbert, who demanded
  finitism only for consistency proofs. 
% \ Moreover, because of \herbrand's
%   rejection of set-based semantics, in \sectref{section herbrand
%     loewenheim skolem} we will have to take \herbrand's statements on
%   \loewenheimindex\citep{loewenheim-1915} with a grain of salt.
\begin{quote}
``\herbrand's negative view of set theory leads him to take,
on certain questions,
a \nolinebreak stricter attitude than \hilbert\ and his collaborators. \ 
He is more royalist than the king. \
\hilbert's metamathematics has as its main goal to establish the consistency 
of certain branches of mathematics and thus to justify them; \hskip.3em
there, one had to restrict himself to finitistic methods. \ 
But in logical investigations other than the consistency problem of
mathematical theories the 
\index{Hilbert!school}%
\hilbert\ school was ready to work with
set-theoretic notions.''\getittotheright
{%
\heijenoortindex\citep[\p 118]{heijenoort-work-herbrand}}%\pagebreak
\yestop\yestop\yestop
\end{quote}\end{enumerate}
%%%%%%%%%%%%%%%%%%%%%%%%%%%%%%%%%%%%%%%%%%%%%%%%%%%%%%%%%%%%%%%%%%%%%%%%%%%%
\section{The Context of \herbrand's Work on Logic}

\noindent Let us now have a look at what was known in \herbrand's time
and at the papers that influenced his work on logic.

\zarembaname\ \zarembalifetime\ \hskip.2em
is mentioned in \citep{herbrand-first}, \hskip.2em
where \herbrand\ cites 
\zaremba's textbook on mathematical logic \citep{zaremba}, \hskip.2em
which clearly influenced \herbrand's notation.\footnote
{\Cfnlb\ \goldfarb's Note to \citep{herbrand-first} on
\p\,32\ff\ in \citep{herbrand-logical-writings}. \ 
\zaremba\ was one of the leading Polish mathematicians in the 1920s. \ 
He had close connections to \Paris, 
but we do not know whether \herbrand\ ever met him.} 

\herbrand's subject, 
\index{logic!first-order}%
\firstorder\ logic, became a field of special interest
not least because of the seminal paper \loewenheimindex\citep{loewenheim-1915},
which singled out \firstorder\ logic in the {\em Theory of
  Relatives}\/ developed by 
\index{Peirce--Schr\"oder tradition}%
\peirce\ and \schroeder.\footnote{\label{note peirce schroeder tradition}%
 \majorheadroom
 For the heritage of 
\peirceindex\peirce\ \cfnlb\ 
%\citep{mitchell-1883},
%\makeaciteoftwo{peirce-1883}{peirce-1885}, 
\citep{peirce-1885}, 
\citep{brady}; for that of 
\schroederindex\schroeder\ \cfnlb\ 
\citep{schroeder-vorlesungen-III}, 
\citep{brady},
\citep{schroeder-handbook}.}
With this paper, \firstorder\ logic became an area of special interest, 
because of the surprising metamathematical properties of this logic,
which was intended to be an especially useful tool with a restricted area of 
application.\footnote{%
 \majorheadroom
 Without set theory, \firstorder\ logic was too poor to serve
 as such a single universal logic as the ones  
 for which 
\fregeindex\frege, 
\peanoindex\peano, and 
\russellindex\russell\ had been searching;
 \cfnlb\ 
\heijenoortindex\citep{heijenoort-absolutism-relativism}. \ 
For the suggestion
of \firstorder\ logic as the basis for set theory, 
we should mention \citep{skolem-1923b}, 
which is sometimes cited as of the year\,1922,
and therefore easily confused with \citep{skolem-1923a}. \ 
For the emergence of \firstorder\ logic as 
the basis for mathematics see \cite{moore-1987}.}

\pagebreak

%\begin{sloppypar}
As the presentation in \loewenheimindex\citep{loewenheim-1915} is opaque, 
\skolemname\ \skolemlifetime\ wrote five clarifying papers
contributing to the substance of 
\loewenheimindex\loewenheim's
{\germanfont Satz\,2}, the now famous \loewenheimskolemtheorem\ 
\skolemindex\makeaciteoffive
{skolem-1920}{skolem-1923b}{skolem-1928}{skolem-1929}{skolem-1941}. \ 
From these papers, \herbrand\ cites \loewenheimindex\citep{loewenheim-1915} and
\citep{skolem-1920}, and the controversy pro and contra \herbrand's
reading of \citep{skolem-1923b} and \citep{skolem-1928} will be
presented in \nlbsectref{section herbrand loewenheim skolem} below.
%\end{sloppypar}

While \herbrand\ neither cites 
\peanoindex\peano\ nor even mentions 
\fregeindex\frege, the
\index{Principia Mathematica}%
{\em\PM} \citep{PM} were influential at his time, 
and he was well aware of \nolinebreak this. \hskip.2em
\herbrand\ cites all editions of the 
\index{Principia Mathematica}%
{\em\PMshort}\/ 
and there are indications that he
studied parts of it carefully.\footnotemark\
% \begin{quote}
% ``The result of this and other features of \russell's formulation 
%   of type theory made it seem quite complicated,
%   so in spite of the enormous influence of\emph\PM\ as a landmark
%   in the development of logic, it was the book that everyone talked about,
%   but practically no one read.''
% \getittotheright{\opt{\citep{andrews-herbrand-award}, \p 171}}
% \end{quote}
% This situation described for the late 1950s, was probably not different 
% from the one in the late 1920s, and, let it be noted in passing, is not 
% different from today: \ 
% The {\em in}\/famous\emph\PM\ is still rare and unaffordable, 
%  and --- as standard notions and notation have changed quite a bit 
%  in the meanwhile --- has 
%  become also quite 
%  incomprehensible for the occasional reader. \ 
%  It is a shame for the social and economical situation of mankind
%  that there is no cheap student edition and no public interactive \WWW\
%  version of the {\em Principia}, 
%  which facilitates look-up by translation into modern notation
%  and online help with obsolete names.
%

\russellindex\russell's influence,
however,
is 
% comparatively 
minor compared to
\hilbert's, as, indeed, \herbrand\ was most interested in proving
consistency, decidability, and 
\index{completeness}%
completeness. \ 
\heijenoortname\ \heijenoortlifetime\ notes on \herbrand:%\pagebreak
\footnotetext{\herbrand\ seems to have studied \math{\ast9} and \math{\ast10} of
\index{Principia Mathematica}%
\citep[\Vol\,I]{PM} \hskip.3em 
  carefully, \hskip.2em
  leaving traces in \herbrand's
\index{Property A}%
  \propertyA\ and in \litchapref 2 of \herbrand's \PhDthesis, \ \cfnlb\
\heijenoortindex\citep[\PP{102}{106}]{heijenoort-work-herbrand}.}%
\begin{quote}
``The difficulties provoked by the 
\russellsparadoxindex\russell\ Paradox, 
stratification, ramification, the problems connected with the 
\index{axiom!of infinity}%
axiom of infinity or the 
\index{axiom!of reducibility}%
axiom of reducibility, nothing of that seems to
retain his
% [\herbrand's] 
attention.
\\  
The reason for this attitude is that \herbrand\ does not share
\russellindex\russell's conception concerning the relation between 
logic and mathematics,
but had adopted 
\hilbertindex\hilbert's. 
In\,1930 
\herbrand\ indicates quite well where he sees the limits of 
\russell's accomplishment:
`So far we have only replaced ordinary language 
with another more convenient one,
but this does not help us \atall\ with respect to the problems regarding 
the principles of mathematics.' \makeanextendedciteoftwo
{\p\,248}{herbrand-hilbert}{\p\,208}{herbrand-logical-writings}. 
And the sentence that follows indicates the way to be followed:
`\hilbert\ sought to resolve the questions which can be raised 
% \opt{posed} 
by applying himself to the study of collections 
of signs which are translations of 
propositions true in a determinate theory.'\,\,''
\getittotheright{%
\heijenoortindex\citep[\p 105]{heijenoort-work-herbrand}.}
% \noitem
\end{quote}

%\begin{sloppypar}
\noindent As \herbrand's major orientation was toward the 
\index{Hilbert!school}%
\hilbert\ school, 
it is not surprising that the majority of his citations\footnote{%
 \majorheadroom
%   herbrand\ provides us with a single list of references. \ It is in
%   his thesis, but does not include all the actual citations of the
%   thesis. \ 
  In his thesis \citep{herbrand-PhD}, \herbrand\
  cites \citep{ackermann-1925}, 
  \citep{artin-schreier},
  \citep{behmann}, \citep{grundlagenvortrag-zusatz},
\bernaysindex\citep{bernays-schoenfinkel},
  \makeaciteofthree{neubegruendung}{unendliche}{grundlagenvortrag},
  \citep{grundzuege}, and
\neumannindex\makeaciteofthree{neumann-1925}{neumann-1927}{neumann-1928}. \ 
  Furthermore,
  \herbrand\ cites \citep{nicod} in \citep{herbrand-PhD}, \hskip.3em
  \citep{ackermann-1928} in \citep{herbrand-fundamental-problem}, \hskip.3em
  and
  \citep{zahlenlehre} and 
\goedelindex\citep{goedel} in
  \citep{herbrand-consistency-of-arithmetic}.}
refer to mathematicians related either to the
\index{Hilbert!school}%
\hilbert\ school or to 
\index{Goettingen@G\"ottingen}%
\Goettingen,
which was the Mecca of mathematicians until its
intellectual 
and 
organizational 
destruction by the Nazis in 1933.

% All in all, although \herbrand\ had a critical mind,
% he was a young scholar and an outright follower of \hilbert's finitism
% and proof theory, which \herbrand\ prefers to call
% ``metamathematics\closequotefullstopextraspace
By 
% Joerg wrote: subsequent 
% But this wrong. CP
% his work on logic and by 
the title of his thesis {\frenchfont\em\herbrandPhDtitle},
\herbrand\ clearly commits himself to King
\hilbert's following, and
%sometimes more royalist than the king.
%\cfnlb\ \sectref{section Subject Area and Methodological Background}. \ 
being the first contributor to 
\hilbert's finitistic \programme\ in
France, he was given the opportunity to write on \hilbert's logic
in a French review journal, and
this paper \citep{herbrand-hilbert} 
is historically interesting because it captures \herbrand's 
personal view of 
\hilbertindexend\index{program@\programme!Hilbert's|)}%
\index{finitism|)}%
\hilbert's finitistic \programme.
%\end{sloppypar}

\vfill\pagebreak

%%%%%%%%%%%%%%%%%%%%%%%%%%%%%%%%%%%%%%%%%%%%%%%%%%%%%%%%%%%%%%%%%%%%%%%%%%%%%%%%
\section{A Genius with some Flaws}\label{section flaws}

\noindent On the one hand, \herbrand\ was a creative mathematician whose 
ideas were truly outstanding, not only for his time. \ 

Besides logic, 
he also contributed 
to class-field theory and to the 
theory of algebraic number fields. \ 
Although this is not our subject here,
we should keep in mind that \herbrand's contributions to algebra 
are as important from a mathematical point of view and as numerous 
as his contributions to logic.\footnote{%
 \Cfnlb\ \ \startacitewithnine
 {herbrand-lost-one}
 {herbrand-lost-two}
 {herbrand-1930b}
 {herbrand-lost-three}
 {herbrand-corps-abstract}
 {herbrand-group}
 {herbrand-1932}
 {herbrand-1932a}
 {herbrand-corps-I}%
 \stopacitewithfour
 {herbrand-discriminant}
 {herbrand-algebraic-functions}
 {herbrand-corps-II}
 {herbrand-1936}, \ 
 \chevalleyindex\citep{chevalley-herbrand-lost-one} \ and \
 \citep{herbrand-chevalley}, as well as 
 \cite{dieudonne-1982}.%
 \majorfootroom% 
}
% \herbrand\ 
% was a creative mathematician whose 
% ideas were truly outstanding at his time. \
% Besides his outstanding contributions to logic, 
% he also contributed an essential simplification of 
% computations pertaining to class field theory
% (\cfnlb\ \citep{herbrand-group,herbrand-1932a,herbrand-1936}),
% to the arithmetic theory of algebraic number fields (\citep
% {herbrand-corps-abstract,herbrand-corps-I,herbrand-corps-II,herbrand-1936}),
% and to the theory of algebraic functions
% (\citep{herbrand-algebraic-functions}). \ 

Among the many statements about \herbrand's abilities as 
a mathematician is 
\weilindex\weil's letter to 
\hasseindex\hasse\ in August\,1931 where he
writes that he would not need to tell him what a loss \herbrand's death means
especially for number theory.\footnote{%
 \majorheadroom
 \Cfnlb\ \citep[\p 119, \litnoteref 6]{noether-hasse-correspondence}.%
 \majorfootroom%
}
As 
\andrewsname\ put it:\begin{quote}
``\herbrand\ was by all accounts a brilliant mathematician.''
%\footnote
\getittotheright{%\Cfnlb\ 
\citep[\p 171]{andrews-herbrand-award}.}
\end{quote}

\noindent
On the other hand, \hskip.2em
\herbrand\ neither had the education nor the
supervision to present his results in proof theory with the technical
rigor and standard, \hskip.2em
say, \hskip.2em
of the 
\index{Hilbert!school}%
\hilbert\ school in the
1920s, \hskip.3em
let alone today's emphasis on formal precision.\footnote{%
 \majorheadroom
%  And the standard of \hilbert's school in the 1920s
%  again falls short of the
%  precision we would want to teach a second year student of formal
%  logic today.\\
%  Furthermore, 
\heijenoortindex\heijenoort\ 
writes in his well-known ``source book\closequotecolon
  \begin{quote}
    ``\herbrand's thesis bears the marks of hasty writing; this is
    especially true of \nolinebreak\litchapref 5\@. \ Some sentences
    are poorly constructed, and the punctuation is haphazard.
    \herbrand's thoughts are not nebulous, but they are so hurriedly
    expressed that many a passage is ambiguous or obscure.  To bring
    out the proper meaning of the text the translators had to depart
    from a literal rendering, and more rewriting has been allowed in
    this translation than in any other translation included in the
    present volume.'' \getittotheright
      {\citep[\p\,525]{heijenoort-source-book}}
  \end{quote}
  Similarly, 
  \goldfarbname, 
  the translator and editor of \herbrand's logical writings,
  writes:\begin{quote}
    ``\herbrand\ also tended to express himself rather hastily,
    resulting in many obscurities;
    in these translations an attempt has been made to balance the demands
    of literalness and clarity.'' \getittotheright
      {\citep[\p V]{herbrand-logical-writings}}\end{quote}} \

Finitistic proof theory sometimes strictly 
demands the disambiguation of form and content
and a higher degree of precision than most other mathematical fields. \ 
Moreover, \hskip.2em
the field was novel at \herbrand's time and probably
hardly anybody in France was able to advise \herbrand\ competently. \

Therefore, \hskip.2em
\herbrand, \hskip.2em
a {\frenchfont\em g\'enie cr\'eateur}, \hskip.2em
as
\heijenoortindex\heijenoort\ called him,\footnote
{\Cfnlb\ \citep[\p 1]{herbrand-ecrits-logiques}.} \hskip.2em
was apt to make undetected errors. \
Well known today is a serious flaw in his thesis
which stayed unnoticed by its reviewers at the time. \ 
Moreover, \hskip.2em
several theorems are in fact conceptually correct, \hskip.2em
but incorrectly formulated. \
% There is not need to say more on this here as it follows below anyway. CP

\pagebreak

\yestop\noindent
Let us have a look at three 
% Joerg wrote: most well-known. I cannot see a justification for this
% statement in the first two case. CP
flaws in \litsectref{3.3} of the 
\litchaprefs{2, \nolinebreak 3}{5}, respectively:
\begin{description}
\item[\litchapwithsectref{2}{3.3}: ] 
%\paragraph{\litchapwithsectref{2}{3.3}: }
A typical instance for an
  incorrectly formulated theorem which is conceptually correct 
  can be found
  in \litchapwithsectref 2{3.3}, on 
\index{inference!deep}%
  deep inference:
  
  {\em From \bigmathnlb{\yields\,B\implies C}{} we can conclude
    \bigmathnlb{\yields\,A[B]\implies A[C]}, provided that %the hole
    \nlbmath{[\cdots]} denotes only positive positions\/\footnote
    {\label{note positive and negative positions}%
    % \majorheadroom
    Note that a position
      in a formula (seen as a tree built-up from logical operators
      `\tightund\closesinglequotecomma
      `\tightoder\closesinglequotecomma
      `\math\neg\closesinglequotecomma
      `\math\forall\closesinglequotecomma and \nolinebreak 
      `\math\exists') \hskip.2em
      is {\em positive} \udiff\ the number of \math\neg-operators preceding
      it on the path from the root position is even, and {\em
        negative} \udiff\ it is uneven.}  
in \nlbmath A.} \
  
  \herbrand, however, states 
  \par\noindent\LINEmaths{\yields\ \inpit{B\implies
      C}\nottight\implies
    \inpit{A[B]\implies A[C]}}{}
  \par\noindent which is not valid; to wit\footnote{%
 \majorheadroom
 \Cfnlb\ also  \citep
 [\goldfarb's \litnoterefs{6 (\p 78)}{A (\p\,98)}]{herbrand-logical-writings}.}
apply the substitution
  \par\noindent\LINEmaths{\{\ A[\ldots]\mapsto\forall x.\,[\ldots]\comma
    B\mapsto\truepp\comma C\mapsto\Pppp x\ \}}. \
\yestop
\item[\litchapwithsectref{3}{3.3}: ] 
%\paragraph{\litchapwithsectref{3}{3.3}: } 
Incorrectly formulated is also
\herbrand's theorem on the 
relativization of quantifiers.\footnote{\label{note relativization}%
 \majorheadroom
 Relativization of quantifiers was first discussed 
 in the elaboration of \firstorder\ logic in 
\peirceindex\citep{peirce-1885}. \ 
 Roughly speaking, \hskip.3em
 it means to restrict quantifiers to a 
 predicate \nlbmath\Ppsymbol\
 by replacing any \bigmathnlb{\forall x\stopq A}{} with
 \bigmathnlb{\forall x\stopq\inpit{\Pppp x\implies A}},
 and, \ 
 dually, \bigmaths{\exists x\stopq A}{} with
 \bigmaths{\exists x\stopq\inpit{\Pppp x\und A}}.}
This error was recognized later by \herbrand\ himself.\footnote{%
 \majorheadroom
 In \litnoteref 1 of \citep{herbrand-consistency-of-arithmetic}.}  

Moreover, note 
that in this context,
\herbrand\ discusses the 
{\em many-sorted}\/ \firstorder\ logic related to the restriction
to the language where all quantifiers are relativized,
extending a similar discussion found already in 
\peirceindex\citep{peirce-1885}.%
\end{description}

\yestop\noindent
All of this is not terribly interesting, 
except that it gives us some clues on how
\herbrand\ developed his theorems: \ 
It seems that he started, like any mathematician,
with a strong intuition of the semantics and used it 
to formulate the theorem. \ 
Then he had a careful look at those parts of the proof
that might violate the finitistic standpoint. \
%% This is a very 
%% efficient way to proceed, which we recommended for examinations.
The final critical check of minor details of the formalism in the
actual proof,\footnote{%
 \majorheadroom
 \index{Hadamard, Jacques S.}%
 \Cfnlb\ \citep[\litchapref V]{hadamard-psychology} for a nice
 account of 
% Joerg wrote: these two ``modi'' in 
% First, \hadamard\ does not speak of ``modi''.
% Second, the two modi of \litchapref V are not the two modi 
% Joerg refers to, but both deal with working out proof plans formally.
% As this is not our subject and a clarification would be very difficult,
% I have removed this. CP
 this mode of mathematical creativity.}
however, hardly played a \role\ in this work.
\begin{description}
\item[\litchapwithsectref{5}{3.3}: ] 
%\paragraph{\litchapwithsectref{5}{3.3}: }
The drawback of his intuitive
  style of work manifests itself in a serious mistake, which concerns
  a lemma that has crucial applications in the proof of the
  \herbrandsfundamentaltheoremindex\fundamentaltheorem, 
  namely the ``\hskip.001em lemma'' of \litchapwithsectref
  5{3.3}, which we will call 
{\em\herbrandsfalselemma}. \ Before we
  discuss this in \nlbsectref{section lemma}, however, we have to define some
  notions.
\end{description}
\vfill\pagebreak
%%%%%%%%%%%%%%%%%%%%%%%%%%%%%%%%%%%%%%%%%%%%%%%%%%%%%%%%%%%%%%%%%%%%%%%%%%%%%%%%
\section
[Champs Finis, \herbrand\ Universe, and   \herbrand\ Expansion]
{\herbrand\ Universe, 
 Champs Finis, 
 and \\\herbrand\ Expansion}\label
{section herbrand expansion}%
\index{champ fini|(}%
\index{Herbrand expansion|(}%
\index{Herbrand universe|(}%

\halftop\halftop\noindent Most students of logic or computer science know
\herbrand's name in the form of {\em\herbranduniverse}\/ or
{\em\herbrandexpansion}. 

\halftop\halftop\indent
Today, the 
\index{Herbrand universe!{\em definition}}%
{\em\herbranduniverse}\/ is usually defined as the 
set of all terms over a given signature, 
% ground terms we do not need here
% please discuss.
and the {\em\herbrandexpansion}\/ of a set
of formulas results from a systematic replacement of all variables 
in that set of formulas with
terms from the 
\index{Herbrand universe|)}%
\herbranduniverse. \  

\halftop\halftop\indent
Historically, however, this is not quite correct. \
First of all, \herbrand\ does not use {\em term structures}\/ for two reasons:
\nopagebreak\begin{enumerate}\nopagebreak\item 
\herbrand\ typically equates terms with objects of the
  universe, and thereby avoids working explicitly with term
  structures.\footnote{As this equating of terms has no
    essential function in \herbrand's works, but only adds extra
    complication to \herbrand's subjects, we will completely ignore it
    here and exclusively use free term structures in what follows.}
\item As a finitist more royal than King \hilbert, 
\herbrand\ does not accept structures with infinite universes.
\end{enumerate}

\vfill\subsection{Champs Finis}
As a finite substitute for a typically infinite full term universe,
\herbrand\ uses what he calls a 
{\frenchfont\em champ fini}\/ of order \nlbmath {n},
which we will denote with 
\index{\termsofdepthnovars n}%
\index{T@\termsofdepthnovars n}%
\nlbmath{\termsofdepthnovars n}.  \ 
Such a 
\index{champ fini|)}%
\index{champ fini!{\em definition}}%
{\frenchfont champ fini}\/ differs from a full term universe in
containing only the terms \nlbmath t with \bigmaths{\CARD t<n}{\,.} \ 
We \nolinebreak use \nlbmath{\CARD t}
to denote the 
\index{height of a term}%
\index{height of a term!{\em definition}}%
{\em height}\/ of the term \nlbmath t,
which is given by
\par\noindent\LINEmaths{\CARD{\anonymousfpp{t_1}{t_{m}}}
  \nottight{\nottight{\nottight=}}1+\max\{0,\CARD{t_1},\ldots,\CARD{t_m}\}}.
\par\noindent 
The terms of \nlbmath{\termsofdepthnovars n} are constructed from the
function symbols and constant symbols
(which we will tacitly subsume under the function symbols in the following)
of a finite signature and from a
finite set of variables. 
We will assume that an additional variable \nlbmath l, \hskip .2em 
the {\em lexicon}, \hskip .2em is \nolinebreak
included in this construction, if necessary
to have \ \maths{\termsofdepthnovars n\tightnotequal\emptyset}.

\vfill\vfill\pagebreak

\subsection{\herbrand\ Expansion}

\halftop\noindent
The first elaborate description of \firstorder\ logic
---~under the name ``first-intentional logic of relatives''~---
was published by 
\peirceindex\citet{peirce-1885} \hskip.2em
shortly after the invention of quantifiers by \citet{begriffsschrift}. \ 
What today we call an {\em\herbrandexpansion}\/ 
was implicitly given 
already in that publication \citep{peirce-1885}. \
\herbrand\ spoke of ``reduction'' ({\frenchfont\em r\'eduite}) instead.

\halftop\halftop\noindent
\herbrand\ prefers to treat all logical symbols besides
`\math\neg\closesinglequotecomma `\math\vee\closesinglequotecomma and
`\math\exists' as defined. \ 
Only in
\index{form!prenex}%
prenex forms, the universal quantifier `\math\forall'
is also treated as a primitive symbol.

\begin{definition}[\herbrand\ Expansion, \math{A^{\mathcal T}}]\\%
\index{Herbrand expansion!{\em definition}}%
\index{\math{A^{\mathcal T}}}%
\index{T@\math{A^{\mathcal T}}}%
For a finite set of terms \nlbmath{\mathcal T}, \ the expansion
\nlbmath{A^{\mathcal T}} of a formula \nlbmath A is defined as
follows:
\bigmaths{A^{\mathcal T}=A}{} if \math A \nolinebreak does
not have a quantifier, \bigmaths{\inpit{\neg A}^{\mathcal T}=\neg
  A^{\mathcal T}}, \bigmathnlb{\inpit{A\oder B}^{\mathcal T}=
  A^{\mathcal T}\oder B^{\mathcal T}}, \bigmathnlb{\inpit{\exists
    x.\,A}^{\mathcal T}= \bigvee_{t\in\mathcal T}A^{\mathcal
    T}\!\{x\tight\mapsto t\}}, and%\footnote
%% {Note that \bigmaths{\{x\tight\mapsto t\}}{} denotes the 
%% substitution which replaces \nlbmath x with \nlbmath t,
%% and that the postfix application of it has higher precedence
%% than the prefix operator \nlbmath\bigwedge,
%% so that \bigmaths{\bigwedge_{t\in\mathcal T}A^{\mathcal
%%     T}\!\{x\tight\mapsto t\}}{} reads
%% \bigmaths{\bigwedge_{t\in\mathcal T}\inparentheses{\inpit{A^{\mathcal T}}
%% \{x\tight\mapsto t\}}}.
%% \Cfnlb\ also \sectref{section where the precedence is explained}}
\bigmaths{\inpit{\forall x.\,A}^{\mathcal T}= 
\bigwedge_{t\in\mathcal T}A^{\mathcal T}\!\{x\tight\mapsto t\}},
where \math{A^{\mathcal T}\!\{x\tight\mapsto t\}} denotes the result
of applying the substitution \nlbmath{\{x\tight\mapsto t\}}
to \nlbmath{A^{\mathcal T}}.
\getittotheright\qed\end{definition}

\begin{example}[\herbrand\ Expansion, \math{A^{\mathcal T}}]%
\index{Herbrand expansion!{\em example}}%
\\
For example, for \bigmaths{{\mathcal T}\nottight{\nottight{:=}}\{ \ 
\threepp,\ \plusppnoparentheses z\twopp
%\zeropp,\spp z
%\spp\zeropp,\spp{\boundvari x{}}
\ \}}, \ and for \math A being the arithmetic formula \\[.5ex]\LINEmaths{
  \forall\boundvari x{}\stopq \inpit{ \boundvari x{}\tightequal\zeropp
    \nottight{\nottight\oder} \exists \boundvari y{}\stopq \boundvari
    x{}\tightequal\plusppnoparentheses{\boundvari y{}}\onepp}
},\\
the expansion \math{A^{\mathcal T}} is \\[.5ex]\LINEmaths{
  \inparenthesesoplist{ 
    \threepp\tightequal\zeropp \oplistoder
    \threepp\tightequal\plusppnoparentheses\threepp\onepp\oplistoder
    \threepp\tightequal\plusppnoparentheses{\pluspp z\twopp}\onepp} 
%%     \zeropp\tightequal\zeropp \oplistoder
%%     \zeropp\tightequal\spp{\zeropp} \oplistoder
%%     \zeropp\tightequal\spp{\spp z} } 
\nottight{\nottight\und}
  \inparenthesesoplist{ 
    \plusppnoparentheses z\twopp\tightequal\zeropp \oplistoder 
    \plusppnoparentheses z\twopp\tightequal
    \plusppnoparentheses\threepp\onepp \oplistoder 
    \plusppnoparentheses z\twopp\tightequal
    \plusppnoparentheses{\pluspp z\twopp}\onepp
%%     \spp z\tightequal\zeropp \oplistoder 
%%     \spp z\tightequal\spp{\zeropp} \oplistoder 
%%     \spp z\tightequal\spp{\spp z}
  }}.\\[-2.0ex]\mbox{}\getittotheright\qed
\end{example}

\noindent
The \herbrandexpansion\ reduces a \firstorder\ formula to sentential
logic in such a way that a sentential formula is reduced to itself,
and that the semantics is invariant if the terms in 
\nlbmath{\mathcal T} range over the whole universe.
If, however, this is not the case
--- \nolinebreak 
such as in our above example and for all infinite universes
\nolinebreak---, then
\index{Herbrand expansion|)}%
\herbrandexpansion\ changes the semantics by relativization of 
the quantifiers\arXivfootnotemarkref{note relativization}
to range only over those elements of the universe to which the 
elements of \nlbmath{\mathcal T}
evaluate.
\vfill\pagebreak
%%%%%%%%%%%%%%%%%%%%%%%%%%%%%%%%%%%%%%%%%%%%%%%%%%%%%%%%%%%%%%%%%%%%%%%%%%%%%%%%
\section{\skolemization, \smullyan's 
Uniform Notation, \\and \math\gamma- and \math\delta-quantification}\label
{section skolemization}
\index{Skolemization|(}%
\index{notation!uniform|(}%

\noindent
A first-order formula may contain existential as well as universal 
quantifiers. %and sometimes this makes it technically cumbersome. \ 
Can we make it more uniform by replacing either of them? 

Consider the formula \bigmaths{\forall x.\,\exists y.\,\Qppp x y}. 
These two quantifiers express a functional dependence between the values for
\nlbmath x and \nlbmath y, which could also be expressed by a (new)
function, say \nlbmath {g}, such that \math{\forall x.\,\Qppp x
  {\app {g} x}}, \ie\ this function \nlbmath {g} chooses for each
\nlbmath x the correct value for \nlbmath y, provided that it exists.
In other words, we can replace any existentially quantified variable \nlbmath x
which occurs in the scope of universal quantifiers for 
\nlbmath {y_1}, \ldots , \nlbmath {y_n} with the new 
\index{term!Skolem|(}%
{\em\skolem\ term} 
\nlbmath {\app {g} {{y_1}, \ldots, {y_n}}}. 

This replacement, carried out for all existential quantifiers, results
in a formula having only universal quantifiers.
Using the convention that all free
variables are universally quantified, 
we may then just drop these quantifiers as well. 
Roughly speaking,
this transformation, {\em\skolemization} as we call it today,
leaves satisfiability invariant. \ 
It occurs for the first time explicitly in 
\skolemindex\cite{skolem-1928}, \hskip.2em
but was already used 
in an awkward formulation in 
\schroederindex\cite{schroeder-vorlesungen-III} 
and 
\loewenheimindex\cite{loewenheim-1915}.

For reasons that will become apparent later, 
\herbrand\ employs a form of \skolemization\ 
that is dual to the one above. \ 
Now the {\em universal}\/ variables are removed first,
so that all remaining variables are existentially quantified. 
How can this be done?
Well, if the universally 
quantified variable \nlbmath x
% If a universal quantifier `\math{\forall x.}' does not occur within
% the scope of an existential quantifier, it \nolinebreak can be removed
% from a formula without further changes. \ Otherwise, if \nlbmath x
occurs in the scope of the existentially quantified variables
\math{y_1,\ldots,y_m}, we can replace \nlbmath x with the 
\index{term!Skolem}%
{\em\skolem\ term} 
\nlbmath{\app{\forallvari x{}}{y_1,\ldots,y_m}}. \ 
The \secondorder\ variable or \firstorder\ function symbol
\nlbmath{\forallvari x{}} in this \skolem\ term 
stands for any function with arguments \nlbmath{y_1,\ldots,y_m}. \ 
Roughly speaking, this dual 
form of {\em\skolemization} leaves validity invariant.

For example, let us consider the formula \bigmathnlb
{\exists y.\,\forall x.\,\Qppp x y}. \
Assuming the 
\index{axiom!of choice}%
\axiomofchoice\ and the
standard interpretation of (higher-order) quantification, all of the
following statements are logically equivalent:
\begin{itemize}
\noitem\item\math{\exists y.\,\forall x.\,\Qppp x y} \ holds.
\noitem\item There is an object \nlbmath y such that 
\bigmaths{\Qppp x y}{} holds for every object \nlbmath x.
%% \noitem\vspace*{-.2ex}\item 
%% It is not the case that we can find, for every object \nlbmath y,
%%   an object \math x such that \\\Qppp x y does not hold. \
%% \noitem\vspace*{-.2ex}\item 
%% It is not the case that we can find a function \nlbmath{\forallvari x{}} 
%% such that, for every object \nlbmath y, \\\Qppp{\app{\forallvari x{}}y}y 
%% does not hold.
\noitem\item 
\math{\exists y.\,\Qppp{\app{\forallvari x{}}y}y} \ holds for every function
\nlbmath{\forallvari x{}}.
\noitem\item 
\math{\forall f.\,\exists y.\,\Qppp{\app f y}y} \ holds.
\end{itemize}

\yestop\noindent
Now \bigmaths{\exists y.\,\Qppp{\app{\forallvari x{}}y}y}{} is called the
{\em\index{form!Skolemized@(outer) Skolemized}%
\index{form!Skolemized@(outer) Skolemized!{\em example}}%
\skolemizedform\ of\/ \bigmaths{\exists y.\,\forall x.\,\Qppp x y}.} \  
The variable or function symbols 
\nlbmath{\forallvari x{}} of increased logical order are
called 
\index{function!Skolem|(}%
{\em\skolem\ functions}.\footnote
{\herbrand\ calls \skolem\ functions 
\index{function!index}%
{\em index functions},
 translated according to \citep{herbrand-logical-writings}. \
Moreover, in \citep{herbrand-style-consistency-proofs}, \ 
\citep{scanlon73:_consis_number_theor_via_theor}, \ and 
\citep{goldfarb-herbrand-consistency}, \ 
we find the term 
\index{function!indicial}%
{\em indicial functions}\/ instead of 
``index functions\closequotefullstopextraspace
The name ``\skolem\ function'' was used in 
\goedelindex\cite{goedel-consistency-continuum},
probably for the first time, \cfnlb\ 
\anellisindex\cite{anellis-skolem-function}.}
The \skolemizedform\ is also called
\index{form!functional}%
{\em functional form} (with several addenda specifying the dualities),
because 
\index{Skolemization|)}%
\skolemization\ turns the object variable \nlbmath x into a
function variable or function symbol 
\nlbmath{\app{\forallvari x{}}\cdots}.

%\pagebreak

Note that
\bigmaths{A\tightimplies B}{} and \bigmaths{\neg A\oder B}{} and
\bigmaths{\neg\inpit{A\und\neg B}}{} are equivalent in two-valued
logic. \ So are \bigmaths{\neg\forall x.\,A}{} and \bigmaths{\exists
  x.\,\neg A}, as well as \bigmaths{\neg\exists x.\,A}{} and
\bigmaths{\forall x.\,\neg A}. \ 

\halftop\halftop\indent
Accordingly, the 
\index{notation!uniform|)}%
{\em uniform notation}\/ (as introduced in \citep{smullyan}) \hskip.3em
is a modern classification of formulas 
%% A modern notation of formulas and
%% inference rules due to \citep{smullyan} is called 
%% \index{notation!uniform|)}%
%% {\em uniform notation}\/ as it 
%% classifies the 
%% % keeps the zoo of all possible 
%% logical
%% % connectives and quantifiers 
%% symbols
%% and their respective inferences
%% % , but
%% % conveniently classifies them 
into only four categories: 
\index{alpha@\math\alpha}%
\math\alpha,
\index{beta@\math\beta}%
\nlbmath\beta, 
\index{gamma@\math\gamma}%
\nlbmath\gamma, and 
\index{delta@\math\delta}%
\nlbmath\delta. 

\halftop\halftop\indent
More important than the classification of formulas
is the associated classification of the 
reductive inference rules applicable to them as 
{\em principal formulas}.

\halftop\halftop\indent
According to \citep{gentzen}, \hskip.2em
but viewed under the aspect of reduction 
(\ie\ the converse of deduction), \hskip.3em
the {\em principal formula}\/ of an inference rule is the one 
which is (partly) replaced by its immediate ``sub''-formulas, depending on 
its topmost operator.
\begin{itemize}\item 
\mbox{An \math\alpha-formula} is one whose validity reduces to
  the validity of a single operand of its topmost operator.\par
  For example, \bigmath{A\oder B} 
  may be reduced either to \bigmath A or to
  \bigmaths B, and \bigmath{A\implies B} may be reduced either to
  \bigmath{\neg A} or to \bigmaths B.\item 
  \mbox{A \math\beta-formula} is one whose validity reduces to the
  validity of both operands of its topmost binary operator,
  introducing two cases of proof (\math\beta\ = \underline branching).\par
  For example, \bigmath{A\und B} 
  reduces to both \bigmath A and \bigmaths B, and
  \bigmath{\neg\inpit{A\implies B}} reduces to both \bigmath{A} and
  \bigmaths{\neg B}.\item 
  \mbox{A \math\gamma-formula} is one whose validity reduces to
  the validity of alternative instances of its topmost quantifier.\par
  For example, \bigmath{\exists y.\,A} \nolinebreak reduces to
  \bigmath{A\{y\tight\mapsto\existsvari y{}\}} in addition to
  \bigmaths{\exists y.\,A}, for a fresh {\em\fev} \nlbmath{\existsvari y{}}. \
  Similarly, \bigmath{\neg\forall y.\,A} reduces to \bigmath{\neg
    A\{y\tight\mapsto\existsvari y{}\}} in addition to
  \bigmaths{\neg\forall y.\,A}. \ 
  \Fev s may be globally instantiated at any time in a reduction proof.\item 
  A \math\delta-formula is one whose validity reduces to
  the validity of the instance  of its topmost quantifier with its
  \index{term!Skolem|)}%
  \skolem\ term.\par
  For example, \bigmath{\forall x.\,A}  reduces to
  \bigmaths{A\{x\tight\mapsto\app
  {\forallvari x{}}{\existsvari y 1,\ldots,\existsvari y m}\}}, 
  where \math{\existsvari y 1,\ldots,\existsvari y m} are
  the \fev s\ already in use.\footnote{%
  % \majorheadroom
  In the game-theoretic semantics of \firstorder\ logic, the
  \mbox{\math\delta-variables} (such as \math{\forallvari x{}} 
  in the above example)
  stand for the unknown choices by our opponent in the game, whereas,
  for showing validity, we have to specify a winning strategy by
  describing a finite number of \firstorder\ terms as alternative
  solutions for the \mbox{\math\gamma-variables} (such as \math{\existsvari y i}
  above), \cf\ \eg\ \citep{hintikkaprinciples}.}

\end{itemize}
For a more elaborate introduction into free 
\index{variable!free gamma-@free \math\gamma-}%
\index{variable!free delta-@free \math\delta-}%
\math\gamma- and \math\delta-variables
see 
\citep{wirth-jal}.

\pagebreak

\herbrand\ considers {\em validity}\/ and 
{\em \skolemizedform}\/ as above in his thesis. \hskip.4em
In \nolinebreak a similar context, which we will have to discuss below,
\skolem\ considers {\em unsatisfiability}, a dual of validity, and
{\em\skolemnormalform}\/ in addition to \skolemizedform. \ 
As \nolinebreak it was standard at his time,
\herbrand\ called the two kinds of quantifiers
--- \nolinebreak \ie\ \nolinebreak for \nolinebreak\math\gamma- \nolinebreak
and \nolinebreak\math\delta-formulas\footnote{%
 % \majorheadroom
 Note that it is obvious how to generalize the definition of 
 \math\alpha-, \math\beta-,
 \math\gamma- and \math\delta-formulas from top positions
 to inner occurrences according to the category into which they 
 would fall in a stepwise reduction. \ 
 Therefore, we can speak of  \math\alpha-, \math\beta-, \math\gamma- and
 \math\delta-formulas also for the case of subformulas
 and classify their quantifiers accordingly.\majorfootroom} 
\nolinebreak --- \ 
\index{quantifier!restricted}%
{\em restricted}\/ and 
\index{quantifier!general}%
{\em general}\/
quantifiers, respectively. \ 
To avoid the problem of getting lost in
several dualities in what follows, we prefer to speak of
\index{quantifier!gamma-@\math\gamma-}%
\mbox{\em\math\gamma-quantifiers\/} and
\index{quantifier!gamma-@\math\delta-}%
\mbox{\em\math\delta-quantifiers\/} instead. \
%the classification and uniform notation of \citep{smullyan}. \ 
The variables bound by \mbox{\math\delta-quantifiers} will be called
\index{variable!bound delta-@bound \math\delta-}%
\mbox{\em bound \math\delta-variables}. \ 
The \nolinebreak variables bound by
\mbox{\math\gamma-quantifiers} will be called 
\index{variable!bound gamma-@bound \math\gamma-}%
\mbox{\em bound \math\gamma-variables}.

%\pagebreak

\halftop\halftop\indent
For a \firstorder\ formula \nlbmath A in which any bound variable is
bound exactly once and does not occur again \freely, 
% (neither as a variable nor as a function symbol)
% We do not need this anymore because now the 
% \math\delta\ marker is not an optional annotation anymore,
% but obligatory!
we define: 

\halftop\halftop\indent
The 
\index{form!Skolemized@(outer) Skolemized}%
\index{form!Skolemized@(outer) Skolemized!{\em definition}}%
{\em outer}\/\footnote{\label{note inner}%
 \majorheadroom
 \herbrand\ has no name for the outer \skolemizedform\
 and he does not use the inner \skolemizedform,
 which is the current standard in two-valued \firstorder\ logic
 and which is required for our discussion in \noteref{note discussion inner}.
 \par
 The 
 \index{form!Skolemized @inner Skolemized}%
 \index{form!Skolemized @inner Skolemized!{\em definition}}%
 {\em inner \skolemizedform\ of \nlbmath A} results from 
 \nlbmath A by repeating the following 
 until all \mbox{\math\delta-quantifiers} have been removed: \ 
 Remove an outermost \math\delta-quantifier
 and replace its bound variable \bigmath x
 with \bigmaths{\app{\forallvari x{}}{y_1,\ldots,y_m}}, \ 
 where \bigmath{\forallvari x{}} is a new symbol and 
 \bigmaths{y_1,\ldots,y_m}, in this order, are the variables of the 
 \mbox{\math\gamma-quantifiers} in whose scope the 
 \mbox{\math\delta-quantifier}
 occurs and which actually occur in the scope of the 
 \mbox{\math\delta-quantifier}. \par
 The {\em inner}\/ \skolemizedform\ 
 is closely related to the {\em liberalized}\/ \math\delta-rule
 (also called \deltaplus-rule) in reductive calculi, such as sequent, tableau,
 or matrix calculi; \ 
 \cfnlb\ \eg\
 \index{Ferm\"uller, Christian G.}%
 \citep{baazdelta}, \
 \citep{strongskolem}, \ 
 \citep[\litsectrefs{1.2.3}{2.1.5}]{wirthcardinal}, \ 
 \citep{nonpermut}, \
 \citep[\litsectref 4]{wirth-jal}.\majorfootroom} 
{\em \skolemizedform\ of \nlbmath A}\/ results from \nlbmath
A by removing any \math\delta-quantifier and replacing its bound
variable \bigmath x with \bigmaths{\app{\forallvari x{}}{y_1,\ldots,y_m}}, \ 
where \bigmath{\forallvari x{}} is a new symbol and 
\bigmaths{y_1,\ldots,y_m}, in this order,\footnote{%
 \majorheadroom
 Contrary to
 our fixation of the order of the variables as arguments to the
 \index{function!Skolem|)}%
 \skolem\ functions (to achieve uniqueness of the notion), \herbrand\
 does not care for the order in his definition of the outer
 \index{form!Skolemized@(outer) Skolemized}%
 \skolemizedform. \ 
 Whenever he takes the order into account,
 however, he orders by occurrence from left to right or else by the
 \index{height of a term}%
 height of the terms \wrt\ a substitution, but never by the names of
 the variables.}  
are the variables of the \math\gamma-quantifiers in
whose scope the \mbox{\math\delta-quantifier} occurs.
\vfill\pagebreak

%%%%%%%%%%%%%%%%%%%%%%%%%%%%%%%%%%%%%%%%%%%%%%%%%%%%%%%%%%%%%%%%%%%%%%%%%%%%%%
\section{Axioms and Rules of Inference}\label
{section herbrands calculi}\label
{section where the precedence is explained}%
\herbrandsfundamentaltheoremindexbegin

\noindent In the following we will present the calculi
of \herbrand's thesis 
(\ie\ the axioms and rules of inference)
as required for our presentation of the \fundamentaltheorem.

When we speak of a term, a formula, or a structure, 
we refer to \firstorder\ terminology without
mentioning this explicitly. \ 
When we explicitly speak of ``first
order\closequotecomma however, 
this \nolinebreak is to emphasize the contrast to
sentential logic. \ 

%% In the notation of our terms and formulas
%% we apply modern standard \firstorder\ preference order.
%% Thus, operator preference decreases from
%% \begin{itemize}\noitem\item
%% {\em postfix}\/ (such as an argument \nolinebreak
%%     ``\math{(x)}'' in \nolinebreak
%%     ``\math{f(x)}'' or a substitution \nolinebreak
%%     ``\math{\{x\tight\mapsto t\}}'')
%%     (associating to the left, \ie\ ``\math{f(x)(y)}'' reads
%%      ``\math{(f(x))(y)}''), over \noitem\item
%% {\em prefix}\/ (such as negation \nolinebreak
%%     ``\math\neg'' or a binder \nolinebreak
%%     ``\math{\forall x.}'' or
%%     ``\math{\varepsilon x.}'') (with minimal scope) 
%%     (associating to the right, \ie\
%%     ``\math{\neg\neg A}'' reads \nolinebreak
%%     ``\math{\neg(\neg A)}''), 
%%     to\noitem\item
%% {\em infix}\/
%%     (where ``\tightund'' and ``\tightoder'' have higher preference than
%%     \nolinebreak``\tightimplies'' (associating to the right) and
%%     \nolinebreak``\tightequivalent'').
%% \end{itemize}
%% % Redundantly, we tend to indicate lower preference by more white space.
%% %%
\begin{description}
\item[\SententialTautology: ] 
\index{tautology!sentential}%
Let \math B be a \firstorder\ formula.
    \ \math B \nolinebreak is a {\em\sententialtautology}\/ \udiff\ it
    is quantifier-free and truth-functionally valid, provided its
    atomic subformulas are read as atomic sentential variables.\footnote
{Note that this notion is more
      restrictive than the following, 
    which is used by \herbrand\ in his thesis only initially, 
    but which is standard for the
      predicate calculi of the 
\index{Hilbert!school}%
\index{Hilbert!school!calculi of}%
\hilbert\ school and the 
\index{Principia Mathematica!calculi of}%
\PM; \ 
\cfnlb\ 
\citep[Editors' Preface to Part\,B of Volume\,I, \p\,lxiii\,\f]
{grundlagen-german-english-edition-volume-one-two}
(or \citep[\litsectfromtoref{3}{5}]{grundlagen-first-edition-volume-one},
\citep[Supplement\,I\,D]{grundlagen-first-edition-volume-two}), \hskip.2em 
\index{Principia Mathematica}\citep[*10]{PM}\@. \ %\\ 
      \math B \nolinebreak is \nolinebreak a
\index{tautology!substitutional sentential}% 
      {\em\substitutionalsententialtautology}\/
      \udiff\ there is a truth-functionally valid sentential formula \nlbmath
      A and a substitution \nlbmath\sigma\ mapping any sentential
      variable in \nlbmath A to a \firstorder\ formula such that \math
      B \nolinebreak is \nlbmath{A\sigma}. \ 
      %\\
      For example, both \bigmathnlb{\Pppp x{\oder}\neg\Pppp x}{} and
      \bigmathnlb{\exists x\stopq\Pppp x\nottight{\oder}\neg\exists
        x\stopq\Pppp x}{} are \substitutionalsententialtautologies,
      related to the truth-functionally 
      valid sentential formula \bigmathnlb{p{\oder}\neg
        p}, but only the first one is a \sententialtautology.}
\yesitem\item[Modus Ponens: ] 
\index{modus ponens}%
\index{modus ponens!{\em definition}}%
\bigmaths{\displaystyle{A\quad\quad\quad
        A\implies B}\over \displaystyle{B}}.
\item[Generalized Rule of \math\gamma-Quantification: ]
\index{Generalized Rule!of gamma-Quantification@of \math\gamma-Quantification}%
\index{Generalized Rule!of gamma-Quantification@of \math\gamma-Quantification!{\em definition}}%
  \bigmaths{\displaystyle{A[B\{x\mapsto t\}]}\over
    \displaystyle{A[\gamma x.\,B]}}, where the free variables of the
  term \nlbmath t must not be bound by quantifiers in \nlbmath B, and
  \math{\gamma} stands for \nlbmath\exists\ if 
  \math{[\ldots]} denotes a positive 
  position\arXivfootnotemarkref{note positive and negative positions}  
  in \nlbmath{A[\ldots]}, and \math{\gamma} stands
  for \nlbmath\forall\ if this position is negative. \ 
  Moreover, we
  require that \nlbmath{[\ldots]} does not occur in the scope of
  any quantifier in \nlbmath{A[\ldots]}. \ This requirement is not
  necessary for soundness, but for the constructions in the proof of
  \herbrandsfundamentaltheorem.
\par\noindent
For example,
we get 
\\\LINEmaths{\displaystyle
{\inpit{t\tightprec t}
\nottight{\oder}\neg\inpit{t\tightprec t}}\over\displaystyle
{\inpit{t\tightprec t}
\nottight{\oder}\exists x\stopq\neg\inpit{x\tightprec t}}}{}
\\and\\\LINEmaths{\displaystyle
{\inpit{t\tightprec t}
\nottight{\oder}\neg\inpit{t\tightprec t}}\over\displaystyle
{\inpit{t\tightprec t}
\nottight{\oder}\neg\forall x\stopq\inpit{x\tightprec t}}}{} 
\par\noindent  
via the meta-level substitutions
\par\noindent\LINEmaths{%\scriptstyle
\{\ \ \ \ A[\ldots]\ 
\mapsto\ \inpit{t\tightprec t}
\oder [\ldots]\comma\ \ \
B\ \mapsto\ \neg\inpit{x\tightprec t}\ \ \ \ \}}{\,\,}\\and\\
\LINEmaths{%\scriptstyle
\{\ \ \ \ A[\ldots]\ \mapsto\ 
\inpit{t\tightprec t}
\oder\neg[\ldots]\comma\ \ \
B\ \mapsto\ \inpit{x\tightprec t}\ \ \ \ \}},\\respectively.

Note that \herbrand\ considers equality of formulas only up to
  renaming of bound variables and often implicitly assumes that a
  bound variable is bound only once and does not occur \freely. \ 
  Thus, if a free variable \nlbmath y of the 
  term \nlbmath t is bound by quantifiers in \nlbmath B, an implicit
  renaming of the bound occurrences of \nlbmath y in 
  \nlbmath{B} is admitted to enable backward application of the
  inference rule.\footroom  

\pagebreak

\notop\item[Generalized Rule of \math\delta-Quantification: ]
\index{Generalized Rule!of delta@of \math\delta-Quantification}%
\index{Generalized Rule!of delta@of \math\delta-Quantification!{\em definition}}%
  \bigmaths{\displaystyle{A[B]}\over\displaystyle{A[\delta x.\,B]}},
  where the variable \math x must not occur in the context
  \nlbmath{A[\ldots]}, and \math{\delta} stands for \nlbmath\forall\
  if \math{[\ldots]} denotes a positive position in
  \nlbmath{A[\ldots]}, and \math{\delta} stands for \nlbmath\exists\ if this
  position is negative. \ 
  Moreover, both for soundness and for the reason mentioned above, we require 
  that \math{[\ldots]} \nolinebreak
  does not occur in the scope of any quantifier in \nlbmath{A[\ldots]}.\\ 
  Again, if \math x occurs in the context \nlbmath{A[\ldots]}, an implicit
  renaming of the bound occurrences of \nlbmath x in 
  \bigmaths{\delta x.\,B}{} is admitted to enable backward application.
% of the inference rule.  
%\pagebreak

\noitem\item[Generalized Rule of Simplification: ]
\index{Generalized Rule!of Simplification}%
\index{Generalized Rule!of Simplification!{\em definition}}%
  \bigmaths{\displaystyle{A[B\circ B]}\over\displaystyle{A[B]}}, where
  \math{\circ} stands for \tightoder\ if \math{[\ldots]} \nolinebreak
  denotes a positive position in \nlbmath{A[\ldots]}, \ 
  and \math{\circ} stands for \tightund\ if this position is negative.\\
  To enable a forward application of the inference rule,
  the bound variables may be renamed such that the two occurrences of
  \nlbmath B become equal.\\
  Moreover, the 
\index{Generalized Rule!of gamma-Simplification@of \math\gamma-Simplification}%
\index{Generalized Rule!of gamma-Simplification@of \math\gamma-Simplification!{\em definition}}%
{\em Generalized Rule of \math\gamma-Simplification} 
  is the sub-rule for the case that
  \math B \nolinebreak is of the form \nlbmath{\exists y.C} 
% or \nlbmath{\neg\forall y.C}, 
% CP: I do not see a reason for these typically redundant cases.
  if \math{[\ldots]} denotes a
  positive position in \nlbmath{A[\ldots]}, and of the form
  \nlbmath{\forall y.C} 
% or \nlbmath{\neg\exists y.C}, 
  if this position is negative.
\noitem\end{description}
\label{section discussion prenex}%
% {\em Prenex form}  is one of several possible {\em normal forms}\/
% by which a \firstorder\ formula can be represented in two-valued logic. 
To avoid the complication of quantifiers within a formula,
% Please do not write In order to. This English working class slang. 
% Maybe appropriate for a technical paper. Not here. CP.
where it is hard to keep track of the scope of  each individual quantification,
%it was noted early in the history of modern logic that in fact 
all quantifiers can be moved to the front,
provided some caution is taken with the renaming of quantified variables. \ 
This is called the 
\index{form!prenex!{\em definition}}%
{\em prenex form}\/ of a formula. \ 
% While a prenex form results from a transformation 
% which gives any quantifier a maximal scope,
% an 
The 
\index{form!anti-prenex!{\em definition}}%
{\em anti-prenex form}\/ results from the opposite transformation,
\ie\ from moving the quantifiers inward as much as possible. \ 
\herbrand\ achieves these transformations with his 
\index{Rule!of Passage}%
{\em Rules of Passage}.
\begin{description}
\item[Rules of Passage: ] 
\index{Rule!of Passage!{\em definition}}%
The following six \nolinebreak logical
  equivalences may be used for rewriting from left to right 
\index{direction!prenex!{\em definition}}%
({\em prenex direction}\/) and from right to left 
\index{direction!anti-prenex!{\em definition}}%
({\em anti-prenex direction}\/), 
resulting in twelve \nolinebreak deep 
\index{inference!deep}%
inference rules:\par\noindent\LINEmath{\begin{array}{
%%         @{}
%%         % l@{~\,\,}
%%         r@{\,\,\,}c@{\,\,}l
%%         @{~~~}|@{~~~}
%%         % l@{~\,\,}
%%         r@{\,\,\,}c@{\,\,}l
%%         @{~~~}|@{~~~}
%%         % l@{~\,\,}
%%         r@{\,\,\,}c@{\,\,}l
%%         @{}}
        l@{~~~~~~}
        r
        c
        l
      }
       (1) 
      &
      \neg\forall x.A
      &\equivalent
      &\exists x.\neg A
     \\(2)
      &\neg\exists x.A
      &\equivalent
      &\forall x.\neg A
     \\(3)%\footnotemark
      &\inpit{\forall x.A}\oder B
      &\equivalent
      &\forall x.\,\inpit{A\,\tightoder B}
     \\(4)
      &B\oder\forall x.A
      &\equivalent
      &\forall x.\,\inpit{B\,\tightoder A}
     \\(5)
      &\inpit{\exists x.A}\oder B
      &\equivalent
      &\exists x.\,\inpit{A\,\tightoder B}
     \\(6)
      &B\oder\exists x.A
      &\equivalent
      &\exists x.\,\inpit{B\,\tightoder A}
      \\\end{array}}%
%% \footnotetext{In higher-order standard,
%% where the bold square dot ``\nlbmath\squaredot''
%% indicates a ``\math{\stopq (}'' plus a closing parenthesis ``\math)'' 
%% as far to the right as possible,
%%     we would write \bigmaths{\inpit{\inpit{\forall x.\,A} \oder
%%         B}\nottight{\equivalent}\forall x\squaredot\,A\,\tightoder B}.}
\par\noindent
  Here, \math B is a formula in which the variable \nlbmath x does not occur. \ 
  As explained above, 
  if \nlbmath x \nolinebreak occurs \freely\ 
  in \nlbmath B, an implicit renaming of
  the bound occurrences of \nlbmath x in \nlbmath A
  is admitted to enable rewriting in
\index{direction!prenex}%
  prenex direction.
\end{description}
If we restrict the ``Generalized'' rules to outermost applications only 
\hskip.15em (\ie, if we restrict \math A to 
\nolinebreak be the empty context), \hskip.3em
we obtain the rules without the attribute 
``Generalized\closequotecommaextraspace
\ie\ the 
\index{Rule|see {{\em also}\/\, Generalized Rule}}%
\index{Rule!of gamma-Quantification@of \math\gamma-Quantification!{\em definition}}%
\index{Rule!of delta-Quantification@of \math\delta-Quantification!{\em definition}}%
{\em Rules of \math\gamma- and \math\delta-Quantification}\/ and the 
\index{Rule!of Simplification!{\em definition}}%
{\em Rule of Simplification}.\footnote
{The {\em Generalized}\/
  Rules of Quantification are introduced (under varying names) in
\heijenoortindex\makeaciteoffour
{heijenoort-tree-herbrand}
{heijenoort-herbrand}
{heijenoort-oeuvre-herbrand}
{heijenoort-work-herbrand}
  and under the names \inpit{\mu^\ast} and \inpit{\nu^\ast} in
  \citep[\p 166]{grundlagen-second-edition-volume-two}, 
  but not in the first edition \cite{grundlagen-first-edition-volume-two}. \ 
  \herbrand\ had only the
  non-generalized versions of the Rules of Quantification and named
  them ``First \nolinebreak and Second \nolinebreak 
\index{Rule!of Generalization}%
Rule of Generalization\closequotecomma translated according to
\citep{herbrand-logical-writings}. \ 
Note that the restrictions of the Generalized Rules of Quantification guarantee
the equivalence of the generalized and the non-generalized versions by the 
\index{Rule!of Passage}%
Rules of Passage; \cfnlb\ 
\heijenoortindex\citep[\p\,6]{heijenoort-tree-herbrand}. \
  \herbrand's name for {\em modus
    ponens}\/ is 
\index{Rule!of Implication}%
``Rule of Implication\closequotefullstopextraspace
  Moreover, 
\index{Rule!of Simplification}%
\index{Generalized Rule!of Simplification}%
``(Generalized) Rule of Simplification'' and 
\index{Rule!of Passage}%
``Rules of Passage'' are \herbrand's names. \ 
  All other names introduced in \nlbsectref{section herbrands calculi} are 
  our own invention to simplify our following presentation.}
\vfill\pagebreak

%%%%%%%%%%%%%%%%%%%%%%%%%%%%%%%%%%%%%%%%%%%%%%%%%%%%%%%%%%%%%%%%%%%%%%%%%%%%%%%%
\section{Normal Identities, Properties A, B, and C, 
and\\\herbrand\ Disjunction and Complexity}\label
{section herbrands properties}%
\index{Herbrand disjunction|(}%
\index{Herbrand complexity|(}%

Key notions of \herbrand's thesis are
\index{identity!normal}%
{\em normal identity}, 
\index{Property A}%
{\em\propertyA}, 
\index{Property B}%
{\em\propertyB}, and 
\index{Property C|(}%
{\em\propertyC}. \ \
\propertyC\ is the most important and 
the only one we need in this account.\footnote{\label{note normal identity}%
 % \majorheadroom
 Here are the definitions for the 
 omitted notions 
 {\em normal identity}, 
 {\em\propertyA}, and
 {\em\propertyB}
 for a formula \nlbmath D. \ 
 \math D is a 
 \index{identity!normal!{\em definition}}%
 {\em normal identity}\/ \udiff\ 
 \math D \nolinebreak
 has a linear proof starting with a \sententialtautology,
 possibly followed by applications of the Rules of Quantification,
 and finally possibly followed by applications of the 
 \index{Rule!of Passage}%
 Rules of Passage. \ 
 \math D has 
 \index{Property A!{\em definition}}%
 {\em\propertyA}\/ \udiff\
 \math D \nolinebreak
 has a linear proof starting with a \sententialtautology,
 possibly followed by applications of the Rules of Quantification,
 and finally possibly followed by applications of the 
 \index{Rule!of Passage}%
 Rules of Passage
 and the 
 \index{Generalized Rule!of Simplification}%
 Generalized Rule of Simplification. \
 \herbrand's original definition of 
 \index{Property A}%
 \propertyA\ is technically more complicated,
 but extensionally defines the same property 
 and is also intensionally very similar. \ 
 Finally, \math D has 
 \index{Property B!{\em definition}}% 
 {\em \propertyB\ of order \nlbmath n}\/
 \udiff\ \math{D'} \nolinebreak has \propertyC\ of order \nlbmath n,
 where \math{D'} results from possibly repeated
 application of the 
 \index{Rule!of Passage}%
 Rules of Passage to \nlbmath D, in 
 \index{direction!anti-prenex}%
 anti-prenex direction
 as long as possible.%
} 

\begin{sloppypar}

\herbrand's \propertyC\ was implicitly used already in 
\loewenheimindex\citep{loewenheim-1915} 
and % --- with a one-to-one correspondence --- %
\skolemindex\citep{skolem-1928}, \hskip.2em
but as an explicit notion, 
it was first formulated in \herbrand's thesis. \ 
It is the main property of
\herbrand's work and may well be called the central property of
\firstorder\ logic, for reasons to be explained in the following.
%\pagebreak

\end{sloppypar}

In essence, \propertyC\ captures the following intuition taken from
\loewenheimindex\citep{loewenheim-1915}:
Assuming the 
\index{axiom!of choice}%
\axiomofchoice, 
the validity of a formula \nlbmath A
is equivalent to the validity of its 
\index{form!Skolemized@(outer) Skolemized}%
\skolemizedform\ \nlbmath F\@. \ 
Moreover, 
the validity of \nlbmath F would be equivalent to the validity
of the 
\index{Herbrand expansion}%
\herbrandexpansion\ \nlbmath{F^{\cal U}} 
for a universe \nlbmath{\cal U},
provided only that this expansion were a finite formula
and did not vary over different universes. \
To provide this, we replace the semantical objects of the 
universe \nlbmath{\cal U} with syntactical objects, 
namely the countable set of all terms,
used as ``place holders'' or names. \ 
To get a {\em finite} formula,
we again replace this set of terms, which is infinite in general,
with the 
\index{champ fini|(}%
{\frenchfont champ fini} 
\nlbmath{\termsofdepthnovars n}, \ 
as defined in \sectref{section herbrand expansion}. \ 
If \nolinebreak we can show 
\nlbmath{F^{\termsofdepthnovars n}} to be a sentential tautology
for some positive natural number \nlbmath n, \hskip.2em
then we know
that the \math\gamma-quantifications in \nlbmath F
have solutions in any structure, 
and so we know that \math F \nolinebreak 
and \nlbmath A \nolinebreak are valid.\footnote{%
 \majorheadroom
 Indeed, we have \bigmaths{F^{\termsofdepthnovars n}\yields A},
 \cfnlb\ \littheoref 4 in 
 \heijenoortindex\citep{heijenoort-herbrand},
 which roughly is our \lemmref{lemma from C to yields a la heijenoort}.} \
Otherwise, the \loewenheimskolemtheorem\ says that \math A is invalid.
\notop\halftop
\begin{definition}[\propertyC, \herbrand\ Disjunction, \herbrand\ Complexity]%
\index{Property C!{\em definition}}% 
\index{Herbrand disjunction!{\em definition}}%
\index{Herbrand complexity!{\em definition}}%
\\
Let \math A be a \firstorder\ formula, in which, 
without loss of generality, any bound variable is
bound exactly once and does not occur again \freely, neither as a
variable nor as a function symbol. \ 
Let \math F be
the outer 
\index{form!Skolemized@(outer) Skolemized}%
\skolemizedform\ of \nlbmath A. \ Let \math n be a positive
natural number. \ Let the 
\index{champ fini|)}%
{\frenchfont champ fini}
\nlbmath{\termsofdepthnovars n} be formed over the function
and free variable symbols occurring in \nlbmath F. \\\math A
{\em has \propertyC\ of order \nlbmath n}\/ \udiff\ the 
\index{Herbrand expansion}%
\herbrandexpansion\ \nlbmath{F^{\,\termsofdepthnovars n}} is a
\sententialtautology. \
\\\indent
The \herbrandexpansion\ \nlbmath{F^{\,\termsofdepthnovars n}}
is sententially equivalent to the so-called {\em\herbrand\ disjunction
of \nlbmath A of order \nlbmath n}, \hskip .3em
which is the finite disjunction
\nlbmath{\bigvee_{\FUNDEF\sigma\Y{\termsofdepthnovars n}}E\sigma}, \ 
\mbox{where \Y\ is} 
the set of bound (\math\gamma-) variables of \nlbmath F, \
and \math E results from \math F by removing all 
(\math\gamma-) quantifiers. 
\\\indent
This form of representation can be used to define 
the {\em\herbrand\ complexity}\/ of \nlbmath A,
which is the minimal number of instances of \nlbmath E
whose disjunction is a \sententialtautology.\footnotemark
\getittotheright\qed\end{definition}

\pagebreak

\footnotetext
{\herbrand\ has no name for {\em\herbrand\ disjunction}\/
and does not use the notion of {\em\herbrand\ complexity},
which, however, is closely related to \herbrandsfundamentaltheorem,
which says that the \herbrand\ complexity 
of \nlbmath A is always defined as a positive natural number,
provided that \bigmaths{\tightyields A}{} holds. \ 
More formally, 
the {\em\herbrand\ complexity of \nlbmath A}
is defined as the minimal cardinality \nlbmath{\CARD S}
such that, for some positive natural number \nlbmath m and some
\ \bigmaths{S\nottight{\nottight{\nottight\subseteq}}
\FUNSET\Y{\termsofdepthnovars m}}, \ \
the finite disjunction
\bigmath{\bigvee_{\sigma\in S}E\sigma} 
is a \sententialtautology. \ 
It is useful in the comparison of logical calculi 
\wrt\ their smallest proofs for certain 
generic sets of formulas, 
\cfnlb\ \eg\ 
\index{Ferm\"uller, Christian G.}%
\citep{baazdelta}.%
}%
\newcommand\termeins{\app{\forallvari m{}}{\forallvari v{},\forallvari w{}}}%
\newcommand\termzwei{\app{\forallvari m{}}{\forallvari u{},\termeins}}%
\newcommand\boxeins{\framebox{\termeins}}%
\newcommand\boxzwei{\framebox{\termzwei}}%
\newcommand\boxu{\framebox{\forallvari u{}}}%
\newcommand\boxv{\framebox{\forallvari v{}}}%
\newcommand\boxw{\framebox{\forallvari w{}}}%
\begin{example}[\propertyC, \herbrand\ Disjunction, \herbrand\ Complexity]\label
{example running herbrand start}%
\index{Property C!{\em example}}% 
\index{Herbrand disjunction!{\em example}}%
\index{Herbrand complexity!{\em example}}%
\\
Let \math A be the following formula, which 
says that if we have transitivity and
an upper bound of two elements, 
then we also have an upper bound of three elements:
%\footnote
%{\label{note redundancy}We add some redundancy by marking bound
% \math\gamma-variables with a \nlbmath\gamma\ 
% and 
% bound \math\delta-variables with a \nlbmath\delta,
% \cfnlb\ \sectref{section herbrands calculi}.} 
%% We have decided not to annotate bound variables!
\par\noindent\LINEmaths{\noparenthesesoplist{
\forall\boundvari a{},\boundvari b{},\boundvari c{}\stopq\inpit{
\boundvari a{}\tightprec\boundvari b{}
\und
\boundvari b{}\tightprec\boundvari c{}
\implies
\boundvari a{}\tightprec\boundvari c{}}
\oplistund
\forall\boundvari x{},\boundvari y{}\stopq
\exists\boundvari m{}\stopq
\inpit{\boundvari x{}\tightprec\boundvari m{}\und 
\boundvari y{}\tightprec\boundvari m{}}
\oplistimplies \forall\boundvari u{},\boundvari v{},\boundvari w{}\stopq
\exists\boundvari n{}\stopq
\inpit{
\boundvari u{}\tightprec\boundvari n{}\und
\boundvari v{}\tightprec\boundvari n{}\und
\boundvari w{}\tightprec\boundvari n{}}}}{}{\Large\math{(A)}}\par\noindent
The outer 
\index{form!Skolemized@(outer) Skolemized}%
\skolemizedform\ \nlbmath F of \nlbmath A is
\par\noindent\LINEmaths{\noparenthesesoplist{
\forall\boundvari a{},\boundvari b{},\boundvari c{}\stopq
\inpit{
\boundvari a{}\tightprec\boundvari b{}
\und
\boundvari b{}\tightprec\boundvari c{}
\implies
\boundvari a{}\tightprec\boundvari c{}}
\oplistund
\forall\boundvari x{},\boundvari y{}\stopq
\inpit{
\boundvari x{}\tightprec\app
{\forallvari m{}}{\boundvari x{},\boundvari y{}}\und 
\boundvari y{}\tightprec\app
{\forallvari m{}}{\boundvari x{},\boundvari y{}}}
\oplistimplies 
\exists\boundvari n{}\stopq
\inpit{
\forallvari u{}\tightprec\boundvari n{}\und
\forallvari v{}\tightprec\boundvari n{}\und
\forallvari w{}\tightprec\boundvari n{}}}}{}{\Large\math{(F)}}\par\noindent
The result of removing 
the quantifiers from \nlbmath F is the formula \nlbmath E:
\par\noindent\LINEmaths{\noparenthesesoplist{\inpit{
\boundvari a{}\tightprec\boundvari b{}
\und
\boundvari b{}\tightprec\boundvari c{}
\implies
\boundvari a{}\tightprec\boundvari c{}}
\oplistund
\boundvari x{}\tightprec\app
{\forallvari m{}}{\boundvari x{},\boundvari y{}}\und 
\boundvari y{}\tightprec\app
{\forallvari m{}}{\boundvari x{},\boundvari y{}}
\oplistimplies 
\forallvari u{}\tightprec\boundvari n{}\und
\forallvari v{}\tightprec\boundvari n{}\und
\forallvari w{}\tightprec\boundvari n{}}}{}{\Large\math{(E)}}\par\noindent
By semantical considerations it is obvious 
that a solution for \nlbmath{\boundvari n{}} is
\termzwei. \ 
This is a term of 
\index{height of a term}%
height \nolinebreak 3,
which suggests that 
\math A \nolinebreak has \propertyC\ of order \nlbmath 4. \ 
Let us show that this is indeed the case
and that the 
\index{Herbrand complexity|)}%
\herbrand\ complexity of \nlbmath A is \nlbmath 2. \ 
Consider the following 2 substitutions: \
\bigmath{\begin{array}[t]{l l@{}l@{}l l@{}l@{}l l@{}l@{}l l}
  \{ 
 &\boundvari a{}&\mapsto&\forallvari v{},
 &\boundvari b{}&\mapsto&\termeins,
 &\boundvari c{}&\mapsto&\termzwei,
\\
 &\boundvari x{}&\mapsto&\forallvari v{},
 &\boundvari y{}&\mapsto&\forallvari w{},
 &\boundvari n{}&\mapsto&\termzwei
 &\};
\\\{
 &\boundvari a{}&\mapsto&\forallvari w{},
 &\boundvari b{}&\mapsto&\termeins,
 &\boundvari c{}&\mapsto&\termzwei,
\\
 &\boundvari x{}&\mapsto&\forallvari u{},
 &\boundvari y{}&\mapsto&\termeins,
 &\boundvari n{}&\mapsto&\termzwei
 &\}.
\\\end{array}}{}\\\smallheadroom
Indeed, if we normalize the \herbrand\ disjunction generated by 
these two substitutions to a disjunctive normal form 
(\ie\ a disjunctive set of conjunctions) 
we get the following \sententialtautology.\smallfootroom
\\%\par\noindent
\LINEmaths{\begin{array}[c]{@{}l@{}l@{}}
  \{\ \ 
 &\forallvari v{}\tightprec\termeins
  \und\termeins\tightprec\termzwei
  \und\forallvari v{}\tightnotprec\termzwei\comma
\\
 &\forallvari w{}\tightprec\termeins
  \und\termeins\tightprec\termzwei
  \und\forallvari w{}\tightnotprec\termzwei\comma
\\
 &\forallvari v{}\tightnotprec\termeins
  \comma
  \forallvari w{}\tightnotprec\termeins
  \comma
\\
 &\forallvari u{}\tightnotprec\termzwei\comma
  \termeins\tightnotprec\termzwei\comma
\\
 &\forallvari u{}\tightprec\termzwei\und
  \forallvari v{}\tightprec\termzwei\und
  \forallvari w{}\tightprec\termzwei
  \ \ \}
\\\end{array}}{}{\begin{tabular}
{@{}r@{}}\\\\{\Large\math{(C)}}\\\\\qed\\\end{tabular}}\par\noindent
\end{example}

\begin{sloppypar}
\noindent
The different treatment of 
\mbox{\math\delta-quantifiers} and \math\gamma-quantifiers
in \propertyC,
namely by \skolemization\ and 
\index{Herbrand expansion}%
\herbrandexpansion, respectively,
as found in 
\skolemindex\citep{skolem-1928} and \citep{herbrand-PhD},
rendered the reduction to 
sentential logic by hand (or \nolinebreak actually today, with a computer)
practically executable for the first time.\footnotemark\ 
This different treatment of the two kinds of quantification
is inherited from the 
\index{Peirce--Schr\"oder tradition}%
\peirce--\schroeder\ 
tradition\arXivfootnotemarkref{note peirce schroeder tradition}
which came on \herbrand\ via 
\loewenheimindex\loewenheim\ and 
\skolemindex\skolem. \ 
\russellindex\russell\ and 
\hilbertindex\hilbert\ had already merged
that tradition with the one of 
\fregeindex\frege,
sometimes emphasizing their \frege\ heritage over 
one of 
\index{Peirce--Schr\"oder tradition}%
\peirceindex\peirce\ and 
\schroederindex\schroeder\fullstopnospace\footnotemark\
It \nolinebreak was \herbrand\ who completed the bridge 
between these two traditions with his \fundamentaltheorem,
as depicted in 
\sectref{section herbrand fundamental theorem} 
% Note that we should not move the picture to this place,
% because one of the functions of the picture is to help
% readers that skip this technical section to continue reading.
below.\pagebreak

\end{sloppypar}
%%%%%%%%%%%%%%%%%%%%%%%%%%%%%%%%%%%%%%%%%%%%%%%%%%%%%%%%%%%%%%%%%%%%%%%%%%%%%%% 
\section{\herbrandsfalselemma}\label{section lemma}%
\addtocounter{footnote}{-1}%
\footnotetext
{For instance, the elimination of both \math\gamma- and
 \math\delta-quantifiers with the help of
 \index{Hilbert!'s epsilon}%
 \hilbert's \mbox{\nlbmath\varepsilon-operator} suffers from an exponential
 complexity in formula size. \ 
 As \nolinebreak a result, already small formulas grow so large 
 that the mere size makes them inaccessible to human inspection; \hskip.3em
 and this is still the case
 for the term-sharing representation of 
 \index{Hilbert!'s epsilon}%
 \mbox{\math\varepsilon-terms} of \nolinebreak\citep{wirth-jal}.%
}%
\addtocounter{footnote}{1}%
\footnotetext
{While this emphasis on \fregeindex\frege\ 
 will be understood by everybody who ever had the 
 fascinating experience of reading \frege,
 it put some unjustified bias to the historiography of modern logic,
 still present in the selection of 
 the famous source book \citep{heijenoort-source-book}; \
 \cf\ \eg\ 
 \anellisindex%
 \citep[\litchapref 3]{anellis-heijenoort-long}.%
 %, \citep{anellis-brady}. 
 % This article of anellis is great, but I can follow the crucial argument
 % on \p\,353 only in so far as that \peirce\ has a 
 % \firstorder\ theory fully developed and articulated,
 % but not regarding the computation of a \skolem\ normal form in 
 % \cite{peirce-1985} (even the prenex normalization is only explained
 % for formulas already in negation normal form, or more precisely
 % for formulas where no negation occurs above a quantifier.
 % There is also no application of the \loewenheimskolemtheorem\
 % in \cite{peirce-1985}!
}%
\noindent 
For a given positive natural number \nlbmath n, \ 
\index{Herbrand's ``False Lemma''!{\em definition}}%
{\em\herbrandsfalselemma} \ 
says that
\propertyC\ of order \nlbmath n \hskip.2em
is invariant under the application of the 
\index{Rule!of Passage}%
Rules of Passage.
%to a formula \nlbmath A.
\par\indent
The basic function of \herbrandsfalselemma\ in 
the proof of \herbrandsfundamentaltheorem\ 
is to establish the logical equivalence of 
\propertyC\ of a formula \nlbmath A
with \propertyC\ of the 
\index{form!prenex}%
{\em prenex}\/ and 
\index{form!anti-prenex}%
{\em anti-prenex forms}\/ of 
\nlbmath A,
% normal forms for a classical \firstorder\ formula 
% which we have discussed in 
\cfnlb\ \sectref{section discussion prenex}. \ 
\par\indent
\herbrand's Lemma is wrong because the 
\index{Rule!of Passage}%
Rules of Passage 
may change the
outer 
\index{form!Skolemized@(outer) Skolemized}%
\skolemizedform. \ 
This happens whenever 
a \mbox{\math\gamma-quantifier} binding \nlbmath{x} is moved
over a binary operator 
whose unchanged operand \nlbmath B 
contains a \mbox
{\math\delta-quantifier}.\footnotemark
\par\indent
To find a counterexample for \herbrand's Lemma for the case of
\propertyC\ of order \nlbmath 2, \hskip .2em
let us consider moving out the \mbox{\math\gamma-quantifier} \nolinebreak
``\math{\exists\boundvari x{}.}'' \ in
the valid formula
%\footnotemark[\ref{note redundancy}]
\par\noindent\LINEmaths{\inpit{\exists\boundvari x{}.\,\Pppp{\boundvari x{}}}
\oder\neg\exists\boundvari y{}.\,\Pppp{\boundvari y{}}}.
\par\noindent The 
\index{form!Skolemized@(outer) Skolemized}%
(outer) \skolemizedform\ of this formula is 
\par\noindent\LINEmaths{\inpit{\exists\boundvari x{}.\,\Pppp{\boundvari x{}}}
\oder\neg\Pppp{\forallvari y{}}}.
\par\noindent The \herbrand\ disjunction over the single substitution
\nlbmath{\{\boundvari x{}\tight\mapsto\forallvari y{}\}}
is a \sententialtautology. \
The 
\index{form!Skolemized@(outer) Skolemized}%
{\em outer}\/ \skolemizedform\ after moving out the 
``\math{\exists\boundvari x{}.}'' is 
\par\noindent\LINEmaths{\exists\boundvari x{}\stopq\inparentheses
{\Pppp{\boundvari x{}}\oder\neg\Pppp{\app{\forallvari y{}}{\boundvari x{}}}}}.
\par\noindent
To get a \sententialtautology\ again,
we now have to take the \herbrand\ disjunction over both 
\nlbmath{\{\boundvari x{}\tight\mapsto\app{\forallvari y{}}{l}\}}
and \nlbmath{\{\boundvari x{}\tight\mapsto l\}} \ 
(instead of the single 
\nlbmath{\{\boundvari x{}\tight\mapsto\forallvari y{}\}}), \ 
for the lexicon \nlbmath l\@. \ 

\begin{sloppypar}

\par\indent
This, however, is not really 
a counterexample for \herbrand's Lemma
because \herbrand\ treated the lexicon \nlbmath l as a variable and 
defined the height of a \skolem\ constant to be \nlbmath 1, 
and the 
\index{height of a term}%
height of a variable to be \nlbmath 0,
so that 
\ \mbox{\math{\CARD{{\forallvari y{}}{}}=1=\CARD{\app{\forallvari y{}}{l}}}}. \ 
As free variables and \skolem\ constants play exactly the same \role,
this definition of 
\index{height of a term}%
\index{height of a term!{\em treatment of the lexicon}}%
height is a bit unintuitive and was possibly introduced
to avoid this counterexample. 
\par\indent
However, for the similar formula
\par\noindent\LINEmaths{
\inparentheses{\inpit{\exists\boundvari x{}.\,\Pppp{\boundvari x{}}}
\und\forall\boundvari y{}.\,\Qpppeins{\boundvari y{}}}
\nottight{\oder}\neg\inpit{\exists\boundvari x{}.\,\Pppp{\boundvari x{}}}
\nottight{\oder}\neg
\forall\boundvari y{}.\,\Qpppeins{\boundvari y{}}}{}%(1)
\par\noindent
% the outer \skolemizedform\ is 
% \\\LINEmaths{\inpit{\exists\boundvari x{}.\,\Pppp{\boundvari x{}}
% \und\Qpppeins{\app{\forallvari y{}}{}}}
% \oder\neg\Pppp{\app{\forallvari x{}}{}}\oder\neg
% \forall\boundvari y{}.\,\Qpppeins{\boundvari y{}}}{}(2)\\
after moving the first \math\gamma-quantifier \nolinebreak
``\math{\exists\boundvari x{}.}'' out 
over the ``\tightund\closequotecommaextraspace
we have to apply 
\par\noindent instead of\LINEmath{
\begin{array}[b]{l l}\{ \ \boundvari x{}\mapsto\forallvari x{},
 &\boundvari y{}\mapsto\app{\forallvari y{}}{\forallvari x{}} \ \}
%  \phantom{\forallvari y{}}
\\\majorheadroom
  \{ \ \boundvari x{}\mapsto\forallvari x{}, \ 
 &\boundvari y{}\mapsto\forallvari y{} \ \}
  \phantom{\app{\forallvari y{}}{\forallvari x{}}}
\\\end{array}}\phantom{instead of}%
\par\noindent
to get a \sententialtautology,
and we have 
\maths{\CARD{\forallvari y{}}=1}{} \ and
\maths{\CARD{\app{\forallvari y{}}{{\forallvari x{}}{}}}=2}, \ 
and thus
\bigmaths{\forallvari y{}
\in\termsofdepthnovars 2}, but
\bigmaths{\app{\forallvari y{}}{{\forallvari x{}}{}}
\notin\termsofdepthnovars 2}. \ \ 
\par\indent This means:
\propertyC\ of order \nlbmath 2 varies under a single application of a
\index{Rule!of Passage}% 
Rule of Passage, and thus
we have a proper counterexample for \herbrandsfalselemma\ \nolinebreak here.%
\pagebreak\par\end{sloppypar}

\subsection{\bernays' Correction}
\footnotetext{\label{footnote same meta}%
 % \majorheadroom
 Here we use the same meta variables
 as in our description of the Rules of Passage in 
 \nlbsectref{section herbrands calculi}
 and assume that \math x \nolinebreak does not occur \freely\ in \nlbmath B.%
}%
In\,1939, 
\bernaysindexbegin\bernays\ remarked that \herbrand's proof is hard to
follow\footnote{%
 \headroom
 In \litnoteref 1 of \citep[\p 158]{grundlagen-first-edition-volume-two}
 (\cite[\p 161]{grundlagen-second-edition-volume-two}), \hskip.2em 
 we read: \
 ``{\germanfontfootnote Die \herbrand sche Bewei\esi f\ue hrung ist
    schwer zu verfolgen}''} 
and 
---~for the first time~--- 
published a sound proof of a version of \herbrandsfundamentaltheorem\ which is
restricted to 
\index{form!prenex}%
prenex form, but more efficient in the number of terms
that have to be considered in a \herbrand\ disjunction than \herbrand's
quite global limitation to {\em all terms \nlbmath t with \
  \math{\CARD t<n}}, \ \ related to \propertyC\ of order 
\nlbmath n.\footnote{\label{footnote bernays}%
 \majorheadroom
 \Cfnlb\ \litsectref{3.3} of \citep{grundlagen-first-edition-volume-two}\@. \ 
 In the second edition \citep{grundlagen-second-edition-volume-two},
 \bernaysindex\bernays\ also indicates how to remove the restriction to
 \index{form!prenex}%
 prenex formulas.}

\subsection{\goedel's and \dreben's Correction}
According to a conversation with 
\heijenoortindex\heijenoort\ in
autumn\,1963,\footnote{%
 \majorheadroom
 \Cfnlb\ \citep[\p\,8, \litnoteref j]{herbrand-ecrits-logiques}\@.}  
\goedelindex\goedel\ noticed the lacuna in the proof
of \herbrandsfalselemma\ in 1943 and wrote a private note,
but did not publish \nolinebreak it. \
While \goedel's documented attempts to construct a counterexample to
\herbrandsfalselemma\ failed,
he had actually worked out a 
\index{correction (of Herbrand's False Lemma)!G\"odel's and Dreben's|(}%
{\em \repair}\/ of 
\herbrandsfalselemmacomma
which is sufficient for the proof of \herbrandsfundamentaltheorem.\footnote{%
 \majorheadroom
 \Cfnlb\ 
 \citep{goldfarb-herbrand-goedel}.}
\par\halftop\indent
In\,1962, when \goedel's \repair\ was still unknown,
a young student, 
\andrewsindexbegin%
\andrewsname, had the audacity to tell
his advisor 
\churchname\ \churchlifetime\ that there seemed to be a gap in 
the proof of \herbrand's (False) Lemma. 
\church\ sent \andrews\ to \drebenname\ \drebenlifetime, who finally came up
with a counterexample. 
And then \andrews\ constructed a simpler counterexample
(essentially the one we presented above)
and joint work found a \repair\ similar to \goedel's,\footnote{%
 \majorheadroom
 \Cfnlb\ 
 \andrewsindex%
 \citep{andrews-herbrand-award},
 \citep{false-lemmas-in-herbrand}, 
 \citep{supplement-to-herbrand}.}
which we will call 
\index{correction (of Herbrand's False Lemma)!G\"odel's and Dreben's!{\em definition}}%
\andrewsindexend%
{\em\goedel's and \dreben's \repair}
in \nlbsectref{section modus ponens elimination}.
\par\halftop\indent
Roughly speaking, the \repaired\ lemma says that --- to keep \propertyC\ of
\nlbmath A invariant under (a single application of) a 
\index{Rule!of Passage}%
Rule of Passage --- we may have to step from 
order \nlbmath n \hskip.2em to order
\par\noindent\LINEmaths{n\cdot\inpit{N^r+1}^n}.
\par\noindent Here \math{r} \nolinebreak is the
number of \mbox{\math\gamma-quantifiers} in whose scope the 
\index{Rule!of Passage}%
Rule of Passage is applied and \math{N} is the cardinality of
\nlbmath{\termsofdepthnovars n} for the function symbols in the outer
\index{form!Skolemized@(outer) Skolemized}%
\skolemizedform\ of \nlbmath A.\footnote {%
 \majorheadroom
 \Cfnlb\ \citep[\p\,393]{supplement-to-herbrand}.}  
\par\halftop\indent
This \repair\ is not particularly elegant 
because --- iterated several times until a 
\index{form!prenex}%
prenex form is reached \nolinebreak--- it can lead to pretty high orders. \ 
Thus, although this 
\index{correction (of Herbrand's False Lemma)!G\"odel's and Dreben's|)}%
\repair\ serves
well for soundness and finitism,
it results in a
% intractable ist der richtige Fachbegriff!
complexity that is unacceptable in practice (\eg\ in automated reasoning)
already for small non-prenex formulas.%
%examples and thus 
%can hardly be considered to be intuitively clear.
\pagebreak

\subsection{\heijenoort's Correction}
The problems with \herbrandsfalselemma\ in the step from \propertyC\ to 
a proof without 
\index{modus ponens}%
\index{modus ponens!elimination}%
{\em modus ponens}\/ in his \fundamentaltheorem\ 
(\cfnlb\ \sectref{section modus ponens elimination}) \hskip.2em
result primarily\footnote{\label{note discussion inner}%
 % \majorheadroom
 Secondarily, 
 the flaw in 
 \herbrandsfalselemma\  is a peculiarity of the 
 \index{form!Skolemized@(outer) Skolemized}%
 {\em outer}\/ \skolemizedform.
 For the 
 \index{form!Skolemized @inner Skolemized}%
 {\em inner}\/ \skolemizedform\ (\cfnlb\ \noteref{note inner}), 
 moving \math\gamma-quantifiers
 with the 
 \index{Rule!of Passage}%
 Rules of Passage
 cannot change the number of arguments of the \skolem\ functions. 
 This does not help, however,
 because, for the inner \skolemizedform,
 moving a \math\delta-quantifier may change the number
 of arguments of its \skolem\ function if the 
 \index{Rule!of Passage}%
 Rule of Passage is 
 applied within the scope of a \mbox{\math\gamma-quantifier} whose bound
 variable occurs in \nlbmath B but not in 
 \nlbmath A.\arXivfootnotemarkref{footnote same meta}
 The inner \skolemizedform\ of 
 \\\LINEmaths{\exists\boundvari y 1\stopq
 \forall\boundvari z 1\stopq\Qppp{\boundvari y 1}{\boundvari z 1} 
 \oder
 \exists\boundvari y 2\stopq
 \forall\boundvari z 2\stopq\Qppp{\boundvari y 2}{\boundvari z 2}}{}
 \\is 
 %\phantom,
 \\\LINEmaths{\exists\boundvari y 1\stopq
 \Qppp{\boundvari y 1}{\app{\forallvari z 1}{\boundvari y 1}}
 \oder
 \exists\boundvari y 2\stopq
 \Qppp{\boundvari y 2}{\app{\forallvari z 2}{\boundvari y 2}}
 },%\phantom{is}
 \\ but the inner \skolemizedform\ of any 
 \index{form!prenex}%
 prenex form
 has a {\em binary}\/ \skolem\ function, unless we use
 \henkin\ quantifiers as found in \hintikka's %independence-friendly 
 \firstorder\ logic, \cfnlb\ \citep{hintikkaprinciples}.\majorfootroom}
from a detour over 
\index{form!prenex}%
prenex form, which 
was standard at \herbrand's time.
\loewenheimindex\loewenheim\ and \skolem\ had always reduced their problems to 
\index{form!prenex}%
prenex forms
of various kinds. 
The reduction of a proof task to prenex form
has several disadvantages, however, such as 
serious negative effects on proof complexity.\footnote{%
 \majorheadroom
 \Cfnlb\ \eg\
 \index{Ferm\"uller, Christian G.}%
 \citep{baazdelta}; \ \citep{baazleitschcolllog}.\majorfootroom}
If \herbrandname\ had known of his flaw, 
he would probably have avoided the whole detour over 
\index{form!prenex}%
prenex forms,
namely in the form of what we will call 
\index{correction (of Herbrand's False Lemma)!Heijenoort's}%
\index{correction (of Herbrand's False Lemma)!Heijenoort's!{\em definition}}%
{\em\heijenoort's \repair},%
\index{Generalized Rule!of gamma-Quantification@of \math\gamma-Quantification}%
\footnote{%
 \majorheadroom
 The first published hints on {\em\heijenoort's \repair}\/ are
 \citep[\litnoteref{77}, \p\,555]{heijenoort-source-book} and
 \citep[\litnoteref{60}, \p171]{herbrand-logical-writings}. \ 
 On page\,99 of 
 \heijenoortindex\citep{heijenoort-work-herbrand}, without giving a definition, 
 \heijenoortindex\heijenoort\ speaks of generalized versions 
 (which \herbrand\ did not have) 
 of the rules of
 ``existen\-tial\-ization and universalization\closequotecomma
 which we have formalized in our 
 Generalized Rules
 of Quantification in \nlbsectref{section herbrands calculi}.
 Having studied \herbrand's \PhDthesis\ \citep{herbrand-PhD} and 
 \heijenoortindex\makeaciteoftwo{heijenoort-tree-herbrand}{heijenoort-herbrand},
 what 
 \heijenoortindex\heijenoort's generalized rules must look like
 can be inferred from the following two facts:
 \herbrand\ has a generalized version
 of his 
 \index{Generalized Rule!of Simplification}%
 \index{Rule!of Simplification}%
 Rule of Simplification in addition to a non-generalized one. 
 Rewriting with the 
 Generalized Rule of \math\gamma-Quantification
 within the scope
 of quantifiers would not permit \herbrand's constructive proof
 of his \fundamentaltheorem.%
%% Without studying the cited papers in addition to
%% \pagebreak
%% \citep{heijenoort-work-herbrand}, however,
%% \heijenoort's non-prenex \repair\ can hardly be understood
%% by the readers of \citep{heijenoort-work-herbrand}. \ 
%% Thus, \heijenoort\ had better made this departure from
%% the historical facts explicit.
\par
Note that 
\index{correction (of Herbrand's False Lemma)!Heijenoort's}%
\heijenoort's \repair\ avoids 
the detour over the Extended 
\index{Hilbert!'s epsilon}%
First \nlbmath\varepsilon-Theorem of 
the proof of 
\bernaysindexend\bernays\ mentioned above; \cfnlb\ \noteref{footnote bernays}. \ 
Moreover, 
\heijenoortindex\heijenoort\ gets along without 
\herbrand's complicated 
\index{form!prenex}%
prenex forms with raised
\math\gamma-multiplicity, which are required for \herbrand's definition of
\index{Property A}%
\propertyA\@. \par 
Note, however, that \goedel's and \dreben's \repair\ is still needed for the 
 step from a proof with 
\index{modus ponens}%
\index{modus ponens!elimination}%
{\em modus ponens}\/ to \propertyC,
 \ie\ from Statement\,4 to Statement\,1 in 
 \theoref{theorem herbrand fundamental two}. \ 
 As the example on top of page\,\,201 
 in \citep{herbrand-logical-writings} shows,
 an intractable increase of the order of \propertyC\ 
 cannot be avoided in general for 
an inference step by {\em modus ponens}.\majorfootroom}
which avoids an intractable and 
unintuitive\footnote{%
 \majorheadroom
 Unintuitive \eg\ in the sense of 
 \citep{tait-2006}.}
rise in complexity, 
\nolinebreak\cfnlb\ \sectref{section modus ponens elimination}.\vfill\pagebreak
%%%%%%%%%%%%%%%%%%%%%%%%%%%%%%%%%%%%%%%%%%%%%%%%%%%%%%%%%%%%%%%%%%%%%%%%%%%%%%%%
\section{The \fundamentaltheorem}%
\label{section herbrand fundamental theorem}%

\noindent The \fundamentaltheorem\ of {\herbrandname} 
is not easy to comprehend at first, because of
its technical nature, but it rests upon a basic intuitive idea, 
which turned out to be one of the 
most profound insights in the history of logic. 

We know --- and so did {\herbrand} --- that sentential logic is decidable: 
for any given sentential formula,
we could, for instance, use truth-tables to decide its validity. \ 
But what about a \firstorder\ formula with quantifiers? 

There is 
%% we combine \peirce's observation that \bigmaths{\exists y\stopq
%%   \Pppp{y}}{} is a shorthand for \bigmaths
%% {\Pppp{a}\oder\Pppp{b}\oder\Pppp{c}\oder\ldots\quad}, where
%% \bigmaths{a,b,c,\ldots}{} \nolinebreak denote the elements of a
%% domain, and 
\loewenheimindex\loewenheim's and 
\skolemindex\skolem's observation that
\bigmaths{\forall x\stopq\Pppp{x}}{} in the context of the
existentially quantified variables \nlbmath{y_1,\ldots,y_n} stands for
\Pppp{\forallvari x{}(y_1,\ldots,y_n)} for an arbitrary \skolem\ function
\math{\forallvari x{}(\cdots)}, 
as outlined in \nlbsectref{section skolemization}. \ 
This gives us a formula with existential quantifiers only. \
Now, taking the \herbrand\ disjunction,
an existentially quantified formula can be shown to be valid, 
if we find a finite set of names denoting elements from the domain to be
substituted for the existentially quantified variables, such that the
resulting sentential formula is truth-functionally valid. \
%% Using the duality $\neg\forall
%% x\stopq\Pppp{x}\equiv\exists x\stopq\Pppp{x}$ and the technical device
%% of {\skolemization} (see \sectref{section skolemization})
Thus, we have a model-theoretic argumentation 
how to reduce a given \firstorder\ formula to
a sentential one. \
The semantical elaboration of this idea is due to 
\loewenheimindexstart\loewenheim\ and 
\skolemindex\skolem, and this was known to {\herbrand}. \ 

But what about the reducibility of an
{\em actual proof}\/ of a given formula within a \firstorder\ calculus? \

The affirmative answer to this question is the essence of \herbrand's
Fundamental Theorem and the technical device, by which we can
eliminate a switch of quantifiers 
(such as \math{\exists y.\,\forall x.\,\Qppp x y}{} 
 of \nlbsectref{section skolemization}) \hskip .2em
is captured in his \propertyC\@.

Thus, if we want to 
cross the river that divides the land of {\em valid}\/ 
\firstorder\ formulas from the
land of {\em provable}\/ ones, it is the sentential \propertyC\ that
stands firm in the middle of that river and holds the bridge, 
whose first half was built by 
\loewenheimindexend\loewenheim\ and 
\skolemindex\skolem\ and 
the other by \herbrand:
\yestop
\begin{center}\includegraphics
%[width=120mm,height=55mm]
[bb=20 20 575 308,width=144mm,height=66mm]
{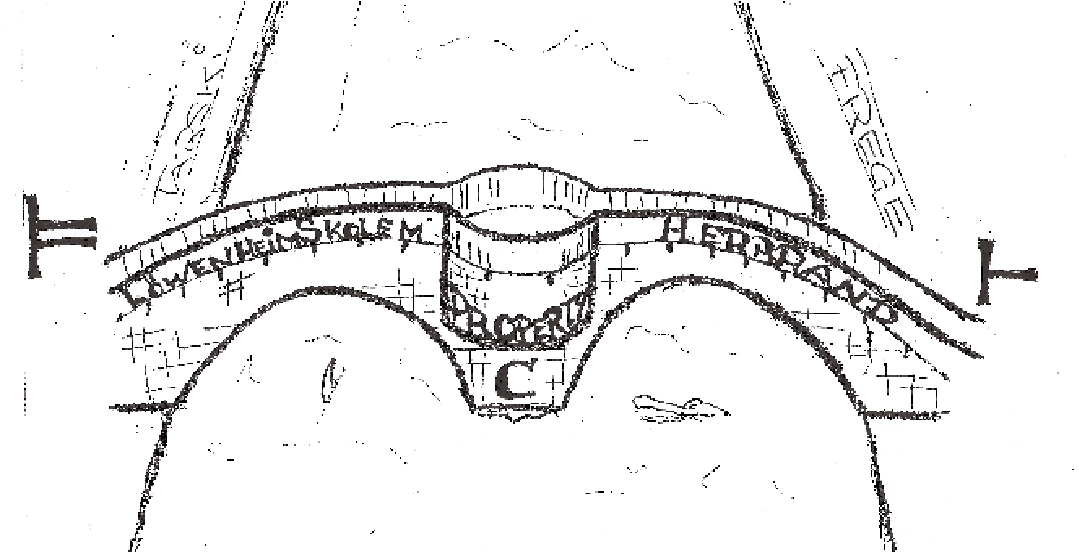}
\end{center}
\pagebreak

\herbrandsfundamentaltheorem\ shows that if a formula $A$ has
\propertyC\ of some order \nlbmath n\, 
--- \nolinebreak\ie, by the \loewenheimskolemtheorem, if $A$ is a valid 
(\math{\models A}) \nolinebreak---
then we not only {\em know of the existence}\/ of 
a proof in any of the standard proof calculi (\math{\tightyields A}), \hskip.3em
but we can actually {\em construct}\/ a 
proof for \nlbmath A in \herbrand's calculus
from a given \nlbmath n. \ 
The proof construction process
is guided by the 
\index{champ fini|(}%
{\frenchfont champ fini} of order \nlbmath n, 
whose size determines the multiplicities of $\gamma$-quantifiers and whose
elements are the terms substituted as witnesses in the 
$\gamma$-Quantification steps. 
That proof begins with a sentential tautology
and may use the Rules of\/ \math\gamma- and\/
\math\delta-Quantification, the 
\index{Generalized Rule!of Simplification}%
Generalized Rule of Simplification, and the 
\index{Rule!of Passage}%
Rules of Passage.

Contrary to what 
\herbrandsfalselemma\ implies,
a detour over a 
\index{form!prenex}%
prenex form of \nlbmath A
dramatically increases 
the order of \propertyC\ and thus the length of that proof,
\cfnlb\ \sectref{section lemma}. \ 
\heijenoortindex\heijenoort, 
however, 
observed that this rise of proof length can be overcome 
by avoiding the
problematic 
\index{Rule!of Passage}%
Rules of Passage
with the help of 
\index{inference!deep}%
deep (or Generalized) quantification rules,
which may introduce quantifiers deep within formulas
\index{correction (of Herbrand's False Lemma)!Heijenoort's}%
(``\heijenoort's \repair''). \
We have included these considerations into our statement of
\herbrandsfundamentaltheorem.

\yestop
\begin{theorem}[\herbrandsfundamentaltheorem]%
\label{theorem herbrand fundamental two}%
\index{Herbrand's Fundamental Theorem!{\em definition}}%
\\Let\/ \math A be a \firstorder\ formula 
in which each bound variable
is bound by a single quantifier and does not occur \freely. 
The following five statements are logically equivalent. \
Moreover, 
we can construct a witness for any statement from a witness of any other
statement.
\begin{enumerate}
\noitem\item[1.]\math A has \propertyC\ of order \nlbmath n for some positive 
natural number \nlbmath n.
\noitem\item[2.]\sloppy We can derive \math A from a \sententialtautology,
starting possibly with applications of 
the 
\index{Generalized Rule!of delta@of \math\delta-Quantification|(}%
\index{Generalized Rule!of gamma-Quantification@of \math\gamma-Quantification|(}%
Generalized Rules of\/ \math\gamma- and\/ \math\delta-Quantification,
which are then possibly followed by applications of the 
\index{Generalized Rule!of gamma-Simplification@of \math\gamma-Simplification|(}%
Generalized Rule of\/ \mbox{\math\gamma-Simplification}.%
\noitem\item[3.]We can derive \math A from a \sententialtautology,
starting possibly with applications of 
the Rules of\/ \math\gamma- and\/ \math\delta-Quantification,
which are then possibly followed by applications of 
the Generalized Rule of\/ \math\gamma-Simplification and the 
\index{Rule!of Passage}%
Rules of Passage.
\noitem\item[4.]We can derive \math A from a \sententialtautology\ with
the 
\index{Rule!of delta-Quantification@of \math\delta-Quantification}%
\index{Rule!of gamma-Quantification@of \math\gamma-Quantification}%
Rules of\/ \math\gamma- and\/ \math\delta-Quantification, the 
\index{Rule!of Simplification}%
Rule of Simplification, the 
\index{Rule!of Passage}%
Rules of Passage, and 
\index{modus ponens}%
Modus Ponens.
\noitem\item[5.]We can derive \math A in one of the standard \firstorder\
calculi of 
\index{Principia Mathematica!calculi of}%
\PM\ or of the 
\index{Hilbert!school!calculi of}%
\hilbert\ school.\footnote
{\Cfnlb\ 
\index{Principia Mathematica}%
\citep[*10]{PM}, \ 
\citep[Editors' Preface to Part\,B of Volume\,I, \p\,lxiii\,\f]
{grundlagen-german-english-edition-volume-one-two}
(or \citep[\litsectfromtoref{3}{5}]{grundlagen-first-edition-volume-one},
\citep[Supplement\,I\,D]{grundlagen-first-edition-volume-two}), \hskip.2em 
respectively.}\getittotheright
\qed\end{enumerate}\end{theorem}

\yestop\yestop\noindent
The following deserves emphasis: \hskip.3em
The derivations in the
above Statements \nolinebreak 2 to \nolinebreak 5 as well as the 
number \nlbmath n \hskip.1em
of Statement\,1 \hskip.05em
can be {\em constructed} \hskip.18em 
from each other; \hskip .3em
and this construction is finitistic in the spirit of
\herbrand's basic beliefs in the nature of proof theory and
metamathematics. \hskip.3em
Statement\,\,2 \hskip.09em 
is due to 
\index{correction (of Herbrand's False Lemma)!Heijenoort's}%
\heijenoort's \repair; \hskip.25em 
\cfnlb\ \sectrefs{section lemma}{section modus ponens elimination}. \ \ \
Statement\,\,3 \hskip.09em  
and \herbrand's 
\index{Property A}%
\propertyA\, are 
extensionally equal and intensionally very close to each other. \

\vfill\pagebreak
%%%%%%%%%%%%%%%%%%%%%%%%%%%%%%%%%%%%%%%%%%%%%%%%%%%%%%%%%%%%%%%%%%%%%%%%%%%%%%%%
\section{{\em Modus Ponens}\/ Elimination}\label
{section modus ponens elimination}%
\index{modus ponens|(}%
\index{modus ponens!elimination|(}%

\noindent The following lemma provides the step from Statement\,1 to
Statement\,2 of \theoref{theorem herbrand fundamental two}
with additional details exhibiting 
an elimination of {\em modus ponens}\/
similar to the 
\index{Cut elimination}%
Cut elimination in 
\index{Gentzen!'s {\germanfont Hauptsatz}}%
\gentzensHauptsatz.

\halftop
\halftop
\noindent
We present the lemma in parallel both 
% in the version of 
% \index{correction (of Herbrand's False Lemma)!G\"odel's and Dreben's}%
% \goedel's and \dreben's \repair\ 
% and of
%CP 20140424:
in a version following   
\index{correction (of Herbrand's False Lemma)!Heijenoort's}%
\heijenoort's \repair\footnote{%
 % \majorheadroom
 \Cfnlb\ \sectref{section lemma}.
 We present \heijenoort's \repair\ actually in form of \littheoref 4 in
 \heijenoortindex\citep{heijenoort-herbrand} with a slight change, 
 which becomes necessary for our use of \herbrand\ disjunction
 instead of the 
 \index{Herbrand expansion}%
 \herbrandexpansion, namely the addition of
 the underlined part of 
 Step\,1 in \lemmref{lemma from C to yields a la heijenoort}.
 %CP 20140424:
 A strongly improved and more elaborate version of this whole
 section on {\it modus ponens} elimination, 
 which uses the much more efficient \herbrand\ expansion 
 and is directly mirroring 
 \index{correction (of Herbrand's False Lemma)!Heijenoort's}%
 \heijenoort's \repair,
 is found in \cite[\litsectrefs{5.2}{5.3}]{wirth-heijenoort}.%
}  
%CP 20140424:
and in a restricted version that gets along without 
\herbrandsfalselemma. \hskip.2em
To melt these two versions into one,
we\begin{itemize}\noitem\item
underline the parts that are just part of 
\index{correction (of Herbrand's False Lemma)!Heijenoort's}%
\heijenoort's \repair\
and\noitem\item
overline the 
% part that result from \goedel's and \dreben's.
%CP 20140424:
restriction that is required to get along without \herbrandsfalselemma.
\noitem
\end{itemize}
Thus, the lemma stays valid if we omit 
either the underlined or else the overlined part of \nolinebreak it, 
but not both.

\yestop\begin{lemma}[{\em Modus Ponens}\/ Elimination\footroom]
\label{lemma from C to yields a la heijenoort}\\\noindent
\index{modus ponens!elimination}%
Let\/ \math A be a \firstorder\ formula 
\index{form!prenex}%
\overline{\mbox{in prenex form}}
%% \overline{\raisebox{.7em}{}\mbox{in}} 
%% \overline{\raisebox{.7em}{}\mbox{prenex}} 
%% \overline{\raisebox{.7em}{}\mbox{form}}
in which each bound variable
is bound by a single quantifier and does not occur \freely. \ 
Let\/ \math F be the 
\index{form!Skolemized@(outer) Skolemized}%
outer \skolemizedform\ of \nlbmath A. \
Let\/ \Y\ be the set of bound (\math\gamma-) variables of \nlbmath F. \ 
Let\/ \math E result from \math F by removing all 
\mbox{(\math\gamma-) quantifiers}. \
Let\/ \math n be a positive natural number. \ 
Let the 
\index{champ fini|)}%
{\frenchfont champ fini} 
\nlbmath{\termsofdepthnovars n} 
be formed over the function and free variable symbols occurring in \nlbmath F. \
\\If\/ \math A has \propertyC\ of order \nlbmath n, 
then we can construct a derivation of \nlbmath A
of the following form, 
in which we read any term starting with a \skolem\ function 
as an atomic variable:
\begin{description}\item[Step\,1: ]\begin{tabular}[t]{@{}l@{}}
  We \nolinebreak start with 
  \underline{a sentential tautology whose disjunctive normal form is a}
\\\underline{re-ordering of a disjunctive normal form of} 
  the \sententialtautology\nlbmath
  {\!\displaystyle\bigvee_{\FUNDEF\sigma\Y{\termsofdepthnovars n}}
   \!\!\!\!\!E\sigma}.
\\\end{tabular}

\item[Step\,2: ]Then we may repeatedly apply the 
\index{Generalized Rule!of delta@of \math\delta-Quantification|)}%
\index{Generalized Rule!of gamma-Quantification@of \math\gamma-Quantification|)}%
\underline{Generalized} Rules of\/ \math\gamma- and 
\math\delta-Quanti\-fi\-cation.

\item[Step\,3: ]\begin{tabular}[t]{@{}l@{}}
  Then, (after renaming all bound \math\delta-variables)
  we may repeatedly apply 
\\the 
\index
{Generalized Rule!of gamma-Simplification@of \math\gamma-Simplification|)}%
  Generalized Rule of\/ \math\gamma-Simplification.
\\[-2ex]\end{tabular}
\\\getittotheright\qed%\halftop
\end{description}\end{lemma}

\yestop\yestop\noindent
Obviously, there is no use of {\em modus ponens}\/ in such a proof,
and thus, it is linear, \ie\ written as a tree, 
 it \nolinebreak has no branching. \ 
Moreover, all function and predicate symbols 
within this proof occur already in \nlbmath A,
and all formulas in the proof are similar to \nlbmath A
in the sense that they have the so-called {\em``sub''-formula property}.

\pagebreak

\yestop\begin{example}[{\em Modus Ponens}\/ Elimination]\label
{example from C to yields}\sloppy\hfill
\index{modus ponens!elimination}%
{\em(continuing \examref{example running herbrand start})}\par\noindent
Let us derive the formula \math A of \examref{example running herbrand start}
in \sectref{section herbrands properties}. \ 
As \math A is not in 
\index{form!prenex}%
prenex form we have to apply the version of 
\lemmref{lemma from C to yields a la heijenoort} without the overlined
part. \ 
As explained in \examref{example running herbrand start},
\math A \nolinebreak has 
\index{Property C|)}%
\propertyC\ of order \nlbmath n
for \nlbmath{n\tightequal 4}, \hskip .2em
and the result of removing 
the quantifiers from the 
\index{form!Skolemized@(outer) Skolemized}%
outer \skolemizedform\ of \nlbmath A
is the formula \nlbmath E:
\par\halftop\noindent\LINEmaths{\noparenthesesoplist{\inpit{
\boundvari a{}\tightprec\boundvari b{}
\und
\boundvari b{}\tightprec\boundvari c{}
\implies
\boundvari a{}\tightprec\boundvari c{}}
\oplistund
\boundvari x{}\tightprec\app
{\forallvari m{}}{\boundvari x{},\boundvari y{}}\und 
\boundvari y{}\tightprec\app
{\forallvari m{}}{\boundvari x{},\boundvari y{}}
\oplistimplies 
\forallvari u{}\tightprec\boundvari n{}\und
\forallvari v{}\tightprec\boundvari n{}\und
\forallvari w{}\tightprec\boundvari n{}}}{}{\Large\math{(E)}}
\par\yestop\noindent
Let \math N denote the cardinality of \termsofdepthnovars n. \ \
Let \bigmaths{\termsofdepthnovars n=\{t_1,\ldots,t_N\}}. 

\yestop\noindent
For the case of \math{n\tightequal 4}, \
we have \bigmaths{N=3+3^2+\inpit{3+3^2}^2=156}, and, \ 
for \math{\Y:=\{
\boundvari a{},\boundvari b{},\boundvari c{},
\boundvari n{},\boundvari x{},\boundvari y{}
\}}, \ 
the \herbrand\ disjunction 
\bigmathnlb
{\bigvee_{\FUNDEF\sigma\Y{\termsofdepthnovars 4}}E\sigma}{}
has \ \math{N^{\CARD Y}} elements, \ie\
more than \nlbmath{10^{13}}. \ 
Thus,
we had better try a reduction proof here,\footnote{%
 % \majorheadroom
 As \herbrand's proof of his version of 
 \lemmref{lemma from C to yields a la heijenoort}
 \citep[\p 170]{herbrand-logical-writings}
 proceeds reductively too,
 we explain \herbrand's general proof in parallel to the development of our 
 special example, 
 in \notefromtoref
 {first note on Herbrand's proof}{last note on Herbrand's proof}. \
 \herbrand's proof is interesting by itself and similar 
 to the later proof 
 of the 
 \index{Hilbert!'s epsilon}%
 Second \nlbmath\varepsilon-Theorem in 
 \citep[\litsectref{3.1}]{grundlagen-first-edition-volume-two}\@.}
applying the inference rules backward,
and be content with arriving at a \sententialtautology\ which is a
sub-disjunction of a re-ordering of a
disjunctive normal form of \bigmathnlb
{\bigvee_{\FUNDEF\sigma\Y{\termsofdepthnovars 4}}E\sigma}. \ 

\yestop\halftop\noindent
\index{Generalized Rule!of gamma-Quantification@of \math\gamma-Quantification}%
As the backward application of the Generalized
Rule of \math\gamma-Quantification
admits only a single (\ie\ linear) application of each 
\math\gamma-quantifier (or each ``lemma''), \hskip .2em
and as we will
have to apply both the first and the second line of \math A twice,
we first increase the 
\math\gamma-multiplicity of the top \math\gamma-quantifiers
of these two lines to two. \ 
This is achieved by 
applying the 
\index{Generalized Rule!of gamma-Simplification@of \math\gamma-Simplification}%
Generalized Rule of \math\gamma-Simplification twice
backward to \nlbmath A, 
resulting in:\footnote{\label{note gamma}\label{first note on Herbrand's proof}%
 \majorheadroom
 To arrive at the full 
\index{Herbrand disjunction|)}%
\herbrand\ disjunction \bigmathnlb
{\bigvee_{\FUNDEF\sigma\Y{\termsofdepthnovars n}}E\sigma},
\herbrand's proof requires us to apply the 
\index{Rule!of Simplification}%
Rule of Simplification
top-down at each occurrence of a \math\gamma-quantifier 
\math N \nolinebreak times,
and the idea is to substitute \nlbmath{t_i}
for the \mth i occurrence of this \math\gamma-quantifier on each branch.}
\par\halftop\noindent\LINEmaths{\noparenthesesoplist{
\forall\boundvari a{},\boundvari b{},\boundvari c{}\stopq\inpit{
\boundvari a{}\tightprec\boundvari b{}
\und
\boundvari b{}\tightprec\boundvari c{}
\implies
\boundvari a{}\tightprec\boundvari c{}}
\oplistund
\forall\boundvari a{},\boundvari b{},\boundvari c{}\stopq\inpit{
\boundvari a{}\tightprec\boundvari b{}
\und
\boundvari b{}\tightprec\boundvari c{}
\implies
\boundvari a{}\tightprec\boundvari c{}}
\oplistund
\forall\boundvari x{},\boundvari y{}\stopq
\exists\boundvari m{}\stopq\inparenthesesoplist{
\boundvari x{}\tightprec\boundvari m{}
\oplistund 
\boundvari y{}\tightprec\boundvari m{}}
\oplistund
\forall\boundvari x{},\boundvari y{}\stopq
\exists\boundvari m{}\stopq\inparenthesesoplist{
\boundvari x{}\tightprec\boundvari m{}
\oplistund 
\boundvari y{}\tightprec\boundvari m{}}
\oplistimplies \forall\boundvari u{},\boundvari v{},\boundvari w{}\stopq
\exists\boundvari n{}\stopq\inpit{
\boundvari u{}\tightprec\boundvari n{}\und
\boundvari v{}\tightprec\boundvari n{}\und
\boundvari w{}\tightprec\boundvari n{}}}}{}

\pagebreak

\yestop\halftop\noindent
Renaming the bound \math\delta-variables to 
some terms from \nlbmath{\termsofdepthnovars n}, 
and applying the 
\index{Generalized Rule!of delta@of \math\delta-Quantification}%
Generalized Rule of \nlbmath\delta-Quantification three times backward
in the last line,
we get:\nopagebreak
\par\noindent\LINEmaths{\noparenthesesoplist{
\forall\boundvari a{},\boundvari b{},\boundvari c{}\stopq\inpit{
\boundvari a{}\tightprec\boundvari b{}
\und
\boundvari b{}\tightprec\boundvari c{}
\implies
\boundvari a{}\tightprec\boundvari c{}}
\oplistund
\forall\boundvari a{},\boundvari b{},\boundvari c{}\stopq\inpit{
\boundvari a{}\tightprec\boundvari b{}
\und
\boundvari b{}\tightprec\boundvari c{}
\implies
\boundvari a{}\tightprec\boundvari c{}}
\oplistund
\forall\boundvari x{},\boundvari y{}\stopq
\exists\boxeins\stopq\inparenthesesoplist{\boundvari x{}\tightprec\boxeins
\oplistund 
\boundvari y{}\tightprec\boxeins}
\oplistund
\forall\boundvari x{},\boundvari y{}\stopq
\exists\boxzwei\stopq\inparenthesesoplist{\boundvari x{}\tightprec\boxzwei
\oplistund 
\boundvari y{}\tightprec\boxzwei}
\oplistimplies 
%\forall\boxu,\boxv,\boxw\stopq
\exists\boundvari n{}\stopq\inpit{
\boxu\tightprec\boundvari n{}\und
\boxv\tightprec\boundvari n{}\und
\boxw\tightprec\boundvari n{}}}}{}

\begin{sloppypar}\noindent
The boxes indicate that the enclosed term actually denotes
an atomic variable whose structure cannot be changed by a substitution. \ 
By this nice trick of taking outermost \skolem\ terms
as names for variables, \herbrand\ avoids the hard task of giving
semantics to \skolem\ functions, 
\cfnlb\ \sectref{section herbrand loewenheim skolem}.\footnote{
 % \majorheadroom
 According to \herbrand's proof we would have to replace 
 any bound \math\delta-variable \nlbmath{\boundvari x{}}
 with its \skolem\ term \nlbmath{\forallvari x{}(t_{i_0},\ldots,t_{i_k})}, 
 provided that
 \math{i_0,\ldots,i_k} denotes the branch on which this 
 \mbox{\math\delta-quantifier}
 occurs \wrt\ the previous step of raising 
 each \math\gamma-multiplicity to \nlbmath N,
 described in \noteref{note gamma}.}
\end{sloppypar}
%\vfill\pagebreak

\begin{sloppypar}\yestop\noindent We apply 
the 
\index{Generalized Rule!of gamma-Quantification@of \math\gamma-Quantification}%
Generalized Rule of \nlbmath\gamma-Quantification four times backward,
resulting in application of \\\linemath{\{
\boundvari x{}\mapsto\boxv
\comma
\boundvari y{}\mapsto\boxw
\}}{} to the third line and 
\\\LINEmaths{\{
\boundvari x{}\mapsto\boxu
\comma
\boundvari y{}\mapsto\boxeins
\}\footroom\headroom}{}\\\headroom
to the fourth \nolinebreak line. \ 
This yields:\footroom\headroom%
\index{Rule!of gamma-Quantification@of \math\gamma-Quantification}%
\footnotemark\end{sloppypar}
\par\noindent\LINEmaths{\noparenthesesoplist{
\forall\boundvari a{},\boundvari b{},\boundvari c{}\stopq\inpit{
\boundvari a{}\tightprec\boundvari b{}
\und
\boundvari b{}\tightprec\boundvari c{}
\implies
\boundvari a{}\tightprec\boundvari c{}}
\oplistund
\forall\boundvari a{},\boundvari b{},\boundvari c{}\stopq\inpit{
\boundvari a{}\tightprec\boundvari b{}
\und
\boundvari b{}\tightprec\boundvari c{}
\implies
\boundvari a{}\tightprec\boundvari c{}}
\oplistund
\exists\boxeins\stopq\inpit{\boxv\tightprec\boxeins\und 
\boxw\tightprec\boxeins}
\oplistund
\exists\boxzwei\stopq\inparenthesesoplist{\boxu\tightprec\boxzwei
\oplistund 
\boxeins\tightprec\boxzwei}
\oplistimplies 
%\forall\boxu,\boxv,\boxw\stopq
\exists\boundvari n{}\stopq\inpit{
\boxu\tightprec\boundvari n{}\und
\boxv\tightprec\boundvari n{}\und
\boxw\tightprec\boundvari n{}}
}}{}\par\yestop\noindent 
Applying (always backward) 
the 
\index{Generalized Rule!of delta@of \math\delta-Quantification}%
Generalized Rule of \nlbmath\delta-Quantification twice and the 
\index{Generalized Rule!of gamma-Quantification@of \math\gamma-Quantification}%
Generalized Rule of \mbox{\nlbmath\gamma-Quantification} seven times,
and then dropping 
the boxes (as they 
are irrelevant for sentential reasoning without substitution) and 
rewriting it all into a disjunctive list of conjunctions,
we arrive at the disjunctive set \nlbmath C of 
\examref{example running herbrand start},
which is a \sententialtautology. \ 
Moreover, as a list, \math C is obviously a re-ordered 
sublist of a disjunctive normal form of  
\bigmathnlb
{\bigvee_{\FUNDEF\sigma\Y{\termsofdepthnovars 4}}E\sigma}.
% \par\noindent\LINEmaths{\noparenthesesoplist{
% \inpit{
% \forallvari v{}\tightprec\boxeins
% \und
% \boxeins\tightprec\boxzwei
% \implies
% \forallvari v{}\tightprec\boxzwei}
% \oplistund
% \inpit{
% \forallvari w{}\tightprec\boxeins
% \und
% \boxeins\tightprec\boxzwei
% \implies
% \forallvari w{}\tightprec\boxzwei}
% \oplistund
% {\forallvari v{}\tightprec\boxeins\und 
% \forallvari w{}\tightprec\boxeins}
% \oplistund
% {\forallvari u{}\tightprec\boxzwei\und 
% \boxeins\tightprec\boxzwei}
% \oplistimplies 
% \forallvari u{}\tightprec\boxzwei\und
% \forallvari v{}\tightprec\boxzwei\und
% \forallvari w{}\tightprec\boxzwei}}{}\par\noindent
% \par\noindent\LINEmaths{\noparenthesesoplist{
% \inpit{
% \forallvari v{}\tightprec\termeins
% \und
% \termeins\tightprec\termzwei
% \implies
% \forallvari v{}\tightprec\termzwei}
% \oplistund
% \inpit{
% \forallvari w{}\tightprec\termeins
% \und
% \termeins\tightprec\termzwei
% \implies
% \forallvari w{}\tightprec\termzwei}
% \oplistund
% {\forallvari v{}\tightprec\termeins\und 
% \forallvari w{}\tightprec\termeins}
% \oplistund
% {\forallvari u{}\tightprec\termzwei\und 
% \termeins\tightprec\termzwei}
% \oplistimplies 
% \forallvari u{}\tightprec\termzwei\und
% \forallvari v{}\tightprec\termzwei\und
% \forallvari w{}\tightprec\termzwei}}{}\par\noindent
\getittotheright\qed\end{example}

\pagebreak

\footnotetext{\label{last note on Herbrand's proof}%
 % \majorheadroom
 Note that the terms 
to be substituted for a bound \math\gamma-variable,
say \nlbmath{\boundvari y{}},
in such a reduction 
step can always be read out from any bound \math\delta-variable 
in its scope: \
If there are \math j \mbox{\math\gamma-quantifiers}
between the quantifier for \nlbmath{\boundvari y{}} inclusively and
the quantifier for the \math\delta-variable, 
the value for \nlbmath{\boundvari y{}} is the
\nonumbermth j argument of the bound \mbox{\math\delta-variable},
counting from the last argument backward.
\\
For instance, in the previous reduction step,
the variable \nlbmath{\boundvari y{}} in the third \nolinebreak
line was replaced with \boxw, the last argument of 
the bound \mbox{\math\delta-variable} \nlbmath\boxeins,
being first in the scope of \nlbmath{\boundvari y{}}.
\\
This property is obvious from \herbrand's proof
but hard to express as a 
property of proof normalization.
\\
Moreover, this property is useful in \herbrand's proof for showing 
that the side condition of the 
Rule of \mbox{\math\gamma-Quantification}
\bigmaths{{B\{x\mapsto t\}}\over{\exists x.\,B}}{}
is always satisfied, even for a certain 
\index{form!prenex}%
prenex form. \ 
Indeed, \mbox{\math\gamma-variables} never occur in the replacement \nlbmath t
and the 
\index{height of a term}%
height of \nlbmath t is strictly smaller than the 
height of all bound \math\delta-variables in the scope \nlbmath B, \hskip.2em
so that no free variable in \nlbmath t can be bound by quantifiers in 
\nlbmath{B}; \
\cfnlb\ \sectref{section herbrands calculi}.}%
%\vfill\pagebreak
% The \hilbert\ school had not realized that \loewenheim\ had a sound
% and complete proof procedure. \ 
% What did the connection between validity and 
% \propertyC\ \hskip .1em 
% (which \citep{loewenheim-1915} had established) \hskip .3em 
% have to to with proof theory? \ 
% Seemingly nothing! \ 
% With \herbrandsfundamentaltheorem, this connection became most obvious.
%%
%\yestop\yestop
\noindent
In the time before \herbrandsfundamentaltheorem, 
a calculus was basically a means to describe a set of theorems 
in a semi-decidable and theoretical fashion. \ 
%% It was hardly a practice at \herbrand's or \hilbert's life times
%% to use a calculus for actually writing down a formal proof 
%% and searching for it. \ 
In \nolinebreak\hilbert's calculi, for instance, the 
\index{proof search|(}%
search for concrete proofs is very hard. 
Contrary to most other 
\index{Hilbert!-style calculi}%
\hilbert-style calculi, 
the normal form of proofs given in
Statement\,2 of 
\herbrandsfundamentaltheoremindex\theoref{theorem herbrand fundamental two}, 
however,
supports the search for reductive proofs: \ 
Methods of 
\index{proof search!human-oriented}%
\index{theorem proving!human-oriented}%
human\footnote{%
 \majorheadroom
 Roughly speaking, we may do a proof by hand, count the lemma applications
 and remember their instantiations, 
 and then try to construct a formal normal form proof accordingly, 
 just as we have done in \examref{example from C to yields}. \
 See \citep{wirthcardinal} for more on this.}
and 
\index{proof search!automatic}%
\index{theorem proving!automated}%
automatic\footnote{%
 \majorheadroom
 Roughly speaking, we may compute the connections and search for a reductive
 proof in the style of say \citep{wallen}, 
 which we then transform into a proof in the normal form
 of Statement\,2 of \theoref{theorem herbrand fundamental two}.}
proof search may help us to find simple proofs in this normal form.

This means that, for the first time in known history,
\herbrand's version of \lemmref{lemma from C to yields a la heijenoort} 
gives us the means to search successfully for
simple proofs in a formal calculus by hand 
(or \nolinebreak actually today, with a computer), \hskip.2em
just as we have done in 
\examref{example from C to yields}.\footnote{%
 \majorheadroom
 Even without the avoidance of the detour over 
\index{form!prenex}%
prenex forms 
due to 
\index{correction (of Herbrand's False Lemma)!Heijenoort's}%
\heijenoort's \repair,
this already holds for the normal form given by 
Statement\,3 of \theoref{theorem herbrand fundamental two},
which is extensionally equal to \herbrand's original 
\index{Property A}%
\propertyA. \
The next further steps to improve this support for proof search would be
sequents and {\em free} \math\gamma- and \math\delta-variables; \ 
\cfnlb\ \eg\ \makeaciteoftwo{wirthcardinal}{wirth-jal}.} 

The normal form of proofs --- as given by
\lemmref{lemma from C to yields a la heijenoort} --- eliminates 
detours via
\index{modus ponens|)}%
\index{modus ponens!elimination|)}%
{\em modus ponens}\/ in a similar fashion
as 
\index{Gentzen!'s {\germanfont Hauptsatz}}%
\gentzensHauptsatz\ eliminates the 
\index{Cut elimination}%
Cut. \ 
It \nolinebreak is remarkable not only because it establishes a 
connection between \skolem\ terms and free variables
without using any semantics for \skolem\ functions
(and thereby, without using the 
\index{axiom!of choice|(}%
\axiomofchoice). \ 
It \nolinebreak also seems to be the first time 
that a normal form of proofs is shown to exist
in which {\em different phases}\/ are considered. \hskip.3em  
Even with 
\index{Gentzen!'s {\germanfont versch\"arfter Hauptsatz}}%
\gentzensverschaerfterHauptsatz\ following \herbrand\ in 
this aspect some years later,
the concrete form of \herbrand's normal form of proofs
remains important to this day, especially in the form of
\index{correction (of Herbrand's False Lemma)!Heijenoort's}%
\heijenoort's \repair,
\cfnlb\ \sectref{section lemma}. \
The manner in which modern sequent, tableau, and matrix calculi
organize proof search\footnote{%
 \majorheadroom
 \Cfnlb\ \eg\ \citep{wallen}, 
\citep{wirthcardinal}, 
\autexierindex\citep{sergecore}.}
does not follow the 
\index{Hilbert!school}%
\hilbert\ school and their 
\index{Hilbert!'s epsilon}%
\math\varepsilon-elimination theorems, but 
\index{Gentzen!'s calculi}%
\gentzen's and \herbrand's calculi. \ 
Moreover, regarding their \skolemization, their 
\index{inference!deep}%
deep inference,\footnote{%
 \majorheadroom
 Note that although the deep inference rules of 
{\em Generalized}\/ Quantification are an 
extension of \herbrand's calculi by 
\heijenoortindex\heijenoort,
the deep inference rules of 
\index{Rule!of Passage}%
Passage and of 
\index{Generalized Rule!of Simplification}%
Generalized  Simplification are
\herbrand's original contributions.}
and their focus on \mbox{\math\gamma-quantifiers} and their multiplicity,
these modern 
\index{proof search|)}%
proof-search calculi
are even more in \herbrand's tradition than in 
\index{Gentzen!'s calculi}%
\gentzen's.
\vfill\pagebreak

\section
[The \loewenheimskolemtheorem\ and
\\\herbrand's Finitistic Notion of Falsehood in an Infinite Domain]
{\sloppy The \loewenheimskolemtheorem\ and 
\herbrand's\protect\linebreak 
Finitistic Notion of Falsehood in an Infinite Domain}%
\label{section herbrand loewenheim skolem}%
\loewenheimindexstart\loewenheimskolemtheoremindexbegin\index{finitism|(}%
\index{Property C|(}%

\noindent
Let \math A be a \firstorder\ formula whose terms 
have a 
\index{height of a term}%
height not greater than \nlbmath m. \ 
\herbrand\ defines that 
{\em\math A \nolinebreak is false in an infinite domain}\/
\udiff\
\math A \nolinebreak does not have \propertyC\ of order \nlbmath p 
for any positive natural number \nlbmath p.

\halftop\indent
If, 
for a given positive natural number \nlbmath p, \hskip.2em the
formula \nlbmath A does not have \propertyC\ of order \nlbmath p,
\hskip.2em then we can % effectively
construct a finite structure over the domain 
\nlbmath{\termsofdepthnovars{p+m}}
which falsifies \nlbmath{A^{\termsofdepthnovars p}}; \hskip .2em
\cfnlb\ \sectref{section herbrand expansion}. \hskip.3em 
Thus, instead of requiring a single infinite structure in which
\nlbmath{A^{\termsofdepthnovars p}} \nolinebreak is false for any
positive natural number \nlbmath p, \hskip.2em 
\herbrand's notion of falsehood in an infinite domain 
only provides us, \hskip.2em 
for each \nlbmath p, \hskip.2em 
with a finite structure in which 
\nlbmath{A^{\termsofdepthnovars p}} is false. \ 
\herbrand\ explicitly points out that these structures 
do not have to be extensions of each other. \hskip.2em
From a given falsifying structure for some \nlbmath p one can, 
of course, generate falsifying structures 
for each \nlbmath{p'\!<p}
by restriction to
\nlbmath{\termsofdepthnovars{p'+m}}. \  
\herbrand\ thinks, however, that to require an infinite sequence of
structures to be a sequence of extensions would necessarily include
some form of the \axiomofchoice, which he rejects out of principle.
Moreover, he writes that the basic prerequisites of the
\loewenheimskolemtheorem\ are generally misunderstood, but does not
make this point clear.

\begin{sloppypar}\halftop\indent
It seems that \herbrand\ reads \loewenheimindex\citep{loewenheim-1915} as if it
would be a paper on provability instead of validity, \ie\
that \herbrand\ confuses \loewenheim's \nolinebreak`\math\models'
with \herbrand's \nolinebreak`\tightyields\closesinglequotefullstopextraspace

\end{sloppypar}

\halftop\indent
All in all, this is, on the one hand, so peculiar and, on the other hand,
so relevant for \herbrand's finitistic views of logic
and proof theory that some quotations may illuminate the controversy.%
\begin{quote}
%\selectlanguage{french}%
  ``\frenchtextonehundredandone''\footnote{%
 % \majorheadroom
 \Cfnlb\ \frenchtextonehundredandonesourcelocationoriginal. \ 
    \frenchtextonehundredandonesourcelocationmodifierforreprint\
    \frenchtextonehundredandonesourcelocationreprint
    %% \selectlanguage{USenglish}%
    \begin{quote}``\englishtextonehundredandone'' \getittotheright
    {\translationnotewithlongcite{\p 165}{herbrand-logical-writings}
     {translation by \dreben\ and \heijenoort}}\notop\end{quote}}
\notop\halftop\end{quote}

\pagebreak

\noindent
After defining the dual notion for unsatisfiability instead of
validity, \herbrand\ continues:\begin{quote}
  %% \selectlanguage{french}%
  ``\frenchtextonehundredandtwo''\footnote 
    {\Cfnlb\
    \frenchtextonehundredandtwosourcelocationoriginal. \
    \frenchtextonehundredandtwosourcelocationmodifierforreprint\ also
    in: \frenchtextonehundredandtwosourcelocationreprint
    % \selectlanguage{USenglish}%
    \begin{quote}``\englishtextonehundredandtwo'' \getittotheright
    {\translationnotewithlongcite{\p 166}{herbrand-logical-writings}
     {translated by \dreben\ and \heijenoort}}\end{quote}}
\end{quote}

\noindent
\herbrandsfundamentaltheorem\ equates {\em provability}\/ 
with \propertyC, \hskip .09em
whereas the \loewenheimskolemtheorem\ 
equates {\em validity}\/ with \propertyC. \ 
Thus, it is not the case that \herbrand\ somehow corrected \loewenheim. \
Instead, 
the \loewenheimskolemtheorem\ and \herbrandsfundamentaltheorem\ 
had better be looked upon as a bridge from validity to provability 
with two arcs and 
\propertyC\ as the eminent pillar in the middle of the river,
offering a magnificent view from the bridge on 
properties of \firstorder\ logic; \hskip .2em
as depicted in \nlbsectref{section herbrand fundamental theorem}. \ \
% such as the \bernaysschoenfinkel\ 
% class (no function symbols, only \math\gamma-quantification). \ 
And this was probably also \herbrand's view when he correctly wrote:\begin{quote}
``\frenchtextonehundred''\footnote{%
 \majorheadroom
 \Cfnlb\ \frenchtextonehundredsourcelocationoriginal. \
 \frenchtextonehundredsourcelocationmodifierforreprint\ also in:
 \frenchtextonehundredsourcelocationreprint.
 \begin{quote}``\englishtextonehundred'' \getittotheright
 {\translationnotewithlongcite
  {\englishtextonehundredsourcelocationpagenumber}
  {\englishtextonehundredsourcelocationwithoutpagenumber}
  {translated by \dreben\ and \heijenoort}}\notop\notop\end{quote}}
\end{quote}
\noindent
Moreover, \herbrand\ criticizes 
\loewenheimindex\loewenheim\ for not showing the 
\index{consistency!of first-order logic}%
consistency of \firstorder\ logic, 
but this, of course, was never 
\loewenheimindex\loewenheim's concern.

The mathematically substantial part of \herbrand's critique of 
\loewenheimindex\loewenheim\
refers to the use of the \axiomofchoice\ in \loewenheim's  proof of the 
\loewenheimskolemtheorem.

\pagebreak

The \loewenheimskolemtheorem\ as found in many textbooks,
such as \cite[\p 141]{enderton}, 
says that any satisfiable set of \firstorder\ formulas
is satisfiable in a countable structure. \ 
In \loewenheimindex\citep{loewenheim-1915}, however, we only find a dual statement,
namely that any invalid \firstorder\ formula has a denumerable counter-model. \ 
Moreover, what is actually proved, read charitably,\footnotemark\
is the following stronger theorem:

\loewenheimindex\footnotetext
{As \loewenheim's paper lacks some minor details,
 there is an ongoing discussion whether its proof 
 of the \loewenheimskolemtheorem\ is complete 
 and what is actually shown. \ 
 Our reading of the proof of the \loewenheimskolemtheorem\ 
 in \citep{loewenheim-1915} \hskip.2em 
 is a standard one. \ 
 Only in 
\skolemindex\citep[\p\,26\ff]{skolem-1941} and 
 \citep[\litsectref{6.3.4}]{badesa-loewenheim}
 we found an incompatible reading, namely that --- to construct
 \nlbmath{\salgebra'} of \lititemref 2
 of \theoref{theorem loewenheim skolem loewenheim} \nolinebreak--- 
 \loewenheim's 
 proof requires an additional falsifying structure of arbitrary cardinality 
 to be given in advance. \
 The similarity of our presentation with
 \herbrandsfundamentaltheorem, however, is in accordance with 
\skolemindex\citep[\p\,30]{skolem-1941}, but not with \citep[\p145]{badesa-loewenheim}. \ 
 The relation of \herbrandsfundamentaltheorem\ to the \loewenheimskolemtheorem\
 is further discussed in 
\anellisindex%
\citep{anellis-loewenheim}. \ 
 \Cfnlb\ also our \noteref{note gap}.}%
\begin{theorem}[\loewenheimskolemtheorem\ \`a la \citet{loewenheim-1915}]%
\label{theorem loewenheim skolem loewenheim}%
\par\noindent
Let us assume the \axiomofchoice. \ 
Let \math A be a \firstorder\ formula.\begin{enumerate}\noitem\item[1.]
If\/ \math A 
\nolinebreak has \propertyC\ of order \nlbmath p 
for some positive natural number \nlbmath p, \ 
then \bigmaths{\models A}.\noitem\item[2.]
If\/ \math A does not have \propertyC\ of order \nlbmath p
for any positive natural number \nlbmath p, \ 
then 
%, without knowing \nlbmath{\salgebra''}, 
we can construct a sequence of partial structures \nlbmath{\salgebra_i}
that converges to a structure\/ \nlbmath{\salgebra'} 
with a 
%(finitely of infinitely) 
denumerable universe such that
\bigmaths{\notmodels_{\salgebra'}\ A}.
\getittotheright\qed\end{enumerate}\end{theorem}
% Note that, in the latter case, by interpreting 
% the term structure \nlbmath{\salgebra'} in \nlbmath{\salgebra''},
% we can construct a sub-structure \nlbmath{\salgebra'''} of
% \nlbmath{\salgebra''} with
% countable (sub-) universe and \bigmaths{\notmodels_{\salgebra'''}\ A},
% but \loewenheim\ does not mention this.
\par\halftop\noindent As 
\index{Property C|)}%
\propertyC\ of order \nlbmath p can be effectively tested for
\math{p=1,2,3,\ldots}, \ 
\loewenheimindex\loewenheim's proof provides us with a
% sound and 
\index{completeness}%
complete proof procedure
which went unnoticed by \skolem\ as well as the 
\index{Hilbert!school}%
\hilbert\ school. \ 
Indeed, there is no mention in the discussion of the 
\index{completeness}%
completeness problem
for \firstorder\ logic
in \citep[\p\,68]{grundzuege}, \
where it is considered 
as an open problem.\footnote{%
 \majorheadroom
 Actually, 
 the completeness problem is slightly ill defined in \citep{grundzuege}. \  
 \Cf\ \eg\ \citep[\Vol\,I, \PP{44}{48}]{goedelcollected}.}

Thus, for validity instead of provability,
\index{completeness}%
\index{Goedel@G\"odel!'s C@'s Completeness Theorem}%
\goedel's Completeness Theorem\footnote{%
 \majorheadroom
 \Cfnlb\ \citep{goedel-completeness}.%
}
is contained already in 
\loewenheimindex\citep{loewenheim-1915}. \ 
\goedel\ has actually acknowledged this
for the version of the proof of the 
\loewenheimskolemtheorem\ in 
\skolemindex\citep{skolem-1923b}.\footnote{%
 \majorheadroom
 Letter of \goedelindex\goedel\ to \heijenoortindex\heijenoort,
 dated \Aug\,14, 1964. \ 
 \Cfnlb\ \citep[\p\,510, \litnoteref i]{heijenoort-source-book}, \ 
 \citep[\Vol\,I, \p\,51; \Vol\,V, \PP{315}{317}]{goedelcollected}.%
}

Note that the convergence of the structures \nlbmath{\salgebra_i} against 
\nlbmath{\salgebra'} in \theoref{theorem loewenheim skolem loewenheim}
is hypothetical in two aspects:
First, as validity is not co-semi-decidable,
in general we can never positively know that
we are actually in Case\,2 of \theoref{theorem loewenheim skolem loewenheim},
\ie\ that a convergence toward \nlbmath{\salgebra'} exists. \ 
%% and that we may not finally end up with
%% Case\,1 of \theoref{theorem loewenheim skolem loewenheim}. \
Second, even if we knew about the convergence toward \nlbmath{\salgebra'},
we would 
have no general procedure to find out which parts of \nlbmath{\salgebra_i}
will be actually found in \nlbmath{\salgebra'} and which will be removed
by backtracking. \ 
This makes it hard to get an intuition for \nlbmath{\salgebra'}
and may be the philosophical reason for \herbrand's rejection
of ``falsehood in \nlbmath{\salgebra'}\,''
as a meaningful notion. \ 
Mathematically, however, we see no
justification in \herbrand's rejection of this notion and will
explain this in the following.

\pagebreak

\herbrand's critical remark concerning the \loewenheimskolemtheorem\
is justified, however, insofar as 
\loewenheimindex\loewenheim\ needs the \axiomofchoice\
at two steps in his proof without mentioning this.
\begin{description}\item[\nth 1 Step: ]
To show the equivalence of a formula to its 
\index{form!Skolemized@(outer) Skolemized}%
outer \skolemizedform,
\loewenheimindex\loewenheim's proof requires the full \axiomofchoice.
\item[\nth 2 Step: ]
For constructing the structure \nlbmath{\salgebra'},
\loewenheim\ would need 
\koenigslemmaindexbegin\koenigslemma,
which is a weak form of the \axiomofchoice.\footnote 
{\koenigslemma\ is Form\,10 in \citep{weakaxiomofchoice}. \ 
This form is even weaker than the well-known \principleofdependentchoice,
namely Form\,43 in \citep{weakaxiomofchoice}; \ 
\cfnlb\ also \citep{axiomofchoice}.}%
\end{description}

\noindent
Contrary to the general 
perception\commanospace\loewenheimindex\footnote{\label{note gap}%
 \majorheadroom
 This perception is partly based on
 the unjustified criticism of 
 \skolemindex\skolem, \herbrand, and 
 \heijenoortindex\heijenoort.

 We are not aware of any negative critique against \citep{loewenheim-1915}
 at the time of publication.
 \citep[\p\,27\ff]{wang-skolem},
 proof-read by 
 \bernaysindex\bernays\ and 
 \goedelindex\goedel,
 after being most critical with the proof in 
 \skolemindex\citep{skolem-1923b},
 sees no gaps in \loewenheim's proof,
 besides the applications of the \axiomofchoice. \ 
 The same holds for \cite[\litsectref 8]{brady},
 sharing expertise in the 
 \index{Peirce--Schr\"oder tradition}%
 \peirce--\schroeder\ 
 tradition\arXivfootnotemarkref{note peirce schroeder tradition}
 with \loewenheim.
 \par
 Let us have a look at the criticism of 
 \skolemindex\skolem, \herbrand, and 
 \heijenoortindex\heijenoort\ in detail:
 %\begin{itemize}\item
 \par
 The following statement of 
 \skolemindex\skolem\ on 
 \citep{loewenheim-1915} 
 is confirmed
 in \citep[\p\,230]{heijenoort-source-book}:
 %\getittotheright{\em(continued on following page)}
 %\pagebreak\par
 %\noindent{\em(\noteref{note gap}, continued from previous page)}\par
 \notop\halftop\begin{quote} 
 %\selectlanguage{ngerman}%
 ``{\germanfontfootnote\germantextskolemeins}''
 \getittotheright{%
 \skolemindex\citep[\p\,220]{skolem-1923b}}
 \notop\end{quote}\begin{quote}
 %\selectlanguage{USenglish}%
 ``\englishtextskolemeins'' \getittotheright
 {\translationnotewithlongcite{\p\,293}{heijenoort-source-book}
  {translation by \bauermengelbergname}}
 \notop\end{quote}
 That detour, however, is not an essential part of the proof,
 but serves for the purpose of illustration only.
 This is clear from the original paper and also the conclusion
 in \citep[\litsectrefs{3.2}{3.3}]{badesa-loewenheim}.
 % As \loewenheim\ lacks an established framework which 
 % distinguishes syntax and semantics,
 % he applies meta formulas for infinite formulas 
 % for illustration of both Steps, just to communicate his intended semantics.
 %\item
 \par
 When \herbrand\ criticizes \loewenheim's proof,
 he actually does not criticize the proof as such, but only 
 \loewenheim's semantical notions; even though \herbrand's verbalization
 suggests the opposite, especially in 
 \citep[\litchapref 2]{herbrand-fundamental-problem}, 
 where \herbrand\ repeats 
 \loewenheim's reducibility results in finitistic style:
 \notop\halftop\begin{quote}
 ``\frenchtextfourhundred'' \getittotheright
 {\citep[\p\,187, \litnoteref 29]{herbrand-ecrits-logiques}}
 \notop\end{quote}\begin{quote}
 ``\englishtextfourhundred'' \getittotheright
 {\translationnotewithlongcite{\p\,237, \litnoteref 33}
  {herbrand-logical-writings}
  {translation by \dreben\ and \heijenoort}}
 \notop\end{quote}
 \heijenoortindex\heijenoort\ realized 
 that there is a missing step in 
 \loewenheim's proof:
 \notop\halftop\begin{quote}
 ``What has to be proved is that, from the assignments thus obtained
   for all \nlbmath i,
  there can be formed one assignment such that \math{\Pi F}
  is true, that is, \bigmaths{\Pi F=0}{} is false. \
  This \loewenheim\ does not do.''\getittotheright
  {\citep[\p\,231]{heijenoort-source-book}}\notop\halftop\end{quote} 
 Except for the \index{choice}principle of choice, 
 however, the missing step is trivial 
 because in \loewenheim's presentation the already fixed part of the assignment
 is irrelevant for the extension. \ 
 Indeed, in the ``Note to the Second Printing\closequotecomma 
 in the preface of the \nth 2 printing,
 \heijenoortindex\heijenoort\ partially corrected himself:\notop\halftop
 \begin{quote} 
 ``I am now inclined to think that \loewenheim\ 
 came closer to 
 \koenigslemma\ than his paper, on the surface, suggests.
 But a rewriting of my introductory note on that point (\p\,231) 
 will have to wait for another occasion.''\getittotheright
 {\citep[\p\,ix]{heijenoort-source-book}}\notop\halftop\end{quote}
 This correction is easily overlooked because no note was inserted into the 
 actual text.
 %\end{itemize}
} 
%commonplace opinions, probably inspired by the hardly justified
% criticism of \skolem, \herbrand, and \heijenoort,
there are no essential gaps in \loewenheim's proof,
with the exception of 
the implicit application of the \axiomofchoice,
which was no exception at his time.
Indeed, fifteen years later,
\goedelindex\goedel\ still applies the \axiomofchoice\ tacitly in the proof 
of his 
\index{completeness}%
\index{Goedel@G\"odel!'s C@'s Completeness Theorem}%
Completeness Theorem.\footnote{% 
 % \majorheadroom
 \Cfnlb\ \citep{goedel-completeness}.}
Moreover, 
as none of these theorems state any consistency properties,
from the point of view of 
\index{finitism}%
\hilbert's finitism
there was no reason to avoid the application of the \axiomofchoice. \ 
Indeed, in the proof of his Completeness Theorem, \
\goedel\ \
``is not interested in avoiding an appeal to the 
\axiomofchoice.''\footnote{% 
 \majorheadroom
 \Cfnlb\ \citep[\p\,24]{wang-skolem}.} \ \
Thus, again,
as we already noted in \lititemref 2 of 
\sectref{section Subject Area and Methodological Background}, regarding 
\index{finitism|)}%
finitism, \herbrand\ is more royalist than King \hilbert.
\\\indent
% While none of the contemporaries ever seems to have complained 
% on \citep{loewenheim-1915},
% only 5 years after the publication 
% it was difficult to accept 
% its proof of the \loewenheimskolemtheorem\ for the following reasons.
% The article is awfully written, violating all rules 
% on mathematical paper writing, \cfnlb\ \citep{writing-mathematics}.
% Moreover, it is 
% written in the \peirce\ --- \schroeder\ style of quantification theory, which 
% had lost acceptance against the presentation of the {\em\PM}.
% % Furthermore, 
% \loewenheim\ lacks a formal set-theoretic \firstorder\ semantics.
% And, finally, \loewenheim\ 
% applies the full \axiomofchoice\ without
% mentioning it.

The proof of the \loewenheimskolemtheorem\ in 
\skolemindex\citep{skolem-1920} already avoids the \axiomofchoice\ in the \nth 1 Step
by using 
\index{form!Skolem normal}%
{\em\skolemnormalform}\/ 
instead of  
\index{form!Skolemized@(outer) Skolemized!vs.\ Skolem normal form}%
\skolemizedform.\footnote{%
 \majorheadroom
 To achieve \index{form!Skolem normal}\skolemnormalform, 
 \skolemindex\skolem\ defines predicates for the subformulas 
 starting with a \mbox{\math\gamma-quantifier},
 and then rewrites the formula
 % the disjunction of the original
 % formula and the negated definitions 
 into a 
 \index{form!prenex}prenex form with a 
 \firstorder\ \math{\gamma^*\delta^*}-prefix.\arXivfootnotemarkref
 {where Skolem normal form and where Skolemized}
 By definition,
 \index{form!Skolemized@(outer) Skolemized!vs.\ Skolem normal form}%
 {\em\skolemizedform s}\/ have a
 \math{\delta^*\gamma^*}-prefix with an implicit 
 higher-order \nlbmath{\delta^*}, and 
 \index{raising}%
 {\em raising}\/ is the dual of \skolemization\ which 
 produces a \math{\gamma^*\delta^*}-prefix with a 
 higher-order \nlbmath{\gamma^*}, 
 \cfnlb\ \citep{miller}. \
 The {\em\skolemnormalform}, however, has a \math{\gamma^*\delta^*}-prefix
 with {\em\firstorder}\/ \nlbmaths{\gamma^*\!}.%
}%
\indent
Moreover, in 
\skolemindex\citep{skolem-1923b},
the choices in the \nth 2 Step of the proof become deterministic,
so that no form of the \axiomofchoice\ 
(such as \koenigslemmaindexend\koenigslemma) \hskip.2em
is needed anymore. \
This is achieved
by taking the universe of the structure \nlbmath{\salgebra'}
to be the natural numbers and by using 
the \wellordering\ of the natural numbers.
% Thus, there is no need to require (any forms of) the \axiomofchoice\
% as prerequisites for construction of the 
% denumerable model in the \loewenheimskolemtheorem,
% contrary to what \herbrand\ thought.
% %

\begin{theorem}[\loewenheimskolemtheorem\ \`a la \citep{skolem-1923b}]%
\label{loewenheim skolem theorem version 1923}%
\loewenheimindexend\loewenheimskolemtheoremindexend
\skolemindex
\par\noindent
Let \math\Gamma\ be a (finite of infinite) denumerable 
set of \firstorder\ formulas. 
Assume \nolinebreak\bigmaths{\notmodels\,\,\Gamma}.
\par\noindent
Without assuming any form of the \axiomofchoice\ 
we can construct a sequence of partial structures \nlbmath{\salgebra_i}
that converges to a structure\/ \nlbmath{\salgebra'} 
with a universe which is a subset of the natural numbers such that
\bigmaths{\notmodels_{\salgebra'}\ \Gamma}.\getittotheright\qed
%\yestop
\end{theorem}

\halftop\noindent
Note that \herbrand\ does not need any form of the \axiomofchoice\
for the following reasons: \ 
In \nolinebreak the \nth 1 Step,
\herbrand\ does not use the semantics of \skolemizedform s \atall,
because \herbrand's \skolem\ terms are just names for free variables,
\cfnlb\ \sectref{section herbrand fundamental theorem}. \
In \nolinebreak the \nth 2 \nolinebreak Step, \herbrand's peculiar notion of
``falsehood in an infinite domain'' makes any choice superfluous. \
This is a device
which --- contrary to what \herbrand\ wrote \nolinebreak--- 
is not really necessary to avoid the
\axiomofchoice, as the above \theoref{loewenheim skolem theorem version 1923}
shows.
\par\indent
In this way, \herbrand\ came close to proving
the 
\index{completeness}%
completeness of 
\russellindex\russell's and 
\hilbertindex\hilbert's calculi
for \firstorder\ logic,
but he did not trust the left arc of the bridge depicted in 
\sectref{section herbrand fundamental theorem}. \ 
And thus \goedel\ proved it first when he
submitted his thesis in 1929, in the same year as \herbrand, and the
theorem is now called 
\index{completeness}%
\index{Goedel@G\"odel!'s C@'s Completeness Theorem}%
{\em \goedel's Completeness Theorem}\/ in all textbooks on logic.\footnote{%
 \majorheadroom
 \Cfnlb\ \citep{goedel-completeness}.}
\vfill\pagebreak\par\indent
It is also interesting to note that \herbrand\ does
not know how to construct a counter-model without using 
% (weak forms of) 
the 
\index{axiom!of choice|)}%
\axiomofchoice, 
as explicitly described in 
\skolemindex\citep{skolem-1923b}. \ \
This \nolinebreak is ---~on the one hand~--- a strong indication
that \herbrand\ was not aware of 
\skolemindex\citep{skolem-1923b}.\footnote{%
 % \majorheadroom
 \Cfnlb\ \p 12 of \goldfarbindex\goldfarb's introduction in
 \citep{herbrand-logical-writings}.}  On the other hand, \herbrand\
names 
\index{Skolem's Paradox}%
\skolemsparadox\ several times and
\skolemindex\makeaciteoftwo{skolem-1923b}{skolem-1929} 
seem to be the only written sources of this at \herbrand's time.\footnote{%
 \majorheadroom
 \index{Skolem's Paradox}%
 \skolemsparadox\ is also briefly mentioned in 
 \neumannindex\citep{neumann-1925}, \hskip.2em 
 not as a paradox, however, 
 but as unfavorable conclusions on set theory drawn by 
 \skolemindex \skolem, who
 wrote about a ``peculiar and apparently paradoxical state of
 affairs\closequotecomma \cfnlb\ \citep[\p\,295]{heijenoort-source-book}.}  
\par\indent
As \herbrand's \index{Property C}\propertyC\ and its use of the outer
\index{form!Skolemized@(outer) Skolemized}\skolemizedform\ 
are most similar to the treatment in \skolemindex\citep
{skolem-1928},\footnote{\label{where Skolem normal form and where Skolemized}%
 \majorheadroom
 Note that the \index{form!Skolemized@(outer) Skolemized}\skolemizedform\ 
 is used in \citep{loewenheim-1915},
 \skolemindex\citep{skolem-1928}, 
 and \citep{herbrand-PhD},
 whereas it is used neither in \skolemindex\citep{skolem-1920},
 nor in \skolemindex\citep{skolem-1923b},
 which use \skolemnormalform\ instead.%
} \hskip .2em 
it \nolinebreak seems likely that \herbrand\
had read \citep{skolem-1928}. \hskip.3em
 Without giving any justification, 
 \heijenoortindex\heijenoort\ assumes,
  however, that 
  \begin{quote}``He \nolinebreak was
  not acquainted either, certainly, with 
 \skolemindex \citep{skolem-1928}.'' \getittotheright{\heijenoortindex
 \citep[\p 112]{heijenoort-work-herbrand}}\end{quote}%
\herbrandsfundamentaltheoremindexend\vfill\cleardoublepage
%%%%%%%%%%%%%%%%%%%%%%%%%%%%%%%%%%%%%%%%%%%%%%%%%%%%%%%%%%%%%%%%%%%%%%%%%%%%%%%%
\section
[\herbrand's First Proof of the Consistency of Arithmetic]
{\herbrand's First Proof of the\\Consistency of Arithmetic}%
\label{section 1 Proof}%
\index{consistency!proof of|(}%
\index{consistency!of arithmetic|(}%

\notop\halftop\noindent
Consider a signature of arithmetic that consists only of zero 
\nolinebreak`\zeropp\closesinglequotecomma 
the successor function 
\nolinebreak`\ssymbol\closesinglequotecomma 
and the equality predicate 
\nolinebreak`\math=\closesinglequotefullstopextraspace
Besides the 
\index{axioms!of equality}%
axioms of equality (equivalence and substitutability), \hskip .3em
\herbrand\ considers several subsets of the following axioms:\footnote{%
 % \majorheadroom
 The labels are ours, not \herbrand's. \ 
 \herbrand\ writes `\math{x\tight+1}' instead of 
 `\spp x\closesinglequotefullstopextraspace
 To save the 
\index{axiom!of substitutability}%
axiom of substitutability,
 \herbrand\ actually uses the biconditional in \nlbmath{(\nat_3)}.}
\par\halftop\noindent\math{\begin{array}{@{}l@{\ \ \ }r@{\ }l@{}}
  \inpit{\ident{S}}
 &%\forall P\stopq
 &%\,\inparentheses{\majorheadroom
\app P \zeropp\nottight\und
\forall y\stopq\inparentheses{\app P y\implies\app P{\spp y}}
{\nottight{\nottight{\nottight\implies}}}
\forall x\stopq\app P  x
%}
%\majorfootroom
\\\majorheadroom
  (\nat_1)
 &%\forall\boundvari x{}\stopq
 &%\inparentheses
  {   \boundvari x{}\tightequal\zeropp
    \nottight{\nottight\oder}
      \exists
      \boundvari y{}\stopq
      \boundvari x{}\tightequal\spp{\boundvari y{}}}
%\majorfootroom
\\\majorheadroom
  (\nat_2)
  &%\forall\boundvari x{}\stopq
  &
  \spp x\tightnotequal\zeropp
%\majorfootroom
\\\majorheadroom
  (\nat_3)
  &%\forall x,y\stopq
  &%\inpit
   {\spp x\tightequal\spp y\nottight\implies x\tightequal y}
%\majorfootroom
\\\majorheadroom
  (\nat_{4+i})
  &%\forall\boundvari x{}\stopq
  &\sppiterated{i+1}x\tightnotequal x
\\\end{array}}
\par\halftop\noindent
Axiom \nlbmath{\inpit{\nat_1}} \nolinebreak
together with the \wellfoundedness\ of the
successor relation `\nlbmath\ssymbol' 
specifies the natural numbers up to isomorphism\fullstopnospace\footnote{%
 \majorheadroom
 \Cfnlb\ 
 \citep[\litsectref{1.1.3}]{wirthcardinal}. \ 
 This idea goes back to \citep{pieri}.}
So do the 
\peanoaxiom s \math{\inpit{\nat_2}} and \nlbmath{\inpit{\nat_3}}  
together with the \peanoaxiom\ of Structural 
Induction \nlbmath{\inpit{\ident S}}, 
provided that the meta variable \nlbmath P is seen as a universally
quantified \secondorder\ variable with the 
standard interpretation\fullstopnospace\footnote{%
 \majorheadroom
 \Cf\ \eg\ \andrewsindex\citep{andrews}.}
\par\indent
Of course, \herbrand, the finitist, 
does not even mention these \secondorder\ properties.  
His discussion is restricted to decidable \firstorder\ axiom sets,
some of which are infinite because of the inclusion of
the infinite sequence \inpit{\nat_4}, 
\inpit{\nat_5}, 
\inpit{\nat_6}, 
%\inpit{\nat_4}, 
\nolinebreak\ldots\ \ \
\par\indent
As \herbrand's axiom sets are first order, \hskip .2em
they cannot specify the natural numbers up to 
isomorphism\fullstopnospace\footnote{%
 \majorheadroom
 For instance, because of the \upwardloewenheimskolemtarskitheorem. \ 
 \Cfnlb\ \eg\ \citep{enderton}.} \
But as the model of arithmetic is infinite,
\herbrand, the finitist, cannot accept it as part of his proof theory. \hskip.1em
Actually, he never even mentions 
% the 
%CP 20140423:
any infinite
model of arithmetic.
\par\indent
\herbrand\ shows that 
(for the poor signature of \zeropp, \ssymbol, and \nlbmath=) \ 
the \firstorder\ theory 
of the axioms
%\smallfootroom
\nlbmath{\inpit{\nat_i}_{i\geq 1}} 
\ (\ie\ \inpit{\nat_i} for any positive natural number \nlbmath i) \ 
% \inpit{\nat_2}, 
% \inpit{\nat_3}, 
% %\inpit{\nat_4}, 
% \nolinebreak\ldots\ \ 
is consistent, 
\index{completeness}%
complete, and decidable. \ 
His constructive proof is elegant, provides a lucid operative
understanding of basic arithmetic, and has been included 
%(freed from the finitistic emphasis)
inter alia into \litsectref{3.1} of \citep{enderton},
one of the most widely used textbooks on logic. \ 
\herbrand's proof has two constructive steps:
\begin{description}\item[\nth 1 Step: ]
He shows how to rewrite any formula into
an equivalent quantifier-free formula without additional free variables. \ 
He proceeds by a special form of quantifier elimination, a technique in the 
\index{Peirce--Schr\"oder tradition}%
\peirce--\schroeder\ 
tradition\arXivfootnotemarkref{note peirce schroeder tradition}
with its first explicit occurrence 
in 
\skolemindex\citep{skolem-1919}.\footnote{%
 \majorheadroom
 More precisely, \cfnlb\ 
 \skolemindex\citep[\litsectref 4]{skolem-1919}. \ 
 For more information on the subject of quantifier elimination in this context,
 \cfnlb\ 
 \anellisindex%
 \citep[\p 120\f, \litnoteref{33}]{anellis-heijenoort-long},
 \citep[\p\,33]{wang-skolem}.}
\pagebreak
\noitem\item[\nth 2 Step: ]
He shows that the quantifier-free fragment is consistent and decidable 
and does not depend on the axiom \nlbmath{\inpit{\nat_1}}. \ \ 
This is achieved with a procedure which rewrites a quantifier-free formula
into an equivalent disjunctive normal form without additional free variables. \
For any quantifier-free formula \nlbmath B, 
this normal-form procedure satisfies:\smallfootroom
\par\noindent\LINEnomath{\bigmaths
{\inpit{\nat_i}_{i\geq 2}\yields B}{}
\ \ \uiff\ \ \ the normal form of \bigmaths{\neg B}{} is 
\bigmaths{\zeropp\tightnotequal\zeropp}.}\end{description}

\halftop\noindent
%Besides textbooks,
This elegant 
work of \herbrand\ is hardly discussed in the secondary literature,
probably because --- as a decidability result --- it became
obsolete before it was published,
because of the analogous result for this theory extended with addition, 
% please do not put this into parentheses!
the so-called 
\index{Presburger Arithmetic}%
{\em\presburgerarithmetic}, as it is known today. \ 
\presburgername\ \presburgerlifetime\footnote{%
 % \majorheadroom
 \presburger's true name is \presburgertruename. \ 
 He was a student of 
 \tarskiname, \lukasiewiczname\ \lukasiewiczlifetime, 
 \ajdukiewiczname\ \ajdukiewiczlifetime,
 and
 \kuratowskiname\ \kuratowskilifetime\ in \Warszawa. \ 
 He was awarded a master (not a \PhD) in mathematics on \Oct\,7, 1930. \ 
 As he was of Jewish parentage, 
 it \nolinebreak is likely that he died
 in the Holocaust (Shoah), maybe in 1943. \ 
 \Cfnlb\ \citep{presburger-life}.} 
% Joerg wrote:
%% He was a student of \tarskiname, and he died in a concentration
%% camp, hence the exact date of his death is unknown.
% According to my documents,
% Prezburger was not a student of Tarski in the sense of any
% supervision etc.
% Does anybody have a proof that he die in a concentration camp? Which one? 
% Both is very likely, of course,
% but also absolutely obvious.
% This, however, is an article on the handbook of HISTORY of logic.
% We should not produce myths here! CP
gave his talk on the decidability of his theory with similar techniques
on \Sep\,24, 1929,\footnote{%
 \majorheadroom
 Moreover, note that \citep{presburger}
 did not appear in print before\,1930. \ 
 Some citations date \citep{presburger} at 1927, 1928, and 1929. \ 
 There is evidence, however, that these earlier datings are wrong,
 \cfnlb\ \citep{presburger-remarks-translation}, 
 \citep{presburger-life}.%
}
five months {\em after}\/
\herbrand\ finished his \PhDthesis. \ 
As \tarskiindex\tarski's 
work on decision methods developed in his 1927/8 lectures in \Warszawa\
also did not appear in print until after World War\,II,\footnote{%
 \majorheadroom
 \Cfnlb\ \citep{presburger-remarks-translation}, \citep{tarski-decision}.}
we \nolinebreak
have to consider this contribution of \herbrand\ as completely original. \ 
Indeed:\begin{quote}\sloppy
  ``{\germanfont\germantextfifty}''\footnote{% 
 \majorheadroom
 \Cfnlb\ \germantextfiftyquotation.\begin{quote}``\englishtextfifty''\notop
 \notop\notop\end{quote}}\end{quote}
\begin{sloppypar}
\noindent In addition, \herbrand\ gives a constructive proof that
the \firstorder\ theories given by the following two axiom sets are identical:
\begin{itemize}\notop\item
  \math{\inpit{\nat_i}_{i\geq 1}}
\noitem\item \inpit{\nat_2}, \nlbmath{\inpit{\nat_3}},
  and the \firstorder\ instances of \nlbmath{\inpit{\ident S}},
  provided that \nlbmath{\inpit{\ident S}} is 
  taken as a \firstorder\ axiom scheme instead of a \secondorder\ axiom.
\noitem\end{itemize}\end{sloppypar}
\vfill\pagebreak
%%%%%%%%%%%%%%%%%%%%%%%%%%%%%%%%%%%%%%%%%%%%%%%%%%%%%%%%%%%%%%%%%%%%%%%%%%%%%%%%
\section
[\herbrand's Second Proof of the Consistency of Arithmetic]
{\herbrand's Second Proof of the\\Consistency of Arithmetic}\label
{section 2 Proof}

\notop\halftop\noindent
\herbrand's contributions to logic discussed so far
are all published in \herbrand's thesis. \ 
In this section,
we consider his journal publication \citep{herbrand-consistency-of-arithmetic}
as well as some material from \litchapref 4 of his thesis.

First, the signature is now \signatureenlarged\ to include the
\index{function!recursive|(}%
recursive functions, \cfnlb\ \sectref{section
  recursive functions} below. \ Second, the axiom scheme
\nlbmath{\inpit{\ident S}} is restricted to just those instances which
result from replacing the meta variable \nlbmath P with {\em
  quantifier-free}\/ \firstorder\ formulas. \ 
For this setting, \herbrand\ again gives a constructive proof of consistency. \
This proof consists of the following two steps:
\begin{description}

\item[\nth 1 Step: ]\sloppy
\herbrand\ defines recursive functions \nlbmath{\fsymbol_P}
such that \fppeinsindex P x is the least natural number \nlbmath {y \leq x}
such that \math{\neg\app P y} holds, provided that such a \nlbmath y exists,
and \zeropp\ otherwise. \ 
%\bigmaths{\fppeinsindex P x\tightequal\mu y.\neg\app P y}{} 
The functions \nlbmath{\fsymbol_P} are primitive recursive
unless the terms substituted for \math P contain a 
non-primitive recursive function. 
These functions imply the instances of \nlbmath{\inpit{\ident S}},
rendering them redundant.
\ %
This is similar to the effect of 
\index{Hilbert!'s epsilon}%
\hilbert's \nth 2 \math\varepsilon-formula:\footnote{% 
 % \majorheadroom
 \Cfnlb\ \eg\ \citep
 [\litsectref{2.3}, \p\,82\ff; Supplement\,V\,B, \p\,535\ff]
 {grundlagen-second-edition-volume-two}.}%
%which may be stated as
\footroom\\\noindent\footroom\LINEmaths{
\varepsilon x.\neg\app P x\nottight{=}\spp y
\nottight{\nottight{\nottight\implies}}
\app P y},\\\noindent
\herbrand's procedure, however, is 
much simpler but only applicable to quan\-ti\-fier-free \nlbmath P.

\item[\nth 2 Step: ] Consider the universal closures of the 
\index{axioms!of equality}%
axioms of equality,
the axioms \inpit{\nat_2} and \inpit{\nat_3},
and an arbitrary finite subset of the axioms for recursive functions. 
Take the negation of the conjunction of all these formulas.
As all quantified variables of the resulting
formula are \mbox{\math\gamma-variables},
this is already in \skolemizedform. \ 
Moreover, for any positive natural number \nlbmath n,
it \nolinebreak is easy to show that this formula 
does not have 
\index{Property C}%
\propertyC\ of order \nlbmath n: \ 
Indeed, we just have to construct a proper finite substructure of arithmetic
which satisfies all the considered axioms for the elements of 
\nlbmath{\termsofdepthnovars n}\@. \hskip.2em
Thus, by
\index{Herbrand's Fundamental Theorem!{\em application}|(}%
\herbrandsfundamentaltheorem, consistency is immediate.

\end{description}
%\pagebreak

\noindent
The \nth 2 step is a prototypical example to demonstrate how 
\herbrandsfundamentaltheorem\ helps to answer seemingly 
non-finitistic semantical questions on infinite structures
with the help of infinitely many
finite sub-structures. \ 
Note that such a semantical argumentation is finitistically acceptable
if and only if the structures are all finite and
effectively constructible. \ 
And the latter is always the case for \herbrand's work on logic.%
\index{Herbrand's Fundamental Theorem!{\em application}|)}%

As the theory of all recursive functions is sufficiently expressive,
 there is the question
 why \herbrand's second consistency proof 
 does not imply the inconsistency of arithmetic 
 by 
\index{Goedel@G\"odel!'s Second Incompleteness Theorem}%
\goedelssecondincompletenesstheorem? \ 
 \herbrand\ explains that we cannot have the theory of {\em all}\/
 total recursive functions because they are not recursively enumerable. \ 
More precisely, 
an evaluation function for an enumerable set of recursive functions
cannot be contained in this set by the standard diagonalization argument.

\pagebreak

%At \herbrand's time,
\index{Hilbert!school}%
\hilbert's school had failed to prove 
the consistency of arithmetic, \hskip .3em
except for the special case that for the axiom \nlbmath{\inpit{\ident S}},
the variable \math x does not occur within the scope of any binder in 
%the instance for of 
\nlbmath{\app P x}.%
\index{Goedel@G\"odel!'s Second Incompleteness Theorem}%
\index{program@\programme!Hilbert's}%
\footnote {%
 % \majorheadroom
 \Cfnlb\ \eg\ \citep[\litsectref{2.4}]{grundlagen-first-edition-volume-two}. \ 

More precisely, \hilbert's school had failed to 
prove the termination of their
first algorithm for computing a valuation
of the 
\index{Hilbert!'s epsilon}%
\mbox{\math\varepsilon-terms} in the
\nth 1 and \nth 2 \math\varepsilon-formulas. \ 
This de facto failure was less spectacular but 
internally more discouraging for 
\hilbertsprogram\ than 
\goedelssecondincompletenesstheorem\ 
with its restricted area of application, 
\cfnlb\ \citep{goedel}. \ 

Only after developing a deeper understanding of the notion of a
{\em basic typus}\/ of an 
\index{Hilbert!'s epsilon}%
\math\varepsilon-term
({\germanfontfootnote Grund\-typu\es}; introduced in 
\neumannindex\citep{neumann-1927};
 called {\em\math\varepsilon-matrix}\/
 in \citep{scanlon73:_consis_number_theor_via_theor}) \hskip.2em
and especially of 
the independence of its valuation of the valuations of
its sub-ordinate \math\varepsilon-expressions,
has the problem been resolved:
The termination problem was cured in 
\citep{ackermann-consistency-of-arithmetic}
with the help of a second algorithm of 
\index{Hilbert!'s epsilon}%
\math\varepsilon-valuation,
terminating within the ordinal number \nlbmath{\epsilon_0},
just as 
\index{Gentzen!'s consistency proof}%
\gentzen's consistency proofs 
for full arithmetic in 
\makeaciteofthree{gentzenfirstconsistent}{gentzenconsistent}{gentzenepsilon}.}
But this fragment of arithmetic is actually equivalent to the 
one considered by \herbrand\ here.\footnote{%
 \majorheadroom
 \Cfnlb\ \citep[\p\,474]{kleene}\@.}
In \nolinebreak this sense, 
\herbrand's result on the consistency of arithmetic was
just as strong as the one of the 
\index{Hilbert!school}%
\hilbert\ school
by 
\index{Hilbert!'s epsilon}%
\math\varepsilon-substitution. \ 
\herbrand's means, however, are much simpler.%
\index{consistency!proof of|)}%
\index{consistency!of arithmetic|)}%
\footnotemark
%\pagebreak
\footnotetext{%
 \majorheadroom
 \herbrand's results on the consistency of arithmetic have little importance, 
however, for today's 
\index{theorem proving!inductive}%
\index{theorem proving!automated}%
\index{theorem proving!human-oriented}%
inductive theorem proving
because the restrictions on \nlbmath{\inpit{\ident S}}
can usually not be met in practice. \ 
\herbrand's restrictions  on \nlbmath{\inpit{\ident S}} require us to
avoid the occurrence of \nlbmath x in the scope of 
quantifiers in \nlbmath{\app P x}. \ 
In practice of inductive theorem proving, 
this is hardly a problem for the \mbox{\math\gamma-quantifiers},
whose bound variables tend to be easily replaceable with witnessing terms. \
There is a problem, however, with the \mbox{\math\delta-quantifiers}. \ 
If \nolinebreak we remove the \mbox{\math\delta-quantifiers},
letting their bound \math\delta-variables become \fuv s,
the induction hypothesis typically becomes too weak for 
proving the induction step. \ 
This is because the now free \mbox{\math\delta-variables} do not admit 
different instantiation in the induction hypotheses and the 
induction conclusion.}%

\vfill\pagebreak
%%%%%%%%%%%%%%%%%%%%%%%%%%%%%%%%%%%%%%%%%%%%%%%%%%%%%%%%%%%%%%%%%%%%%%%%%%%%%%%%
\section{Foreshadowing Recursive Functions}\label
{section recursive functions}

\noindent\herbrand's notion of 
a recursive function is quite abstract: \
A recursive function 
is given by any new \math{n_i}-ary
function symbol \nlbmath{\fsymbol_i} plus a set of 
quantifier-free formulas for its specification (which \herbrand\ calls
{\em the hypotheses}\/), such that,
for any natural numbers \nlbmath{k_1,\ldots,k_{n_i}}, 
there is a constructive proof of the unique existence of
a natural number \nlbmath l such that\footroom
\\\LINEmaths{\yields\ \fppeinsindex i
{\sppiterated{k_1}\zeropp,\ldots,\sppiterated{k_{n_i}}\zeropp}
=\sppiterated{l}\zeropp}.\label{section where text is}
\yestop\begin{quote}
  \frenchtextonehundredandthirteenwithquotes\footnote{%
 % \majorheadroom
 \frenchtextonehundredandthirteensourcelocationmodifierforreprint\begin{quote}%
   ``\englishtextonehundredandthirteen''\end{quote}%
% (Note that an {\em apparent}\/ variable is a bound one.)
}%
\end{quote}

\noindent
In the letter to 
\goedelindexbegin\goedel\ dated \herbrandtogoedeldate,
mentioned already in \nlbsectref{sec:life}, \hskip.2em
\herbrand\ added the requirement
that the hypotheses defining \nlbmath
{\fsymbol_i} contain only function symbols \nlbmath
{\fsymbol_j} with \bigmaths{j\tight\leq i},
for natural numbers \math i and \nlbmath j.\footnote{%
 \majorheadroom
 \Cfnlb\ \citep[\Vol\,V, \PP{14}{21}]{goedelcollected}, \ 
 \citep{sieg05:_only}\@.}

\begin{sloppypar}
\yestop\noindent\goedel's version of \herbrand's notion of a recursive function
is a little different: \ 
He \nolinebreak speaks 
of quantifier-free equations instead of quantifier-free formulas
and explicitly lists the already known functions,
but omits the computability of the functions:
\end{sloppypar}

\begin{quote}``\englishtexttwohundred''\footnote{%
 \majorheadroom
 \Cfnlb\ \citep[\p\,26]{goedel-lecture-notes}, \ 
also in: \citep[\Vol\,I, \p\,368]{goedelcollected}.}
\end{quote}

\pagebreak

\noindent\goedel\ took this 
paragraph on page\,\,26 of his 1934 \Princetonnostate\ lectures
from the above-mentioned letter from \herbrand\ to \goedel,
which \goedel\ considered to be lost, but which was rediscovered
in February\,1986\@.\footnote{%
 % \majorheadroom
 \Cf\ \dawsonindex\citep{goedel-herbrand}.}
In\,1963, 
\goedelindex\goedel\ wrote to 
\heijenoortindex\heijenoort:\begin{quote}
``\englishtextgoedelonherbrand''\footnote{%
 \majorheadroom
 \label{note intuitionism}%
 \index{intuitionism}%
 Letter of \goedel\ dated \Apr\,23,\,1963. \ 
 \Cfnlb\ \citep[\p\,283]{herbrand-logical-writings}, \ 
also in: 
\heijenoortindex\citep[\p 115\f]{heijenoort-work-herbrand}, \ 
also in: \citep[\Vol\,V, \p\,308]{goedelcollected}\@.
\par
It is not clear whether \goedel\ refers to 
\brouwerindex\brouwer's intuitionism
or 
\index{finitism}%
\hilbert's finitism 
when he calls \herbrand\ an ``intuitionist'' here; \hskip .2em
\cfnlb\ \sectref{section Subject Area and Methodological Background}.
\par
And there is more confusion regarding the meaning of
two occurrences of the word ``intuitionistically''
on the page of the above quotation
from \citep{herbrand-consistency-of-arithmetic}. \ 
Both occurrences carry the same note-mark,
probably because \herbrand\ realized that this time 
he uses the word with even a another meaning, different from 
his own standard and different from its meaning for the other occurrence
in the same quotation: \ 
It neither refers to \brouwerindex\brouwer's intuitionism nor to 
\index{finitism}%
\hilbert's finitism, but actually
to the working mathematician's meta level as opposed to the 
object level of his studies; 
\neumannindex\cfnlb\ \eg\ \citep[\p\,2\f]{neumann-1927}.
% Note that proof theory as a subject of mathematical study is still
%  on the object level of the working mathematician and that 
%  the discussion of \herbrand's use of ``intuitionism''
%  in \citep{heijenoort-work-herbrand}, \PP{114}{118}, does not 
%  include our meta level interpretation here.
}
\end{quote}
As we have seen, however,
\goedel's memory was wrong insofar as he had added the restriction to 
equations and omitted the computability requirement.

\yestop\noindent
Obviously, \herbrand\ had a clear idea of 
our current notion of a total recursive function. \ 
\herbrand's characterization, however, 
just transfers the recursiveness 
of the meta level to the object level. \ 
Such a transfer is of little epistemological value. \ 
While there seems to be no way to do much more than such a transfer
for consistency
of arithmetic in \goedel izable systems
(because of 
\index{Goedel@G\"odel!'s Second Incompleteness Theorem}%
\goedelssecondincompletenesstheorem), \hskip.2em
it is well possible to do more than that for 
the notion of recursive functions. \
Indeed, 
in the later developments of 
the theory of term rewriting systems and the today 
standard recursion theory 
for total and partial recursive functions,
we find constructive definitions and 
\index{consistency!proof of}%
\index{consistency!of arithmetic}%
consistency proofs practically useful in programming and 
\index{theorem proving!inductive}%
\index{theorem proving!automated}%
\index{theorem proving!human-oriented}%
inductive theorem proving.\footnote{%
 \majorheadroom
 Partial recursive functions were introduced in \citep{kleene-1938}. \ 
 For consistency proofs and admissibility conditions for 
 the practical specification of partial recursive functions
 with \pnc\ term rewriting systems \cfnlb\ \citep{wirth-jsc}.}
\ %
Thus, as suggested by 
\goedelindexend\goedel,\footnote{%
 \majorheadroom
 Letter of \goedelindex\goedel\ to \heijenoortindex\heijenoort, 
 dated \Aug\,14, 1964. \ 
 \Cfnlb\ \heijenoortindex\citep[\p 115\f]{heijenoort-work-herbrand}.%
}
we \nolinebreak may say that \herbrand\ {\em foreshadowed}\/ 
the notion of a 
\index{function!recursive|)}%
recursive function,
although he did not {\em introduce}\/ \nolinebreak it.\footnote{%
 \majorheadroom
 \Cfnlb\ 
 \heijenoortindex\citep[\p 115\ff]{heijenoort-work-herbrand} for more on this. \ 
 Moreover, note that a general 
 definition of (total) recursive functions was not required 
 for \herbrand's second
 consistency proof because \herbrand's function \nlbmath{\fsymbol_P}
 of \sectref{section 2 Proof}
 is actually a {\em primitive} recursive one, 
 unless \math{P} \nolinebreak contains a non-primitive recursive function.}
\vfill\pagebreak
%%%%%%%%%%%%%%%%%%%%%%%%%%%%%%%%%%%%%%%%%%%%%%%%%%%%%%%%%%%%%%%%%%%%%%%%%%%%%%%%

\section{\herbrand's Influence on Automated Deduction and\\\herbrand's
Unification Algorithm}
\label{sec:atp}

\yestop\yestop\noindent
\index{proof search|(}%
\index{proof search!automatic|(}%
\index{theorem proving!automated|(}%
In the last fifty years the field of automated deduction, 
or automated reasoning as it is more generally called today, 
has come a long way:
modern deduction systems are among the most sophisticated and complex
human artifacts we have, they can routinely 
search spaces of several
million formulas to find a proof.  
Automated theorem proving systems
have solved open mathematical problems and these deduction engines are
used nowadays in many subareas of computer science and artificial
intelligence, including software development and verification as well
as security analysis. The application in industrial software and
hardware development is now 
% (2007) 
% This is a the game of too high significance. 
% Physicians, such as Joerg and CP, cannot accept that.
standard practice in most high
quality products. 
The handbook 
\robinsonindex\citep{HandbookAR} gives a
good impression of the state of the art \nolinebreak today.

%% Although the \AI\ and Automated Deduction communities have failed to 
%% reach their original goal to include the problem-solving ability
%% of a freshman in mathematics into their systems,
%% there is ample evidence
%% for the superiority of the machine in very restricted areas,
%% exemplified by the \mbox{4-Color} and the \robbins\ Algebra problems.
%% Moreover, a new paradigm of interaction between man and machine
%% has recently been established, in which an 
%% anytime search process knows about the 
%% \index{proof search!human-oriented}%
%% \index{theorem proving!human-oriented}%
%% humans'
%% semantic strength and asks the human users for advice in their
%% area of expertise before getting lost in complexity.
%% With a human-oriented main-stream integration following this paradigm,
%% we hope to make man and machine a winning team.\footnote
%% {For example with our \OMEGA\ and \QUODLIBET\ systems,
%%  \cfnlb\ \eg\ 
%% \siekmannindex%
%% \cite{omegacadetwo},
%%  \cite{quodlibet-cade},
%%  \cite{sergecore},
%%  \cite{jancl}.}

\herbrand's work inspired the development of the first computer
programs for automated deduction and mechanical theorem proving,
for the following reason:
The actual test for \herbrand's 
\index{Property C}%
\propertyC\ is
very mechanical in nature and thus can be carried out on a computer,
resulting in a mechanical semi-decision procedure for any mathematical
theorem! \ 
This insight, first articulated in the 1950s,
turned out to be most influential in automated reasoning,
artificial intelligence, and computer science.

\yestop\yestop\noindent
Let us recapitulate the general idea as it is common now in most
monographs and introductory textbooks on 
\index{theorem proving!automated}%
automated theorem proving.

Suppose we are given a conjecture \nlbmath A\@. \
%  Let us assume it to be valid. \ 
Let \math F be its \skolemizedform; \hskip.2em
\cfnlb\ \sectref{section skolemization}. \ 
% which has only \math\gamma-quantifiers, \ie, roughly speaking,
% universal quantifiers. \
We then eliminate all quantifiers in \nlbmaths{F\!}, \hskip.2em
and the result is a quantifier-free
formula \nlbmath E\@. \hskip.3em
We \nolinebreak now have to show that the 
\index{Herbrand disjunction|(}%
\herbrand\ disjunction 
over some possible values of the free variables of \nlbmath E
is valid\@. \
See also \examref{example running herbrand start}
in \sectref{section herbrands properties}.

%% Now, if we can find a
%% finite set of objects from the universe of discourse such that the
%% formula, \ie\ the negated theorem, is false for these objects we are
%% done: the stipulated theorem is true!

How do we find these values? \hskip.1em 
Well, we do not actually have these
``objects in the domain\closequotecomma
but we can use their {\em names}, \ie\ we
take all terms from the 
\index{Herbrand universe|(}%
{\em\herbranduniverse}\/ and substitute
them in a systematic way into the variables and wait what happens:
every substituted formula obtained that way is sentential, so we can
just check whether their disjunction 
is valid or not with one of the many available
decision procedures for sentential logic. \ 
If \nolinebreak it \nolinebreak is valid, 
we are done: the original formula \nlbmath A
must be valid by \herbrandsfundamentaltheorem. \ 
If the disjunction of instantiated sentential formulas turns out to
be invalid, well, then bad luck and we continue the process of
substituting terms from the {\em\herbranduniverse}. \

This process must terminate, if indeed the original theorem is valid. \ 
%%  --- as its negation must be false at least \emph{for some} objects
%% from the domain.
%
% One more trick will lead us from here to the modern world of computers
% and automated deduction: Let us assume the given theorem $T$ is true,
% then we negate it and $\neg T$ must be false now. Using \herbrand's
% construction we can find out if this is indeed the case: after a
% finite number of instantiations we have found a ``counterexample\closequotecomma
% \ie\ $\neg T$ is false and $T$ is valid.
%
But what happens if the given conjecture is in fact not a theorem? \ 
In that case 
%the negated conjecture may be true and 
the process will either run forever or sometimes, if we are
lucky, we can nevertheless show this to be the case.

In the following we will present these general ideas a little more
technically.\footnote{%
 % \majorheadroom
 Standard textbooks covering the early period of automated deduction
 are \citep{chang-lee} and \citep{loveland}. \ 
 \citet{chang-lee} present this and other algorithms in more detail and rigor.} 

\vfill\pagebreak

\yestop\yestop\yestop\yestop\noindent
Arithmetic provided a testbed for the first automated theorem
proving program: In 1954 the program of 
\davisname\ \davislifetime\ proved the exciting
theorem that the sum of two even numbers is again even. \ 
This date is
still considered a hallmark and was used as the date for the \nth{50}
anniversary of automated reasoning in 2004. 
The system that proved this and other theorems was based on
\index{Presburger Arithmetic}%
\presburgerarithmetic, a decidable fragment of \firstorder\ logic,
mentioned already in \sectref{section 1 Proof}.

\yestop\yestop\yestop\yestop\noindent
Another approach, based directly on
\herbrand's ideas, was tried by 
\gilmorename\ \gilmorelifetime.\footnote{%
 % \majorheadroom
 \Cfnlb\ \citep{gilmore-1960}. \ 
 The idea is actually due to \loewenheim\ and \skolem\ besides \herbrand,
 as discussed in detail in \sectref{section herbrand loewenheim skolem}
%{sec:logic}
.}
His program worked as follows: 
%% Given a set of
%% axioms $A$ and the negated theorem $\neg T$, the set of formulas 
%% $A\cup\{\neg T\}$ was Skolemized, 
%% \ie\ in this case all existentially quantified variables were
%% replaced by Skolem functions.
A preprocessor 
% then 
generated
the 
\index{Herbrand disjunction|)}%
{\em\herbrand\ disjunction}\/ in the following sense. \
The formula \nlbmath E
% $A\cup\{\neg T\}$ 
contains finitely many constant and function
symbols, which are used to systematically generate the 
{\em\herbranduniverse}\/ for this set; say 
\par\noindent\LINEmaths{a, b, f(a, a), f(a, b), f(b, a), f(b, b), 
g(a, a), g(a, b), g(b, a), g(b, b), 
f(a, f(a, a)),
\ldots}{}\par\noindent
for the constants $a, b$ and the binary function symbols \nlbmath{f,g}. \ 
The terms of this universe were enumerated and systematically
substituted for the variables in 
\nlbmath E
% $A\cup\{\neg T\}$
such that the program generates a sequence of propositional
formulas \math{E\sigma_1, E\sigma_2, \ldots}
%$(A\cup\{\neg T\})\sigma_1, (A\cup\{\neg T\})\sigma_2, \ldots$ 
where $\sigma_1, \sigma_2, \ldots$ are the substitutions. \
Now each of
these sets can be checked for truth-functional validity, for which 
\gilmoreindex\gilmore\ 
used the ``multiplication method\closequotefullstopextraspace
This method computes the conjunctive normal form\footnote{%
 \majorheadroom
 Actually, the historic development of 
 automated theorem proving
 did not follow \herbrand\ but 
 \skolemindex\skolem\ in
 choosing the duality validity--unsatisfiability.
 Thus, instead of proving validity of a conjecture,
 the task was to show unsatisfiability of the negated conjecture.
 For this reason, \gilmoreindex\gilmore\ actually used the {\em disjunctive}\/
 normal form here.}
%% {Using \herbrand's Rules of Passage, any formula
%%   can be transformed into a disjunctive normal form (see also
%%   Example~\ref{example running herbrand start} in~\sectref{section
%%     herbrands properties}).} 
and checks the individual elements of this conjunction in turn: \
If any element contains the disjunction of an atom and its negation, it
must be true and hence can be removed from the overall conjunction. \
As soon as all disjunctions have been
removed, the theorem is proved --- else it goes on forever.

This method is not particularly efficient and served to prove a few
very simple theorems only. \ 
Such algorithms became known as {\em British Museum
  Algorithms}.
That name was originally justified as follows:
\begin{quote}``Thus we reflect the basic nature of theorem proving;
that is, its nature prior to building up sophisticated proof techniques.
We will call this algorithm the 
\label{British Museum Algorithm}%
British Museum Algorithm,
in recognition of the supposed originators of this type.''\getittotheright
{\citep{newell-shaw-simon}}
\end{quote}
The name has found several more popular explanations since.
The nicest is the following: If monkeys are placed in front of typewriters
and they type in a guaranteed random fashion,
they will reproduce all the books of the 
library of the British Museum,
provided they could type long enough.

\vfill\pagebreak

\yestop\yestop\yestop\yestop\noindent
A few months later \davisname\ and \putnamname\ \putnamlifetime\ 
experimented with a better idea, 
where the multiplication method is replaced by what is now known as the 
\index{Davis--Putnam procedure}%
{\em \davis--\putnam\ procedure}.\footnote{%
 % \majorheadroom
 \Cfnlb\ \citep{davis-putnam-procedure}.}
It works as follows: the initial
formula \nlbmath E is
%formulas $A\cup\{\neg T\}$ are 
transformed (once and for all) into
disjunctive normal form and then the variables are systematically
replaced by terms from the \herbranduniverse\ as before. \hskip.2em
But now the truth-functional check is replaced 
with a very effective procedure,
%% CP: The Davis-Putnam procedure is not our subject here.
%% CP: Moreover the description is bugged!
%% by the following procedure:
%% %%
%% \begin{enumerate}
%% \item \textbf{Tautology Rule:} delete all disjuncts that are tautologies
%% \item \textbf{One Literal Rule:} if there is a disjunct with only one
%%   atom $L$, delete all disjuncts that contain $L$.  If the conjunct is
%%   empty (\ie\ all disjuncts have been removed), we are done:
%%   $A\cup\{\neg T\}$ is unsatisfiable. Otherwise remove all occurrences
%%   of $\neg L$ and continue.
%% \item \textbf{Pure Literal Rule:} If there is an atom $L$ in some disjunct, but no negation $\neg L$ anywhere else, 
%%   delete all disjuncts containing $L$. 
%% \item \textbf{Splitting Rule:} If the conjunctive normal form of
%%   $A\cup\{\neg T\}$ can be transformed into one part containing all
%%   disjuncts with an occurrence of $\neg L$, then split the formula
%%   into two parts and continue with each part.
%% \end{enumerate}
%% %%
which constituted a huge improvement and is still used today
in many applications involving propositional logic.\footnote{%
 \majorheadroom
 \Cf\ The international SAT Competitions web page
 \url{http://www.satcompetition.org/}.} 
However, the
most cumbersome aspect remained: the systematic but blind replacement
of variables by terms from the \herbranduniverse. 
Could we not
find these replacements in a more goal-directed and intelligent way?

\yestop\yestop\yestop\noindent
The first step in that direction was done by 
\citet{davis-1963} in a
method called linked conjuncts, where the substitution was cleverly
chosen, so that it generated the desired tautologies more directly. 
And this idea finally led to the seminal resolution principle
discovered by 
\robinsonname\ \robinsonlifetime, 
which dominated the field ever since.\footnote{%
 \majorheadroom
 \Cfnlb\ \citep{resolution}.}

This technique --- called a 
\index{logic!machine-oriented}%
{\em machine-oriented}\/ logic 
by 
\robinsonindex\robinson\ --- dispenses with the systematic replacement from the
\herbranduniverse\ altogether and finds the proper
substitutions more directly by an ingenious combination of Cut
and a 
\index{unification algorithm|(}%
unification algorithm. 
It works as follows: first transform
the formula \nlbmath E
% $A\cup\{\neg T\}$ 
into disjunctive normal form, 
\ie\ into a disjunctive set of conjunctions. \ 
%% CP: The following is nonsense! Sorry for the emotions!
%% \fullstopnospace\footnote{Essentially this is still based on
%%   \herbrand's Rules of Passage, albeit far more efficient variants
%%   are known today; \Cfnlb\ \citep{HandbookAR}.} 
Now suppose that the following elements are in this disjunctive set:
\par\noindent\LINEnomath{\bigmaths{K_1\und\ldots\und K_m\und L}{} \ and \ 
\bigmaths{\neg L\und M_1\und\ldots\und M_n}.}\par\noindent
%A Cut (\ie\ a generalization of {\em modus ponens})
%allows us to infer the following 
Then we can add their {\em resolvent} 
\par\noindent\LINEmaths{K_1\und\ldots\und K_m\und M_1\und\ldots\und M_n}{}
\par\noindent
to this disjunction,
simply because one of the previous two must be true if the resolvent is true.

Now suppose that the literals $L$ and $\neg L$ 
% are not sentential, 
% CP: ground?
are not 
% already
%CP 20140424:
yet 
complementary 
because they still contain variables, for example
such as 
\math{P(x, f(a, y))} and \math{\neg P(a, f(z, b))}. \ 
It is easy to see, that these two atoms can be made equal, if we
substitute $a$ for the variables $x$ and $z$ and the constant $b$ for
the variable $y$. \ 
The most important aspect of the 
\index{proof search|)}%
\index{proof search!automatic|)}%
\index{theorem proving!automated|)}%
resolution principle is
that this substitution can be computed by an algorithm, which is
called \emph{unification}. \ 
Moreover, there is always \emph{at most
  one} (up to renaming) most general substitution which unifies two
atoms, and this single unifier stands for the potentially infinitely many 
instances from the 
\index{Herbrand universe|)}%
\herbranduniverse\ that would be generated otherwise. 
 
\yestop\yestop\yestop\noindent
\robinsonindex\robinson's original unification algorithm is exponential in
time and space. 
The race for the fastest \emph{unification algorithm}
lasted more than a quarter of a century and resulted in a linear algorithm
and unification theory became a (small) subfield of computer science,
artificial intelligence, logic, and universal algebra.\footnote{%
 \majorheadroom
 \Cfnlb\ \siekmannindex\citep{siekmann-1989} for a survey.} 

\pagebreak

Unification theory had its heyday in the late 1980s, 
when the Japanese challenged the Western
economies with the 
\index{program@\programme!Fifth Generation Computer}%
``Fifth Generation Computer \Programme'' which was
based among others on logical programming languages. \ 
The 
%CPU (central processing unit) 
processors
of these machines realized an ultrafast unification
algorithm cast in silicon, whose performance was measured not in MIPS
(machine instructions per second), as with standard computers, but in
LIPS (logical inferences per second, which amounts to the number of
unifications per second). \ 
A \nolinebreak myriad of computing machinery was built
in special hardware or software on these new concepts and most
industrial countries even founded their own research laboratories to
counteract the Japanese challenge.\footnote{%
 Such as the European Computer-Industry Research Center (ECRC),
 which was supported by the French company Bull, 
 the German Siemens company, and the British company ICL.}

\yestop\yestop\yestop\yestop\noindent
Interestingly, {\herbrandname} had seen the concept of a unifying
substitution and an algorithm already in his thesis in\,1929. \ 
Here is his account in the original French idiom:
\nopagebreak
\yestop\yestop\begin{quote}
\frenchtextfivehundredwithfrontandend
{``}
{''\footnote{\label{note translation unification algorithm}%
 \majorheadroom
 \Cfnlb\ 
 \frenchtextfivehundredsource. \ 
 \frenchtextfivehundredsourcemodifier
 \begin{quote}\begin{enumerate}\item[``\relax 1.]
 \englishtextfivehundredwithoutfrontandend''
 \notop\notop\end{enumerate}\end{quote}}}%
\index{unification algorithm|)}%
\end{quote}

\vfill\pagebreak

\section{Conclusion}

With regard to students interested in logic,
in the previous lectures we have
presented all major contributions of \herbrandname\ to logic, 
and our 150~notes give hints on where to continue studying.

With regard to historians,
we uncovered some parts of the historical truth on \herbrand\
which was varnished by contemporaries such as \goedel\ and \heijenoort.
It was already well-known that \goedel's memories on 
\herbrand's recursive functions were incorrect,
but to the best of our knowledge 
the errors of the reprint of \herbrand's \PhDthesis\ 
\cite{herbrand-PhD} \hskip.2em
in \cite{herbrand-ecrits-logiques} have not been noted before. \ 
The English translation in \cite{herbrand-logical-writings}
is based on this contorted reprint: \
The advantage of working with the original prints
should become obvious from 
a \nolinebreak comparison of our translation of \herbrand's unification
algorithm in \noteref{note translation unification algorithm}
with the translation in \cite{herbrand-logical-writings}.

With regard to logicians, however, notwithstanding our above critique,
the elaborately commented 
book \cite{herbrand-logical-writings} \hskip.2em 
is a great achievement and still the best
source on \herbrand\ as a logician 
(\cfnlb\ \noteref{note on herbrand-logical-writings}), \
and 
our lectures would not have been possible
without \heijenoort's most outstanding and invaluable contributions
to this subject. \
To the best of our knowledge,
what we called
{\em\heijenoort's \repair} of \herbrandsfalselemma\
has not been published before, 
and we have included it into our version of 
\herbrandsfundamentaltheorem.
The consequences of this \repair\ on \herbrand's
\index{modus ponens!elimination}%
{\em Modus Ponens}\/ elimination
(as described in \nlbsectref{section modus ponens elimination}) \hskip.2em
are most relevant still today and should become part of the 
standard knowledge on logic, just as \gentzen's 
\index{Cut elimination}%
Cut elimination.

While \herbrand's important work on 
decidability and consistency of arithmetic was
soon to be topped by \presburger\ and \gentzen,
his \fundamentaltheorem\ will remain of outstanding historical
and practical significance. \
Even under the critical assumptions 
(\cfnlb\ the discussion 
 in \nlbsectref{section herbrand loewenheim skolem}) \hskip.3em
that \herbrand\ took
the outer \skolemizedform\ from \citep{skolem-1928} \hskip.2em
and that he had realized that the presentation in 
\loewenheimindex\citep{loewenheim-1915}
included a sound and complete proof procedure, \hskip.3em
\herbrandsfundamentaltheorem\ remains a 
truly remarkable creation.

All in all, \herbrandname\ has well 
deserved to be the idol that he actually is. \
And thus we were surprised to find out 
how little is known on his personality and life, and
that there does not seem to be anything like a 
\herbrand\ memorial or museum,
nor even a \nolinebreak street named after him,
nor a decent photo of him available in the 
Internet.\footnote{%
 % \majorheadroom
 The best photos of \herbrand\ currently to be found in the Internet
 seem to be the one of 
 \figurefss{figure herbrand}{figure herbrand two}{figure herbrand three}. \
 Outside mathematics, Google hits on \herbrand\ typically refer to 
 P. Herbrand \& Cie., 
 a \nolinebreak historical street-car production company in \Cologne; \hskip.2em
 or else to 
 \herbrand\ Street close to \russell\ Square in \London, 
 probably named after {\namefont\herbrand\ Arthur \russell}, 
 the \nth{11} Duke of Bedford.} \
Moreover, a \nolinebreak careful bilingual edition of \herbrand's complete works
on the basis of the elaborate previous editorial achievements
is in high demand.\footnote{%
 \majorheadroom
 \Cfnlb\ also \noteref{note on herbrand-logical-writings}.}
%% In our humble opinion, {\em la grande nation}\/ should 
%% save no efforts on an a better representation 
%% and appreciation of \herbrandname.

\vfill\pagebreak

\appendix
\vfill\begin{ack}\noindent
We would like to thank \anellisname,
\index{Bussotti, Paolo}%
\bussottiname, and
\index{Paulson, Lawrence C.}%
\paulsonname\
---~the second readers of our previous 
handbook article \cite{herbrand-handbook} on the same subject~---
for their numerous, extensive, deep,
and most valuable and helpful suggestions for improvement
of previous versions of this paper.
%We were able to improve the quality of this paper considerably with their 
%invaluable help.
As always, 
the second readers are, of course,
not responsible for the remaining weaknesses.

Furthermore, we would like to thank 
\dawsonname, 
% Note that the following name made problems with multiple entries
% in the index when used in the bib file.
% Thus, I have made sure otherwise that all references occur properly.
\index{Ferm\"uller, Christian G.}%
\fermuellername, 
\gramlichname, 
\index{Hallmann, Andreas}%
\hallmannname,
\index{Moser, Georg Ch.}%
\mosername,
\roquettename,
\index{Sattler-Klein, Andrea}%
\sattlername, 
\index{Smith, James T.}%
\smithname, and
\wolskaname\
for substantial help.\end{ack}
\vfill\vfill

\begin{figure}
\begin{center}
  \includegraphics
[bb=0 0 114 167,width=0.4\linewidth]
{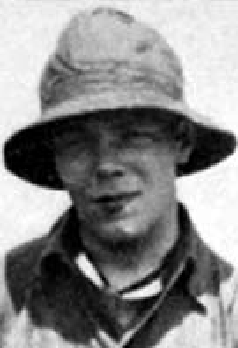}
\caption
{\label{figure herbrand}%
Photo of \herbrandname\ on the expedition during which he found his death.}
\yestop\yestop
\end{center}\end{figure}

\vfill\pagebreak

\nocite{herbrand-second,herbrand-third,herbrand-fourth,herbrand-note-hadamard}
% These are hardly useful abstracts of \herbrand's work on logic
\nocite{herbrand-note-PhD,chevalley-herbrand-lost-one,herbrand-lost-one,herbrand-lost-two,herbrand-lost-three}

%\bibliography{herbrandbib}

\yestop\yestop\yestop\yestop\mysection{Bibliography of Jacques Herbrand}

\begin{itemize}
\yestop\item
To the best of our knowledge,
the following bibliography of \herbrandname\ 
% (\jr)
is complete, with the exception of 
his letters in \citep[\Vol\,V, \PP{3}{25}]{goedelcollected}
and \citep{hasse-herbrand-correspondence}.

\yestop\item
Note that for labels of the form
{\mbox{}\raise.2ex\hbox{[}}Herbrand \ldots{\mbox{}\raise.2ex\hbox{]}},
we have maintained the standard labeling as
introduced by \heijenoortname\ in \makeaciteoftwo
{herbrand-ecrits-logiques}
{herbrand-logical-writings}
and \citep
{heijenoort-source-book}
as far as appropriate. \hskip.1em
For instance, \herbrand's thesis is cited as \cite{herbrand-PhD}
and not as 
{\mbox{}\raise.2ex\hbox{[}}\herbrand, 1930a{\mbox{}\raise.2ex\hbox{]}}, 
\hskip.2em
% which would be the standard style of this handbook and 
which would be in general preferable because of
the additional redundancy.
\end{itemize}

\yestop\yestop\yestop\yestop\mysubsection
{Joint Publications of \herbrandname}\notop\notop\notop\notop

\renewcommand\refname{}

\yestop\yestop\yestop\yestop\mysubsection
{Publications of \herbrandname\ as Single Author}\notop\notop\notop\notop

% ARIANA

\vfill\pagebreak
\mysection{References}

\footnotesize
\notop\notop\notop\notop\renewcommand\refname{}

\cleardoublepage

\index{Herbrand's ``False Lemma''|see{{\em also}\/\, correction}}%
\index{Heijenoort's correction|see{correction}}%
\index{Goedel@G\"odel!'s a@'s and Dreben's correction|see{correction}}%
\index{Hilbert!'s program@'s \programme|see{\programme, Hilbert's}}%
\index{Hilbert!'s finitism|see{finitism}}%
\index{intuitionism|see{{\em also}\/\, finit\-ism {\em and}\/\, logic, intuitionistic}}%
\index{logic!intuitionistic|see{{\em also}\/\, intuitionism}}%
\index{axiom!of choice!weak forms|see{choice,Principle of Dependent Choice, K\oe nig's Lemma}}%
\principleofdependentchoiceindexsee%
\koenigslemmaindexsee%
\index{consistency!proof of!Herbrand-style|see{Herbrand-style consistency proof}}%
\index{consistency!proof of!Gentzen's|see{Gentzen's \sloppy consistency proof}}%
\index{prenex|see{direction, prenex {\em and}\/\, form, prenex}}%
\index{anti-prenex|see{direction, anti-prenex {\em and}\/\, form, anti-prenex}}%
\index{Skolem|see{{\em also}\/\, form, Skolem normal {\em and}\/\, form, Skolemized {\em and}\/\, function, Skolem {\em and}\/\, L\oe wenheim--Skolem Theorem {\em and}\/\, L\oe wenheim--Skolem--Tarksi Theorem {\em and}\/\, Skolemization {\em and}\/\, term, Skolem}}%
\index{identity|see{{\em also}\/\, tautology}}%
\index{Schr\"oder|see{{\em also}\/\, Peirce--Schr\"oder tradition}}%
\index{paradox|see{Russell's Paradox {\em and}\/\, Skolem's Paradox}}%
\index{delta@\math\delta!-variable|see{variable, \math\delta-}}%
\index{delta@\math\delta!-quantifier|see{quantifier, \math\delta-}}%
\index{gamma@\math\gamma!-variable|see{variable, \math\gamma-}}%
\index{gamma@\math\gamma!-quantifier|see{quantifier, \math\gamma-}}%
\index{Hilbert!-style calculi|see{{\em also}\/\, Hilbert, school, calculi of}}%
\index{uniform notation|see{notation, uniform}}%
\index{Smullyan!'s uniform notation|see{notation, uniform}}%
\index{inference!rules of|see{Rule}}%
\index{sentential tautology|see{tautology, sentential}}%
\index{deep|see{inference, deep}}%
\index{elimination|see{Modus Ponens elimination, Cut elimination}}%
\small
\addcontentsline{toc}{section}{Index}
\printindex

\vfill\pagebreak
\begin{figure}
\begin{center}
  \includegraphics
[bb=0 0 421 619,width=0.9\linewidth]
{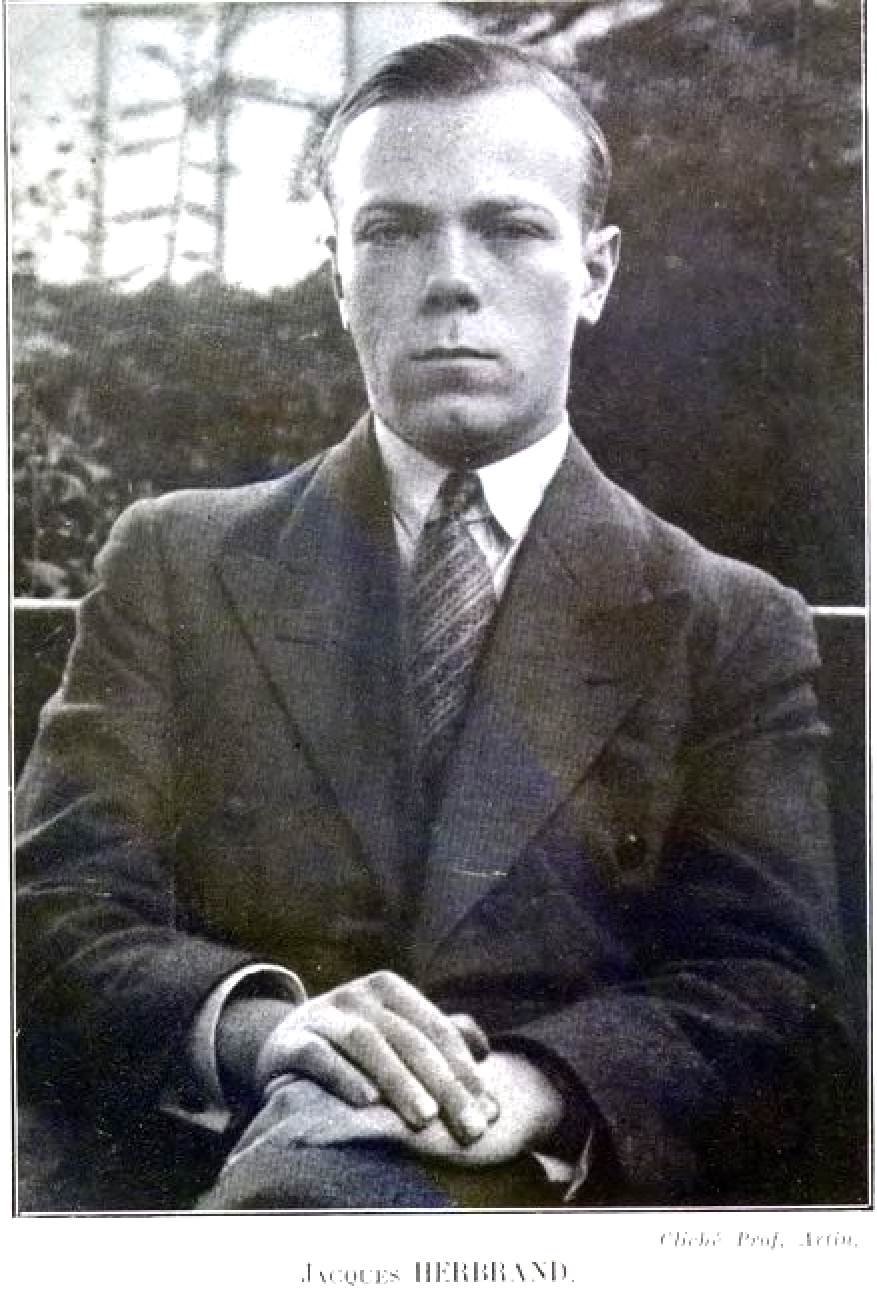}
\caption
{\label{figure herbrand two}%
Photo of \herbrandname\ by \artinname.}
\herbrandindexend
\yestop\yestop
\end{center}\end{figure}

\begin{figure}
\begin{center}
  \includegraphics
[bb=0 0 494 759,width=0.87\linewidth]
{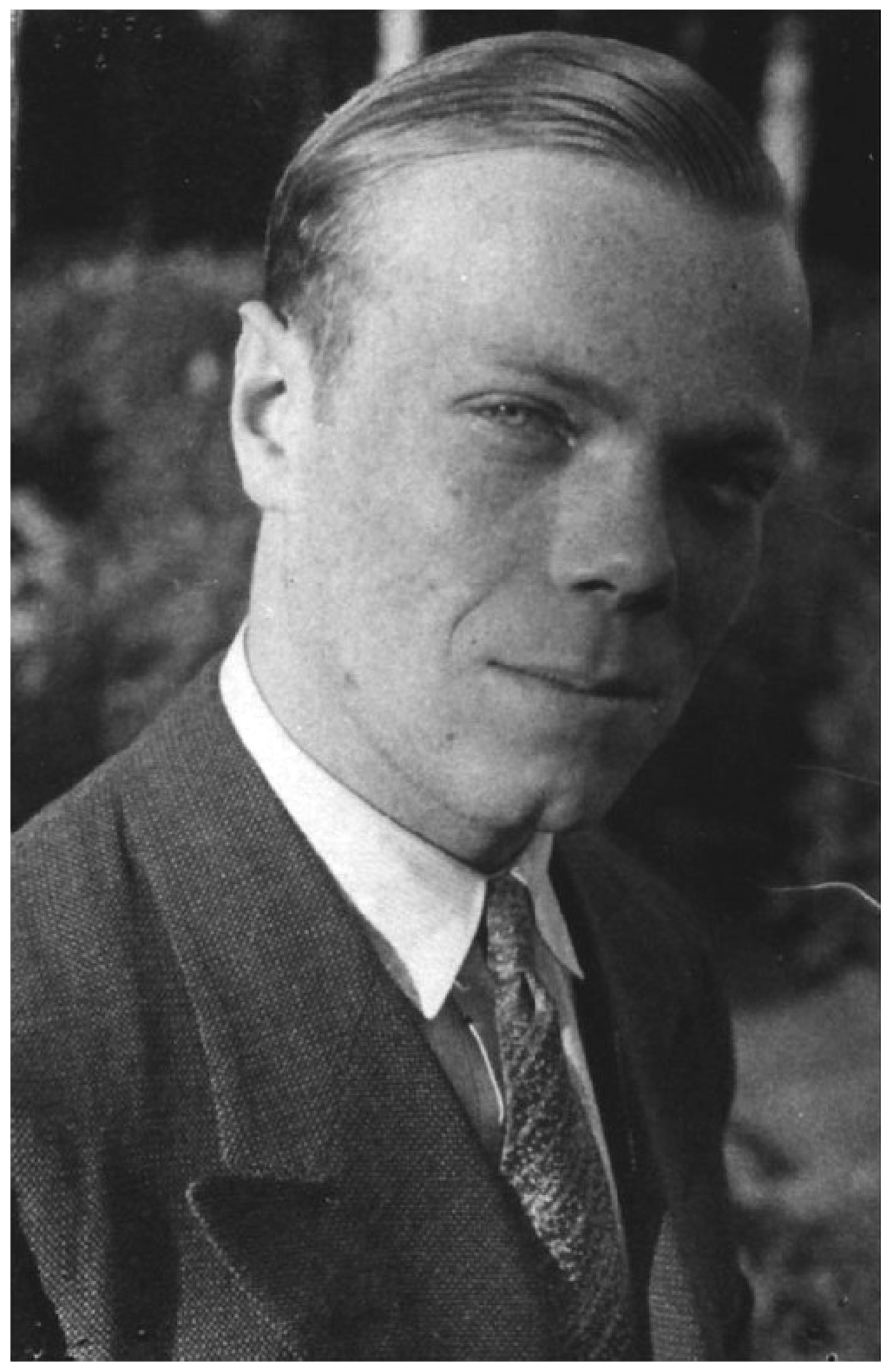}
\caption
{\label{figure herbrand three}%
Photo of \herbrandname, probably by \artinname.}
\yestop\yestop
\end{center}\end{figure}

\end{document}